\documentclass[prb,twocolumn,superscriptaddress,citeautoscript]{revtex4-1}
\usepackage{cmap}           
\usepackage{graphicx}       
\usepackage{amsmath}
\usepackage{amstext}
\usepackage{amssymb}
\usepackage[caption=false]{subfig}
\usepackage{dcolumn}       
\usepackage{bm}             
\usepackage{lipsum}
\usepackage{nameref}
\usepackage{dsfont}
\usepackage{float}
\usepackage{standalone}
\usepackage{mathtools}
\usepackage{color}
\usepackage{epstopdf}
\usepackage[normalem]{ulem}
\usepackage[breaklinks=true,colorlinks=true,linkcolor=blue,urlcolor=blue,citecolor=blue]{hyperref}
\hypersetup{%
	bookmarksopen=true, %
	bookmarksopenlevel=1, %
	pdfstartview={XYZ null null 1.25}
}

\newcommand{\ra}{\rangle}
\newcommand{\la}{\langle}
\newcommand{\ua}{\uparrow}
\newcommand{\da}{\downarrow}

\newcommand{\sgn}{\operatorname{sgn}}
\renewcommand{\Im}{\operatorname{Im}}

\newcommand{\ii}{\mathrm{i}}
\newcommand{\bk}{\mathbf{k}}
\newcommand{\bq}{\mathbf{q}}

\newcommand{\br}{\mathbf{r}}
\newcommand{\brm}[1]{\bm{\mathrm{#1}}}
\newcommand{\pdag}{{\phantom{\dagger}}}
\newcommand{\e}{\mathrm{e}}
\newcommand{\D}{\mathrm{d}}

\newcommand{\ep}{$e$-$ph$}

\newcommand{\matlab}{\mbox{\textsc{Matlab}}}

\newcommand{\CodeRef}{The {\matlab} code is released at \url{https://github.com/johnstonResearchGroup/Migdal}}

\begin{document}
	
	
	\title{Temperature-filling phase diagram of the two-dimensional Holstein model 
		in the thermodynamic limit by self-consistent Migdal approximation}
	
	\author{P. M. Dee}
	\affiliation{Department of Physics and Astronomy, University of Tennessee, Knoxville, Tennessee 37996, USA}
	\affiliation{Joint Institute for Advanced Materials, University of Tennessee, Knoxville, Tennessee 37996, USA}
	
	\author{K. Nakatsukasa}
	\affiliation{Department of Physics and Astronomy, University of Tennessee, Knoxville, Tennessee 37996, USA}
	\affiliation{Joint Institute for Advanced Materials, University of Tennessee, Knoxville, Tennessee 37996, USA}
	
	\author{Y. Wang}
	\affiliation{D{\'e}partement de Physique and Institut Quantique, Universit{\'e} de Sherbrooke, Sherbrooke, Qu{\'e}bec J1K 2R1, Canada}
	
	\author{S. Johnston}
	\affiliation{Department of Physics and Astronomy, University of Tennessee, Knoxville, Tennessee 37996, USA}
	\affiliation{Joint Institute for Advanced Materials, University of Tennessee, Knoxville, Tennessee 37996, USA}
	
	\date{\today}
	
	\begin{abstract}
	We study the temperature-filling phase diagram of the single-band Holstein model in two dimensions using the self-consistent Migdal approximation, where \emph{both} the electron and phonon self-energies are treated on an equal footing.  By employing an efficient numerical algorithm utilizing fast Fourier transforms to evaluate momentum and Matsubara frequency summations, we determine the charge-density-wave (CDW) and superconducting transition temperatures in the thermodynamic limit using lattice sizes that are sufficient to eliminate significant finite size effects present at lower temperatures.  We obtain the temperature-filling phase diagrams for a range of coupling strengths and phonon frequencies for the model defined on a square lattice with and without next-nearest neighbor hopping.  We find the appearance of a superconducting dome with a critical temperature that decreases before reaching the $\bq_\text{max} = (\pi,\pi)$ CDW phase boundary.  For very low phonon frequencies, we also find an incommensurate CDW phase with the ordering vector $\bq_\text{max} \approx (\pi,\pi)$ appearing between the commensurate CDW and superconducting phases.  Our numerical implementation can be easily extended to treat momentum-dependent electron-phonon coupling, as well as dispersive phonon branches, and has been made available to the public. 
	\end{abstract}
	
	\pacs{Valid PACS appear here}
	
	\maketitle
	
	\section{Introduction} \label{sec:intro}
	The electron-phonon ({\ep}) interaction drives many physical phenomena and plays a central role in a wide range of solids.    For example, it leads to the formation of lattice polarons at large {\ep} coupling~\cite{Polaron}; it is a significant factor in determining the electronic and thermal transport properties of many functional materials; it can drive broken symmetry states such as charge-density-wave (CDW) order~\cite{Gruner1988} or conventional superconductivity~\cite{Bardeen1957,Scalapino1966,Bardeen1973,Marsiglio2008} with low~\cite{Hamlin2015} and high $T_c$\cite{Buzea2001,Nagamatsu2001,Drozdov2015,Mazin2015a,Drozdov2018,Kruglov2018}; and, as recently demonstrated, it can even stabilize and control the location of Dirac cones in certain materials \cite{Moller2017}.  
	
	There has been remarkable progress in the accurate modeling of {\ep} interactions in realistic materials using \textit{ab-initio} methods~\cite{Baroni2001,Giustino2017} based on density-functional theory and density-functional perturbation theory.  Notwithstanding predictions for several superconducting transition temperatures $T_c$~\cite{Choi2002a,Luders2005,Marques2005,Floris2005,Margine2013}, \textit{ab initio} methods usually lack the capability of describing ordered phases or resolving competing orders and are often hindered by large computational costs.  Due to these limitations, many researchers turn to model Hamiltonian approaches, which capture the essential physics of the problem while remaining tractable and often easier to interpret.  For {\ep} coupled systems, the simplest model Hamiltonian is the Holstein model~\cite{Holstein1959}, which treats the motion of the ions using independent harmonic oscillators and the electron-lattice interaction as a purely local coupling between the electron density and lattice displacement.  
	
	Except for a couple of extreme cases, e.g., the use of a two-site system~\cite{Han2002,Berciu2007,Zhang2009b} or the atomic limit (with hopping $t=0$)~\cite{Mahan1990}, there are no exact analytical solutions for Holstein model.  Nevertheless, it has been widely studied using approximate analytical methods including the Lang-Firsov canonical transformation~\cite{Grzybowski2007}, diagrammatic expansions~\cite{Migdal1958, Eliashberg1960, Engelsberg1963, Scalapino1966, Marsiglio1990,Berger1995} based on many-body perturbation theory (MBPT)~\cite{Abrikosov1963},	and  variational methods.~\cite{Berciu2006,Berciu2007,Ebrahimnejad2012} The Holstein model has also been studied using several exact or approximate numerical techniques including quantum Monte Carlo (QMC)~\cite{Hirsch1982, Scalettar1989, Marsiglio1990, Noack1991, Vekic1992, Niyaz1993, Berger1995, Hohenadler2004, Huang2003, Johnston2013, Li2015b, Esterlis2018, Costa2018}, variational Monte Carlo (VMC)~\cite{Alder1997, Ohgoe2014, Ohgoe2017}, and dynamical mean-field theory (DMFT)~\cite{Freericks1993, Ciuchi1997, Freericks1998, Meyer2002, Capone2003, Hague2008, Bauer2011, Murakami2014}.  (Many of these numerical studies were conducted in the context of the Hubbard-Holstein model, or some other extension, where results for the pure Holstein model were obtained as a limiting case.) At half-filling, these studies find that the Holstein model is dominated by a $\bq = (\pi/a,\pi/a)$ CDW phase, while doping away from half-filling leads to a competition between CDW and superconducting instabilities.  Moreover, the transition temperatures for both phases vary as a function of the filling $n$, the phonon frequency $\Omega$, the dimensionless {\ep} coupling strength $\lambda$, and the Fermi surface (FS) topology.  Detailed phase diagrams for Holstein(-Hubbard) model for these parameters have been obtained by nonperturbative numerical methods in the spatial dimension $d=1$ by density matrix renormalization group (DMRG)~\cite{Tezuka2007}, $d=2$ by VMC~\cite{Ohgoe2017}, and $d=\infty$ Bethe lattice by DMFT~\cite{Murakami2014}.
	
	Owing to its simplicity and lower computational cost, the Migdal approximation~\cite{Migdal1958,Eliashberg1961} is routinely used to capture the effects of the {\ep} interaction in many materials.  For instance, it is often used to estimate the superconducting transition temperatures in metals with the {\ep} coupling matrix elements from \textit{ab initio} calculations~\cite{Giustino2017, Choi2002a, Margine2013, Aperis2015, Ponce2016}.  It is  also widely employed to estimate electronic structure renormalization in several materials.~\cite{Engelsberg1963,Cuk2004,Shai2013,Yang2016,Margine2016,Verdi2017,Swartz2018} In the Migdal approximation, the vertex corrections to the {\ep} coupling vertex are neglected.  This approximation is typically justified by arguing that such corrections scale as $\mathcal{O}\left(\lambda\frac{\hbar \Omega}{E_\text{F}}\right)$, where $E_F$ is the Fermi energy, and $ \lambda $ is a dimensionless factor quantifying the {\ep} coupling strength.  While this condition is satisfied by most metallic systems (because $\hbar\Omega / E_\text{F} \ll 1$, i.e., adiabatic limit), there is a growing number of materials where it is not, such as the fullerides~\cite{Grimaldi1995,Gunnarsson1997}, solid picene~\cite{Subedi2011,Casula2011}, $ n $-type SrTiO$_{3}$~\cite{Gorkov2016PNAS,Wolfle2018}, and monolayer FeSe on SrTiO$_{3}$~\cite{Gorkov2016PRB,Lee2014}.  Interest in these systems, as well as the wide-ranging applications of the Migdal approximation, has invited intense scrutiny on its	range of validity, resulting in many studies comparing its predictions with those of nonperturbative numerical methods such as QMC and DMFT~\cite{Marsiglio1990, Meyer2002, Capone2003, Hague2008, Bauer2011, Esterlis2018}.  It is found that the Migdal approximation can still break down at large $\lambda$~\cite{Benedetti1998, Alexandrov2001, Capone2003, Hague2008, Bauer2011, Esterlis2018}, even within the adiabatic limit, near half filling.  The effort to map out the validity of the Migdal approximation as a function of the {\ep} coupling strength, phonon frequency, and the electron filling is ongoing~\cite{Marsiglio1990,Bauer2011,Esterlis2018}.
	
	The studies comparing exact numerical methods and MBPT are usually limited to relatively small lattice sizes for QMC or $d=\infty$ (Bethe lattice) for DMFT and frequently are done at or close to half filling.  It is, therefore, essential that we assess the finite-size effects for both the nonperturbative numerical methods and MBPT, and then extrapolate to the thermodynamic limit.  But despite the large body of work surrounding the Holstein model, there is (to the best of our knowledge) no comprehensive study of the general temperature-filling phase diagrams by the full-fledged MBPT in the thermodynamic limit.  Motivated by this, we carried out a detailed study of the single-band two-dimensional Holstein model on a square lattice calculated with the self-consistent Migdal approximation, where the electron and phonon self-energies are both determined self-consistently~\cite{Marsiglio1990}.

	Our implementation is based on the fast Fourier transform (FFT) between the momentum-frequency coordinates and the position-time coordinates, which is often used in fluctuation-exchange (FLEX)~\cite{Serene1991,Deisz2002} and DMFT calculations~\cite{Georges1996}.  We show that the predicted CDW transition temperature $T_c^\text{CDW}$ exhibits significant finite-size effects.  To mitigate this problem, we determined the transition temperatures using large lattice sizes well beyond those used in previous studies to obtain a reliable extrapolation to the thermodynamic limit.  We also found that the superconducting transition temperature $T_c^\text{SC}$ exhibits non-monotonic behavior as a function of filling.  Specifically, we found that it increases gradually as a function of filling $n$ until it approaches a $\bq_\text{max} = (\pi,\pi)/a$ CDW phase boundary, where it is suppressed by competition with the CDW, leading to dome-like behavior.  
	
	The paper is organized as follows.  Section~\ref{sec:model_method} provides details of the Holstein model and its extension for momentum dependent coupling, inclusive of the expressions for the self-energies in the Migdal approximation.  Section~\ref{sec:computation} documents the computational details and the implementation of our numerical algorithm.  Section~\ref{sec:results} presents the results for the temperature-filling phase diagrams for the Holstein model and other related cases. Section~\ref{sec:conclusions} summarizes our conclusions.  Finally, we have made our code publicly available as a set of {\matlab} functions and scripts~\footnote{\CodeRef}.  The details provided in this article should be sufficient to implement the algorithm in other programming languages. 
	
	\section{Models and Methods} \label{sec:model_method}
	\subsection{Holstein Model}
	
	The Holstein Hamiltonian describes the electronic degrees of freedom using a single band tight-binding model.  The lattice degrees of freedom are modeled using independent harmonic oscillators at each site with a spring constant $K = M\Omega^2$, where $M$ is the ion mass, and $\Omega$ is the bare frequency of the oscillator.  The {\ep} interaction is introduced as a purely local coupling between the electrons and the atomic displacement.  In real space, the Hamiltonian is
	\begin{align}
	\hat{H} &= -\sum_{i\neq j,\sigma} t^\pdag_{ij} \hat{c}^\dagger_{i,\sigma}\hat{c}^\pdag_{j,\sigma}
	-\mu\sum_{i,\sigma} \hat{n}_{i,\sigma} \notag\\
	&\quad + \sum_i \left[\frac{\hat{P}_i^2}{2M} + \frac{K \hat{X}^2_i}{2} \right]
	+ \alpha \sum_{i,\sigma} \hat{X}_i \left(\hat{n}_{i,\sigma} - \frac{1}{2}\right),
	\label{Eq:HamiltonianRealSpace}
	\end{align}
	where $\hat{c}^\dagger_{i,\sigma}$ ($\hat{c}^\pdag_{i,\sigma}$) creates (annihilates) an electron with spin $\sigma={\ua}$ or $\da$ on site $i$, $t_{ij}$ is the hopping integral between sites $i$ and $j$, $\mu$ is the chemical potential, $\hat{n}_{i,\sigma} = \hat{c}^\dagger_{i,\sigma}\hat{c}^\pdag_{i,\sigma}$ is the electron number operator, $\hat{X}_i$ and $\hat{P}_i$ are the lattice displacement and momentum operators, respectively, and $\alpha$ is the {\ep} coupling strength.  Throughout this work, we restrict the range of the hopping to nearest-neighbor (NN) ($ t$) and next-nearest-neighbor (NNN) hopping ($t^\prime$), only. For $t^\prime=0$, the Hamiltonian is particle-hole symmetric about half-filling, so we only consider $0\leq n \leq 1$; for $t^\prime \neq 0$, we consider $0\leq n \leq 2$. Here, $n = \sum_{\sigma}\la \hat{n}_{i,\sigma} \ra$ is  the electron filling.  
	
	Fourier transforming the operators and introducing second quantized forms for the lattice operators $\hat{X}_i$ and $\hat{P}_i$ yields
	\begin{align}
	\hat{H} &=   \sum_{\bk,\sigma} \xi^\pdag_{\bk} \hat{c}^\dagger_{\bk,\sigma}\hat{c}^\pdag_{\bk,\sigma}
	+ \hbar\Omega \sum_{\bq}\left( \hat{b}^\dagger_\bq \hat{b}^\pdag_\bq + \frac{1}{2} \right) \notag \\
	&\quad + \frac{1}{\sqrt{N}}\sum_{\bk,\bq,\sigma} g \hat{c}^\dagger_{\bk+\bq,\sigma}\hat{c}^\pdag_{\bk,\sigma}
	\left(\hat{b}^\dagger_{-\bq}+\hat{b}^\pdag_\bq\right),
	\label{Eq:Hamiltonian}
	\end{align}
	where $N$ is the number of lattice sites, $g = \alpha\sqrt{\frac{\hbar}{2M\Omega}}$, and $\xi_{\bk} = \epsilon_{\bk}- (\mu-\tilde{\mu})$ is the band dispersion measured relative to the chemical potential. The additional constant $\tilde{\mu} = \frac{\alpha^2}{K} = \frac{2g^2}{\hbar \Omega}$ arises from the fact that the displacement is coupled to $\hat{n}_{i,\sigma} - \frac{1}{2}$ instead of $\hat{n}_{i,\sigma}$ in Eq.~(\ref{Eq:HamiltonianRealSpace}).  This shift restores the condition that $\mu = 0$ corresponds to  half-filling when $t' = 0$, even in the {\ep} coupled case. Physically, it amounts to shifting the zero of the lattice displacement operator $\hat{X}_i - \frac{\alpha}{K}\to \hat{X}_i$ when going from Eq.~(\ref{Eq:HamiltonianRealSpace}) to Eq.~(\ref{Eq:Hamiltonian}). 
	
	In what follows, we work on a two-dimensional (2D) square lattice with NN hopping $t$ and NNN hopping $t^\prime$.  The resulting electron band dispersion is $\epsilon_{\bk} = -2t\left[\cos(k_xa)+\cos(k_ya)\right] - 4t'\cos(k_xa)\cos(k_ya)$, where the bandwidth $W = 8t$ when $ |t^{\prime}|\leq 0.5t $.  We use a conventional definition for the dimensionless {\ep} coupling $\lambda = 2g^{2}/(W\Omega)$ to facilitate	easy comparisons with QMC calculations.  Finally, we set our choice of units so that $\hbar = k_\text{B} = a = M = 1$, where $ \hbar $ is the reduced Planck constant, and $k_\text{B}$ is the Boltzmann constant.
	
	\subsection{Momentum-Dependent Interactions} \label{subsec:qDependentInteractions}
	Our algorithm can treat momentum dependent interactions, wherein the {\ep} coupling $g$ in Eq.~(\ref{Eq:Hamiltonian}) depends on the phonon wavevector $\bq$ with $|g(\bq)|^{2} = g^{2}f(\bq)$, where  $ f(\bq) $ is a shape function.  Physically, such coupling constants arise when the {\ep} interaction couples the electron density to neighboring atomic displacements.  Motivated by the {\ep} coupling to the oxygen phonon modes in the high-$T_c$ cuprates~\cite{Bulut1996,Sandvik2004,Johnston2010b}, we will consider three different cases 
	\begin{align}
	f(\bq) &=
	\begin{cases}
	1  &(\text{Isotropic}), \\
	\cos^{2}\left(\frac{q_{x}}{2}\right)+\cos^{2}\left(\frac{q_{y}}{2}\right) &(\text{Buckling}),\\
	\sin^{2}\left(\frac{q_{x}}{2}\right)+\sin^{2}\left(\frac{q_{y}}{2}\right) &(\text{Breathing}).
	\end{cases}
	\end{align}
	The ``isotropic'' case corresponds to the conventional Holstein model. The ``buckling'' case approximates the {\ep} vertex expected for $c$-axis polarized Cu-O bond-buckling modes for in the cuprates while the ``breathing'' case approximates the momentum dependence expected for the Cu-O bond-stretching modes.
	
	For a general momentum dependent {\ep} coupling, we define the dimensionless {\ep} coupling constant as $ \lambda=[2g^{2}/(W\Omega)]\langle f(\bq)\rangle$, where the buckling and breathing mode cases share the following symmetry:  $\langle f(\bq)\rangle= N^{-1}\sum_{\bq\in\text{FBZ}}f(\bq)\approx \frac{1}{(2\pi)^{2}}\int_{-\pi}^{\pi}\D^{2}q\,f(\bq) =1 $. 
	
	\subsection{Self-consistent Migdal Approximation}
	In this section we describe how the electron and phonon self-energies are computed self-consistently~\cite{Marsiglio1990}. For convenience, we adopt the 4-vector notation $k \equiv (\bk,\ii\omega_{n})$ and $q \equiv (\bq,\ii\nu_{m})$ for the momentum-(Matsubara) frequency coordinates and $x \equiv (\br,\tau)$ for the position-(imaginary) time coordinates.  The fermionic and bosonic Matsubara frequencies are given by $\omega_{n}=(2n+1)\pi T$ and  $\nu_{m}=2m\pi T$, respectively, with $n,m\in\mathds{Z}$.  The imaginary time is constrained to the range $ \tau\in[0,\beta] $, where $ \beta=1/T $ is the inverse temperature.
	
	The dressed single-particle electron Green's function $G(k)$ can be expressed using Dyson's equation as
	\begin{align} \label{Eq:G}
	G(k) &= \left[G_0^{-1}(k) - \Sigma(k)\right]^{-1} \notag \\
	&= \left[\ii\omega_{n} - \xi_\bk - \Sigma(k)\right]^{-1}, 
	\end{align}
	where $G_0(k) = (\ii\omega_{n} - \xi_\bk)^{-1}$ is the bare electron Green's function and $\Sigma(k)$ is the
	electron self-energy. The dressed phonon Green's function $D(q)$ is similarly given by
	\begin{align}
	D(q) &= \left[D_0^{-1}(q) - \Pi(q)\right]^{-1} \notag \\ 
	&= -\left[\frac{\nu_m^2+\Omega^2}{2\Omega}  + \Pi(q)\right]^{-1},
	\end{align}
	where $D_0(q) = -2\Omega/(\nu_m^2+\Omega^2)$ is the bare phonon Green's function and $\Pi(q)$ is the
	phonon self-energy.
	
    		\begin{figure}
				\includegraphics[width=\columnwidth]{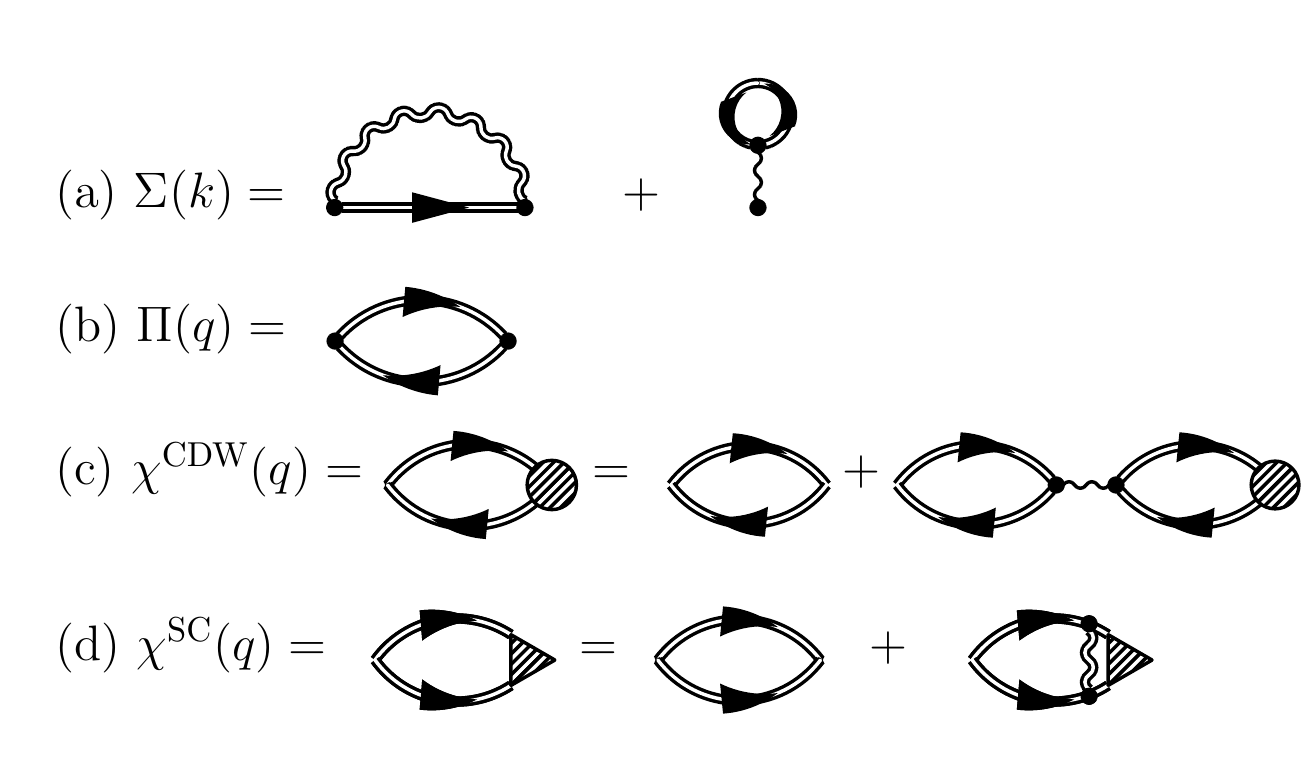}
				\caption{The Feynman diagrams for (a) the electron self-energy $\Sigma(k)$,  (b) phonon self-energy $\Pi(q)$, (c) CDW
					susceptibility $\chi^\text{CDW}(\bq)$, (d) and pairing susceptibility $\chi^{SC}$, evaluated within the
					self-consistent Migdal approximation. The lines (double lines) represent bare (dressed) electron propagators
					$G_0$ ($G$); the wiggly lines (double-wiggly lines) represent bare (dressed) phonon propagators $D_0$ ($D$).
					The black dot represents the bare {\ep} coupling vertex. The equation in (c) defines a series of particle-hole 
					ring diagrams and the recurrence equation in (d) defines a series of particle-particle ladder
					diagrams. }
				\label{fig:Dressed_GF}
			\end{figure}
	
	The skeleton diagram for the electron self-energy within Migdal approximation is shown in Fig.~\ref{fig:Dressed_GF}(a).  Since the vertex corrections are neglected, the full electron self-energy includes only two terms $\Sigma(k)=\Sigma^{\text{F}}(k)+\Sigma^\text{H}$.  The Fock term $\Sigma^{\text{F}}(k)$ includes all non-crossing ``rainbow'' Feynman diagrams if the dressed $G$-skeleton is fleshed out with the self-energy diagrams, and the Hartree term $\Sigma^{\text{H}}$ includes the dressed $G$-skeleton but with the bare phonon propagator $D_0$ instead of $ D $ to avoid double counting.  According to rules for the Feynman diagrams, we have
	\begin{equation}
	\Sigma^{\text{F}}(k) = -\frac{1}{N\beta}\sum_{q} |g(\bq)|^{2}D(q)G(k-q), \label{Eq:MigdalSE}
	\end{equation}
	and
	\begin{eqnarray}\nonumber
	\Sigma^{\text{H}}  &=&\frac{2|g(\mathbf{0})|^{2}}{N\beta}D_{0}(0)\sum_{k'} G(k') \e^{\ii\omega_{n'}0^{+}} \notag \\
	&=&|g(\mathbf{0})|^{2}D_{0}(0) n, \label{Eq:HartreeSE}
	\end{eqnarray}
	where
	\begin{align} \label{Eq:fill}
	n = \frac{2}{N\beta}\sum_{k'} G(k') \e^{\ii\omega_{n'}0^{+}} = 2 G(\mathbf{r=0},\tau=0^{-}).
	\end{align}
	
	The Hartree term is independent of $k$ and thus a constant that is typically absorbed into the definition of the chemical potential $\mu - \tilde{\mu} \to \mu - \tilde{\mu} - \Sigma^\text{H}$.  Here, we refrain from this practice to facilitate easier comparisons to the chemical potentials used in QMC methods,	which include all Feynman diagrams.  Note that at half-filling ($n=1$), $\Sigma^\text{H} = -\tilde{\mu} = -2|g(\mathbf{0})|^2/\Omega$.
	
	The skeleton diagram for the phonon self-energy within Migdal approximation is shown in
	Fig.~\ref{fig:Dressed_GF}(b) and has the analytical form
	\begin{align}
	\Pi(q) &= \frac{2|g(\bq)|^2}{N\beta}\sum_{k}G(k)G(k+q). \label{Eq:PhononSE}
	\end{align}
	It is useful to use $ \Pi(q) $ to define the irreducible charge susceptibility
	\begin{align}
	\chi_0(q) &= -\frac{\Pi(q)}{|g(\bq)|^2} = -\frac{2}{N\beta}\sum_{k}G(k)G(k+q),
	\end{align}
	which diagrammatically corresponds to Fig.~\ref{fig:Dressed_GF}(b) with two {\ep} coupling vertices removed
	[i.e., the first term on the right-hand side of the equation in Fig.~\ref{fig:Dressed_GF}(c)].

	\subsection{Charge-density-wave and Pairing Susceptibilities}
	The charge-density correlation at a wavevector $\bq$ is measured by the CDW susceptibility
	\begin{equation}
	\chi^\text{CDW}(\bq) = \frac{1}{N} \int_{0}^{\beta} \D\tau
	\left\la \hat{\rho}_\bq^{\pdag}(\tau)\hat{\rho}_\bq^{\dagger}(0) \right\ra_\text{c},
	\label{Eq:CDW_Susceptibility}
	\end{equation}
	where
	\begin{equation}
	\hat{\rho}_\bq(\tau) \equiv \sum_{i,\sigma} \e^{-\ii\bq\cdot\mathbf{R}_{i}}
	\hat{c}_{i,\sigma}^{\dagger}(\tau)\hat{c}_{i,\sigma}^{\pdag}(\tau).
	\end{equation}
	In Eq.(\ref{Eq:CDW_Susceptibility}), we have used the notation for the connected correlation function defined as $\la\hat{x}\hat{y}\ra_\text{c} =\la\hat{x}\hat{y}\ra - \la\hat{x}\ra\la\hat{y}\ra$.  When significant charge-density correlations are present on the lattice, the CDW susceptibility becomes strongly peaked at an ordering vector $\bq = \bq_\text{max}$.  For this reason, we will primarily use and discuss the momentum-space representation $ \chi^\text{CDW}(\bq) $.  For spin-singlet $s$-wave pairing due to {\ep} coupling, the superconducting correlations are measured by the pairing susceptibility
	\begin{equation}
	\chi^\text{SC} = \frac{1}{N} \int_{0}^{\beta} \D\tau
	\left \la \hat{\Delta}(\tau)\hat{\Delta}^{\dagger}(0)\right \ra,
	\label{Eq:SP_Susceptibility}
	\end{equation}
	where
	\begin{equation}
	\hat{\Delta}(\tau) \equiv \sum_{i}\hat{c}_{i,\ua}(\tau)\hat{c}_{i,\da}(\tau).
	\end{equation}
	
	In the thermodynamic limit, the temperatures at which the pair field and CDW susceptibilities diverge correspond to the transition temperatures $T_c^\text{SC}$ and $T_c^\text{CDW}$, respectively.  In the case of a $\bq = (\pi,\pi)$ CDW order, the temperature dependence of $ \chi^\text{CDW} $ should follow the 2D Ising universality class, which can be used to find $T_c^\text{CDW}$ [see Sec.~(\ref{subsec:TC_Extrapolation})].  By comparison, $ \chi^\text{SC} $ diverges much more sharply as a function temperature. In the latter case, we can obtain an accurate measure of $T_c^\text{SC} $ by extrapolating $1/\chi^\text{SC}(T)$ to zero.
	
	The CDW susceptibility within the Migdal approximation is obtained by summing the particle-hole ring diagrams shown in Fig.~\ref{fig:Dressed_GF}(c), which is formally identical to the random-phase approximation (RPA) or the $GW$ approximation for the Coulomb interaction in the electron gas.  The CDW susceptibility is largest at zero-frequency, so to determine $T_c^\text{CDW}$ we calculate
	\begin{align}
	\chi^{\text{CDW}}(\bq) &= \frac{\chi_{0}(\bq,0)}{1+|g(\bq)|^2 D_0(\bq,0) \chi_{0}(\bq,0)} \notag \\
	&= \frac{\chi_{0}(\bq,0)}{1-\lambda Wf(\bq)\chi_{0}(\bq,0)}. \label{Eq:Xcdw_q}
	\end{align}
	Here, we have used $D_0(\bq,0) = -2/\Omega$, $|g(\bq)|^2 = g^2f(\bq)$, and $\lambda = 2g^2/(W\Omega)$. In principle, the momentum dependence of a dispersive phonon mode $\Omega_\bq$ can be included in the function $f(\bq)$. This is not the case, however, for any nonzero Matsubara frequencies~\cite{Costa2018}.
	
	The pairing susceptibility within the Migdal approximation is obtained by summing the particle-particle ladder diagrams shown in Fig.~\ref{fig:Dressed_GF}(d) and is given by
	\begin{align}
	\chi^{\text{SC}} = \frac{1}{N\beta}\sum_{k}G(k)G(-k)\Gamma(k), \label{Eq:Xsc}
	\end{align}
	where the vertex function $\Gamma(k)$ is obtained by solving the vertex equation
	\begin{align}
	\Gamma(k) = 1 - \frac{1}{N\beta}\sum_{k^\prime}
	|g(\bq)|^{2}  G(k^\prime) G(-k^\prime) D(q) \Gamma(k^\prime), \label{Eq:scVertex}
	\end{align}
	where $q = k-k^\prime$.

	\section{Computational Details } \label{sec:computation}
	\subsection{Self-consistent Iterations with FFT}
	To obtain the dressed electron and phonon Green's functions, $G$ and $D$, we self-consistently solve Eqs. (\ref{Eq:G})--(\ref{Eq:PhononSE}), while the chemical potential $\mu$ is adjusted to fix	the filling $n$ after every iteration. Once the self-consistent solutions for $G$ and $D$ are obtained, we then evaluate the CDW and pairing susceptibilities using Eq.~(\ref{Eq:Xcdw_q}) and Eq.~(\ref{Eq:Xsc}), respectively. An independent self-consistency loop is performed to solve for the pairing vertex function in Eq.~(\ref{Eq:scVertex}) after the converged Green's functions are obtained. Both the momentum and Matsubara frequency summations in these equations can be viewed as convolutions and thus are evaluated efficiently using FFTs. 
	
\begin{figure}[tbh]
		\includegraphics[width=\columnwidth]{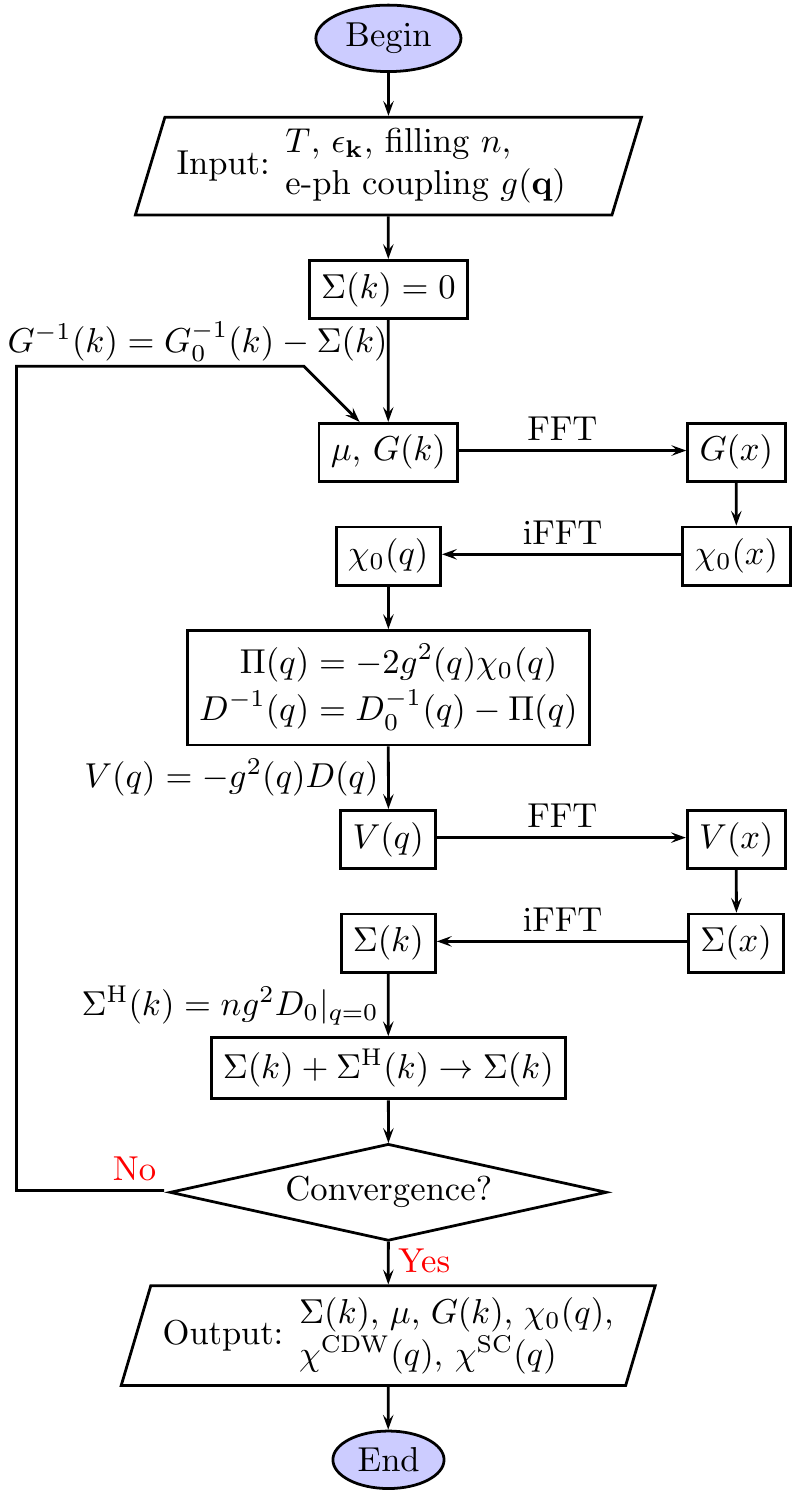}
		\caption{Algorithm for self-consistent iterations of Green's functions and self-energies with FFT.  The
			Hartree self-energy $\Sigma^\text{H} = n |g(\bq = \mathbf{0})|^2 D_0(q=0) = -2n |g(\mathbf{0})|^2/\Omega$ is
			a constant during iterations, and thus in practice the computation of $\Sigma^\text{H}$ and the shift of
			electron self-energy $\Sigma(k) \to \Sigma(k) + \Sigma^\text{H}$ can be moved outside the iteration loop and
			done after the convergence. }
		\label{fig:FlowChart}
	\end{figure}

	Our algorithm for self-consistent calculations of the Green's functions is summarized in the flowchart shown in Fig.~\ref{fig:FlowChart}. The input parameters include the temperature $T$, the filling $n$, the energy dispersion $\epsilon_{\bk}$, the phonon frequency $\Omega$, and the {\ep} coupling function $|g(\bq)|^2$ [or equivalently the dimensionless coupling strength $\lambda$ and the momentum dependent part of the coupling function $f(\bq)$].  We discretized the first Brillouin zone using a uniform $N=n_\bk \times n_\bk$ momentum grid, which corresponds to a square lattice with $N$ sites and periodic boundary condition in real space.  For computational purposes, the fermionic and bosonic Matsubara frequencies $ \omega_{n}=2\pi(2n+1)/\beta $ and $ \nu_{m} = 2\pi m/\beta $ are defined over a range defined by $-N_c \leq n,m \leq N_c -1$.  This cutoff corresponds to evenly dividing the imaginary time interval $0\leq \tau \leq \beta$ into $2N_c$ parts, with $\tau_l = (l-1)\beta/(2N_c)$, where $1\leq l\leq 2N_c+1$. For the electron Green's function $G(\tau)$, the end points should be understood as $0^+$ and $\beta^-$ due to the discontinuities of $G(\tau)$ at these points.  Here, $N_c$ is determined by $N_c =\omega_c\beta / (2\pi)$, where $\omega_c$ is an energy cut-off that is much larger than the band width $W$.  Most of the results obtained here used a cutoff $\omega_c\geq100\Omega $ which implies that the cutoff was close to $ W $ for $ \Omega=0.1t $.  We have checked that larger cutoffs produce no changes in the results.  When in doubt, we recommend a more conservative cutoff $ \omega_c\geq 10W  $. 
	
	The iteration loop begins with an initial value for electron self-energy $\Sigma(k) = 0$ (or the converged $\Sigma$ 
	at the previous temperature $T$ data point).
	The iteration loop then continues through the following steps in sequence: 
	\begin{enumerate}
		\item The dressed electron Green's 
		function $G(k)$ is computed by Dyson's equation with the chemical potential $\mu$ adjusted to fix 
		the filling $n$ at the input value. 
		\item The irreducible susceptibility is computed by 
		$\chi_0(x) = - G(x) G(-x) = G(\br,\tau) G(\br,\beta - \tau)$ with $G(x)$ obtained from $G(k)$ by
		FFT and the spatial inversion symmetry assumed. 
		\item The dressed phonon Green's function $D(q)$ is computed by the Dyson's
		equation with $\chi_0(q)$ obtained from $\chi_0(x)$ by the inverse FFT (iFFT) and then an effective
		interaction $V(q) = -|g(\mathbf{q})|^2 D(q)$ is computed and transformed to $V(x)$ by FFT.  
		\item A new electron self-energy $\Sigma(x) = V(x) G(x)$ is computed and transformed back to $\Sigma(k)$ by iFFT.  
	\end{enumerate}
	Note that the addition of the constant Hartree term $\Sigma^\text{H}$ to the self-energy is optional since $\mu$ is adjusted in every iteration.  The final step of the iteration loop checks for convergence using $\max|\Sigma_\text{new}(k) - \Sigma_\text{old}(k)| < \varepsilon$, where the absolute error is typically $\varepsilon =10^{-8}t $.  If the self-energy $\Sigma(k)$ is converged, final values for the Green's functions $G$ and $D$ are recomputed and used to find the CDW and pairing susceptibilities. If not, a new dressed electron Green's function is computed by Dyson's equation, and the iteration loop continues. The self-consistency condition can usually be achieved within 20--100 iterations.
	
	For the algorithm described above, three issues regarding our implementation deserve further remarks.  The first remark concerns the Fourier transform of variables $k=(\bk,\ii\omega_n) \xrightleftharpoons[\text{iFFT}]{\text{FFT}} x=(\br,\tau)$, where we need to consider the fact that the FFT only applies to discrete variables.  In our case, the transform between $\bk$ and $\br$ is straightforward for the lattice model, but the transform between $\omega_n$ and $\tau$ requires special care, especially for the electron Green's function. When transforming from $G(\omega_n)$ to $G(\tau)$, an infinite number of Matsubara frequencies must be summed to reproduce the discontinuity of $G(\tau)$ between $\tau = 0^{+}$ and $\tau = 0^{-}$.  We accomplish such a feat by approximating $G(\omega_n)$ with the bare electron Green's function $G_0(\omega_n)$ for $|\omega_n| > \omega_c$, and thus the sum of Matsubara frequencies $\omega_n$ with  $n\to \pm \infty$ can be carried out analytically.  Transforming $G(\tau)$ to $G(\omega_n)$ requires a Fourier integral transform over the continuous variable $0<\tau<\beta$. This Fourier integral is evaluated exactly following the interpolation of $G(\tau)$ using a continuous function such as spline or piecewise polynomial on the discrete $\tau$ grid.  Further technical details on this procedure are provided in Appendix~\ref{sec:FFT}.
	
	On a related note, if the effective interaction $V(\omega_n)$ is known and has a simple form, and if the self-energy $\Sigma(\omega_n)$ were to be calculated by a direct frequency sum with a frequency cut-off $\omega_c$, the analytical form of the high-frequency tail~\cite{Schrodi2018} of $\Sigma(\omega_n)$ can be obtained and should be added to the sum to minimize or eliminate the effect of the frequency cut-off. Our algorithm with FFT/iFFT offers two advantages to this scheme: not only is the high frequency tail of $\Sigma(\omega_n)$ generated correctly for any effective interaction $V(\omega_n)$ but the computation time is also significantly reduced by the FFT.

	The second remark concerns the use of the Anderson acceleration (also called Anderson mixing) algorithm~\cite{Anderson1965,Eyert1996,Walker2011}, which can be used to improve and accelerate the convergence of the simple fixed-point iteration scheme presented above.  Conceptually, Anderson mixing is a generalization of and an improvement to the simple mixing iteration method.  In terms of our problem, the simple mixing dictates that $\tilde{G}_\text{new} = \alpha G_\text{new} + (1-\alpha)G_\text{old}$ is used in the next iteration instead of $G_\text{new}$, where $0<\alpha<1$ is a constant. Using Anderson mixing, values of $G_\text{old}$ from the previous $M$ iteration steps are mixed according to the coefficients $\{\alpha_i \}_{i=1}^{M}$, which are optimized for each iteration. For a detailed description of the algorithm, see Ref.~\onlinecite{Walker2011}. 
	
	The final remark concerns the behavior of our algorithm close to a phase transition $T_c$ or sometimes simply at low $T$, where the initial input self-energy at the first iteration sometimes yields a diverged charge susceptibility.  This premature divergence occurs when $\lambda Wf(\bq)\chi_{0}(\bq,0) > 1$ for a few $\bq$ points around the CDW ordering vector in the denominator of Eq.~(\ref{Eq:Xcdw_q}). To circumvent this problem we impose the condition $\lambda Wf(\bq)\chi_{0}(\bq,0) \leq \tilde{\chi}$,  where $\tilde{\chi}$ is a constant that is close to but less than unity.  More specifically, we selectively change the value of $\chi_{0}(\bq,0)$ such that $\lambda Wf(\bq)\chi_{0}(\bq,0) = \tilde{\chi}$ at the offending $\bq$ points. Once the calculation is stabilized after a few iterations, this condition is removed.  In practice, we allow $\tilde{\chi}$ to take values increasingly closer to unity, and usually in the range 0.995--0.99999.  This cutoff is particularly important   as the calculation approaches $T_c^\text{CDW}$, where we expect the charge density instabilities to be present.  Establishing a proper cutoff sequence of $ \tilde{\chi} $ can improve the quality of $ T_{c}^{\text{CDW}} $ extrapolation. 
	\begin{figure}[tbh]
		\includegraphics[width=0.95\columnwidth]{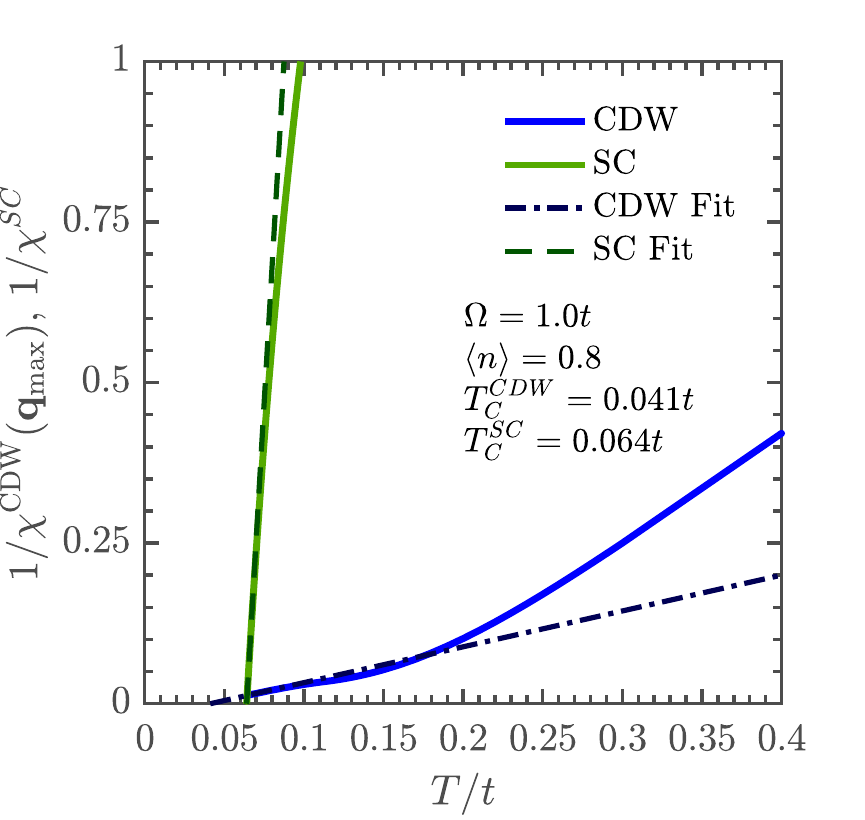}
		
		\caption{An example of extrapolating the inverse CDW and pairing  susceptibilities to zero to obtain $T_c$ on a lattice 
		    with $ N=64\times 64 $ and a coupling constant $ \lambda=0.3 $.
		    The dashed and dashed-dotted lines are linear fits to the three lowest temperature points and their intercepts on $T$-axis are the respective estimated $T_c$'s in this procedure. 
			This extrapolation is accurate for the superconducting $T_c^\text{SC}$; however, the more gradual approach of
			$1/\chi^\text{CDW}(\bq_\text{max})$ to the $T_c$ region makes $T_{c}^\text{CDW}$ more susceptible to
			over(under)-estimation. 
			Therefore, we use a different procedure to estimate $T_{c}^\text{CDW}$ by fitting the low temperature $ \chi^\text{CDW}(T) $ to the $ T $-dependence of the 2-D Ising Universality class (See Sec. \ref{subsec:TC_Extrapolation}).
			}
		\label{fig:Extrap}
	\end{figure}

	\subsection{Determination of Phase Transition Temperatures} \label{subsec:TC_Extrapolation}
	To determine the phase transition temperatures $T_c^\text{CDW}$ and $T_c^\text{SC}$, we compute the CDW and pairing susceptibilities $\chi^\text{CDW}(\bq,T)$ and $\chi^\text{SC}(T)$ with decreasing temperatures from above the phase transitions. In the thermodynamic limit ($N\to \infty$), the susceptibilities should diverge as $T\to T_{c}$. For a finite-size lattice, this divergence is limited once the correlation length $ \xi $ becomes comparable to the lattice size.  Our goal is to determine $T_{c}$ in the thermodynamic limit by using sufficiently large lattice sizes such that a reliable extrapolation to $N \rightarrow \infty$ limit can be performed.
	
	\begin{figure}[tbh]
		\includegraphics[width=0.95\columnwidth]{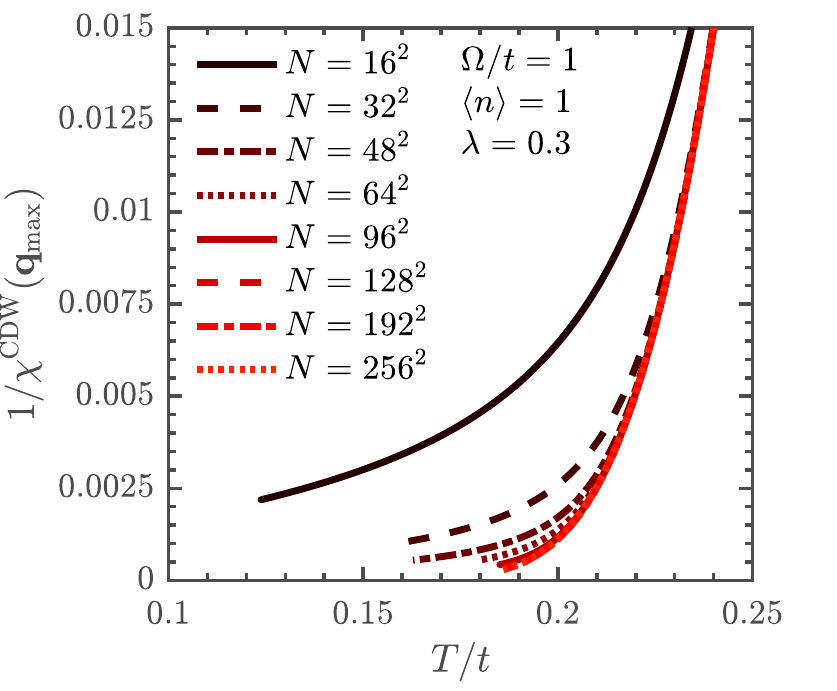}
			
		\caption{The dependence of $1/\chi^\text{CDW}$ on the size of the lattice near   $T^\text{CDW}_{c}$.  
	    For smaller lattice sizes, the temperature dependence deviates strongly from what is expected in
	    the thermodynamic limit ($N\to \infty$). 
	    Consequently, the critical temperatures $T^\text{CDW}_{c}$ found by fitting 
	    $\chi^\text{CDW}(\bq_\text{max})$ with a model function $\chi^\text{2D-Ising}(T)$, 
	    depends strongly on the value of $N$ (see Fig.~\ref{fig:Tc_finite_size}).}
		\label{fig:Tc_finite_size_IXcdw}
	\end{figure}
	
	 These values can be extracted in one of two ways: (i) by extrapolating the low $ T $ behavior of the inverse susceptibilities $1/\chi^\text{CDW}(\bq_\text{max})$ and $1/\chi^\text{SC}$ to zero as a function of temperature (e.g. Fig.~\ref{fig:Extrap}); or (ii) by fitting the susceptibilities near the transition temperature with the appropriate asymptotic forms expected for their universality class in 2D.

	\begin{figure}[tbh]
		\includegraphics[width=0.95\columnwidth]{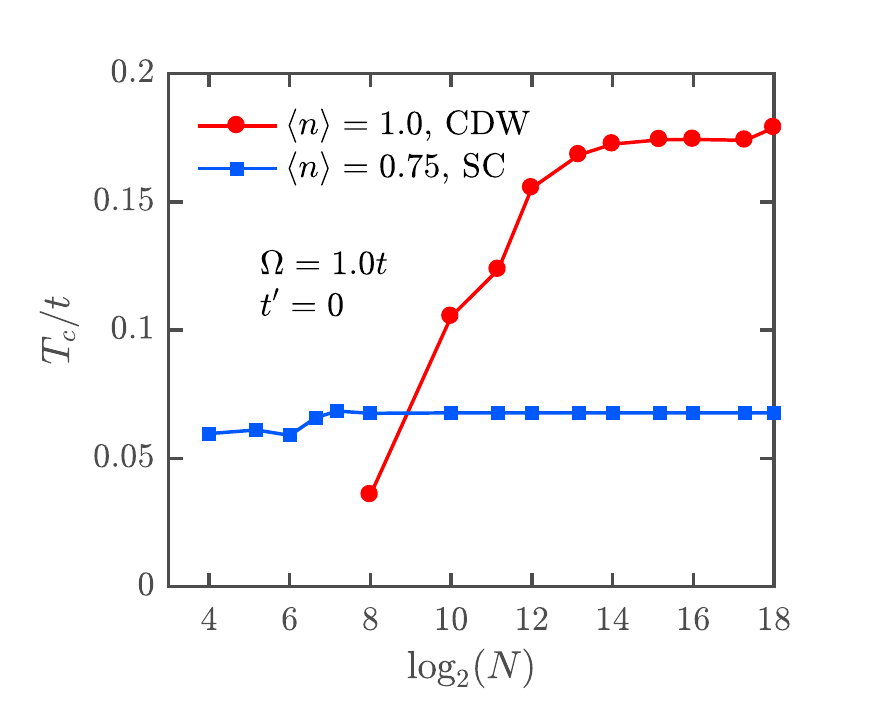}
		
		\caption{The finite size dependence of $T_c$ on the logarithm of the total number of sites $ \log_{2}(N) $ for the Holstein model (isotropic {\ep} coupling).  The fillings $n = 1.0$ and $n = 0.75$ correspond to parts of the phase
		diagram that yield CDW and SC phases, respectively, for $\Omega/t = 1.0$, $\lambda = 0.3$, and $t'=0$. The
		value of $T_c^\text{SC}$ has converged for $ N=16\times16 $, whereas $T_c^\text{CDW}$ has
		significant dependence on lattice size $N$. The value of $T_c^\text{CDW}$ was obtained from the model fit
		$\chi^\text{2D-Ising}(T)$.  For $ \log_{2}(N)\geq8 $, the values of $ N  $  are identical to those shown in the legend of Fig.~\ref{fig:Tc_finite_size_IXcdw}.  Five additional smaller lattice sizes ($ \sqrt{N}=4,6,8,10,12 $) are plotted for $T_c^\text{SC}$ to show the variations due to finite size effects.        
		}
		\label{fig:Tc_finite_size}
	\end{figure}

	It is well known~\cite{Scalettar1989} that the pairing susceptibility in a 2D system has a similar behavior to the Kosterlitz-Thouless~\cite{Kosterlitz1973} phase transition.  However, we do not use this universality class in practice to find $T_{c}^\text{SC}$ because we have found that the extrapolation of $1/\chi^\text{SC}$ to the temperature axis provides a reliable estimate for $T_{c}^\text{SC}$ (e.g., see Fig.~\ref{fig:Extrap}) because of the relatively sharp divergence of the pairing susceptibility within the Migdal approximation. On the other hand, this same extrapolation procedure offers a $ T_{c}^{\text{CDW}} $ with a rather tenuous justification. As seen in Fig.~\ref{fig:Extrap}, extrapolating the last few points in $1/\chi^\text{CDW}(\bq_\text{max})$ tends to estimate the $T_{c}^{\text{CDW}}$ significantly lower than the last obtainable point in the numerical calculation. Close to the transition, $\chi^\text{CDW}(\bq_\text{max}=(\pi,\pi))$ is expected to follow the Ising universality class~\cite{Noack1991}, which follows $\chi^\text{2D-Ising}(T) = A\left| \frac{T-T_c}{T_c} \right|^{-\gamma}$, where $\gamma = 7/4$.  The critical temperatures can thus be obtained by fitting the CDW susceptibility results to this model.  Notice that the slope $1/\chi^\text{2D-Ising}(T)$ approaches zero as $T\to T_c^{+}$.  This partially explains the potential inaccuracy of linear extrapolation for $T_c^\text{CDW}$ since it is expected that $1/\chi^\text{CDW}(\bq_\text{max})$ will not sharply cross the temperature axis.
	
	The accuracy of the Ising fit to $\chi^\text{CDW}$ significantly depends on the lattice-size. In fact, at smaller sizes such as $4\times4$ and $8\times8$, the model above is a remarkably poor descriptor of the $\chi^\text{CDW}$ results.  As we increase the lattice to $128\times128$ and above, we see that the susceptibility more closely follows the expected Ising form (See Fig.~\ref{fig:Tc_finite_size_IXcdw}).  The resulting finite size dependence of $T_{c}^\text{CDW}$ is shown in Fig.~\ref{fig:Tc_finite_size} for the isotropic coupling case, with $\Omega/t=1$, at half-filling.  Here, the changes in $T_{c}^\text{CDW}$ are pronounced until $ N\sim128^{2} $ while $T_{c}^\text{SC}$ is approximately flat for $ N\geq 16^{2} $.
	Furthermore, as shown in Fig.~\ref{fig:Extrap}, for fillings very close to the CDW and superconductivity phase boundary, the comparable values for $T_c^\text{CDW}$ and $T_c^\text{SC}$ make the careful extrapolation more important.  We adhere to a convention that when the CDW and superconducting phases have comparable transition temperatures, the winning phase is determined by the larger of the extrapolated $T_c^\text{CDW}$ and $T_c^\text{SC}$. 
	
	\begin{figure*}[tbh]
		\includegraphics[width=\textwidth]{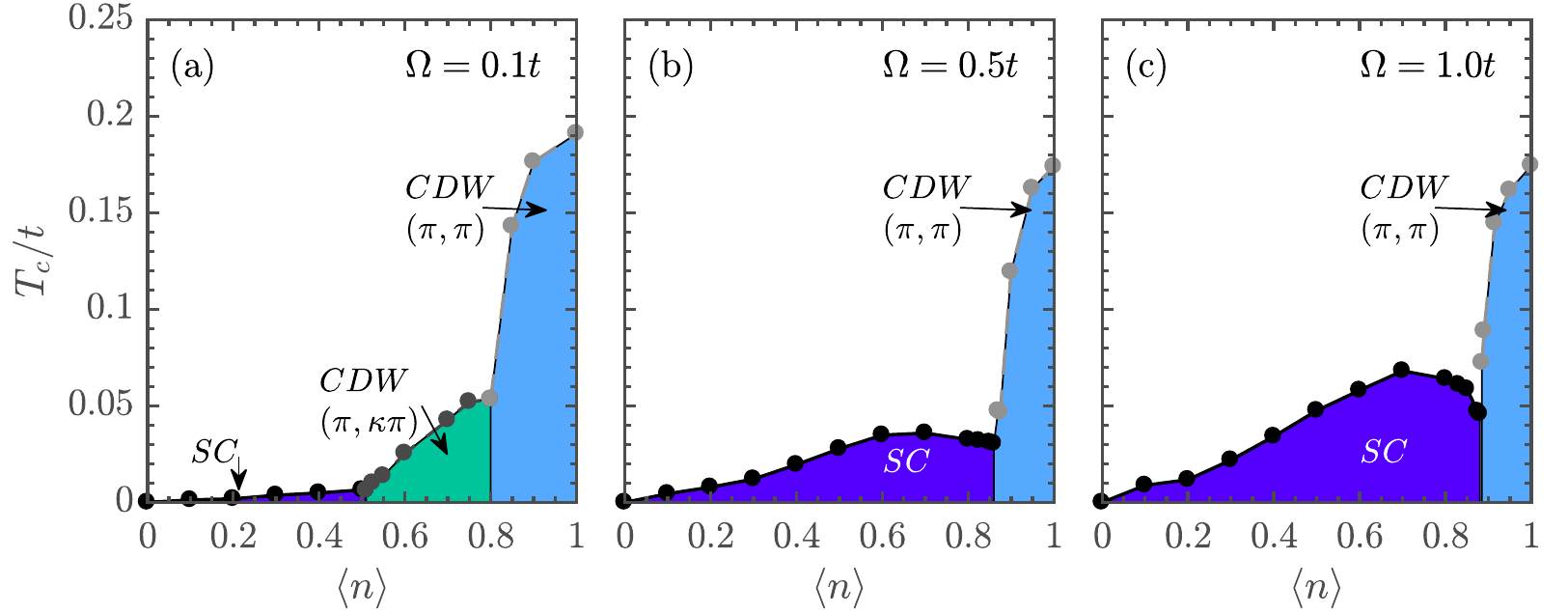}
		
		\caption{(Color online) The temperature-filling phase diagram for the Holstein model for (a) $\Omega = 0.1t$,
			(b) $\Omega = 0.5t$, and (c) $\Omega = t$. Results were obtained on a $N = 128^2$-site lattice and for a
			dimensionless coupling $\lambda = 0.3$.}
		\label{fig:PD_Holstein}
	\end{figure*}
	
	\begin{figure}[tbh]
		\includegraphics[width=\columnwidth]{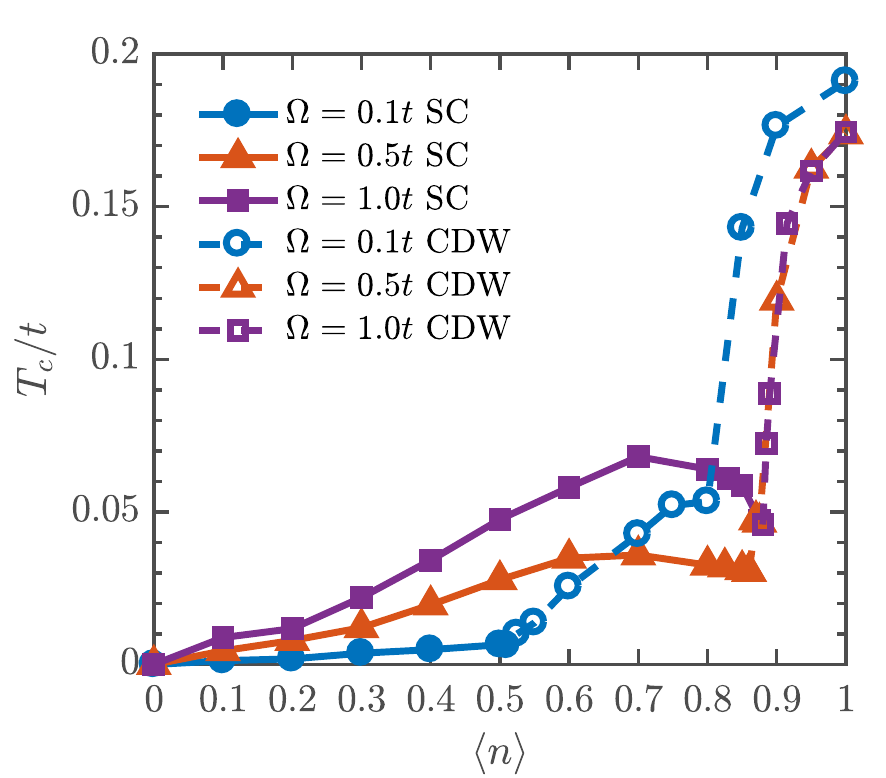}
		
		\caption{(Color online) A comparison of the three phase diagrams shown in Fig.~\ref{fig:PD_Holstein}.}
		\label{fig:PD_comparison}
	\end{figure}

	\section{Results} \label{sec:results}
	\subsection{Temperature-filling Phase Diagram}\label{sec:PD}
	First, we report the full temperature-filling phase diagram within the Migdal approximation for the 2D Holstein model with isotropic {\ep} coupling $g(\bq) = g$ and only the NN hopping $t$.  Fig.~\ref{fig:PD_Holstein} shows results for hole-doped case with filling $0\leq n \leq 1$ and three different phonon frequencies. The electron-doped side is identical due to particle-hole symmetry. The phase boundaries in all three panels in Fig.~\ref{fig:PD_Holstein} are plotted together in Fig.~\ref{fig:PD_comparison} for an easy comparison.
	
	As expected, we observe the competition between CDW and superconducting (SC) ground states, where the CDW phase dominates close to half-filling. For large phonon frequencies, the CDW phase always appears at $\bq_\text{max} = (\pi,\pi)$ but is rapidly suppressed for a small degree of doping away from half-filling.  For the smallest phonon frequency ($\Omega =0.1t$), we find larger values of $ T_{c}^{\text{CDW}} $ indicating a slight decrease in $\chi_{0}$.  Moreover, we observe an incommensurate CDW phase for filling levels below $n \approx 0.8$.  This incommensurate ordering can be distinguished from the commensurate case by noting that when $T\to T_c^\text{CDW}$, the function $\chi^\text{CDW}(\bq)$ develops  peaks at $\bq_\text{max} = (\pi,\kappa\pi)$ with $ 0<\kappa\leq 1 $ (and its symmetry equivalent positions).  These four peaks split off from a single broader peak originally centered at $(\pi,\pi)$ at high temperatures as $T\to T_c^\text{CDW}$.  At intermediate temperatures, these four peaks largely overlap resulting in a plateau centered at $ (\pi,\pi) $.  In the commensurate case, $\chi^\text{CDW}(\bq)$ has a single global maximum at $(\pi,\pi)$ for nearly all the temperatures examined. We also observe strong incommensurate charge correlations when $\Omega$ is large but in this case the superconducting phase forms before the long-range incommensurate charge order forms. 
	
	Once the CDW correlations are sufficiently suppressed at lower values of $n$, the $s$-wave superconducting correlations dominate. Our results show that superconducting transition temperature $T_c^\text{SC}$ depends strongly on the phonon frequency $\Omega$. The fact that $T_{c}^{\text{SC}}$ strongly depends on $\Omega$ while $T_{c}^{\text{CDW}}$ depends more on electronic properties is in qualitative agreement with solutions to the Holstein model in the infinite dimensional limit~\cite{Freericks1993}.  Interestingly, we observe non-monotonic behavior in the superconducting $T_c$, where the maximum value of $T_c$ occurs for fillings away from the CDW phase boundary.  This ``dome'' in the superconducting region of the phase diagram becomes more pronounced as the phonon frequency increases. We will discuss the origin of this behavior in Sec. \ref{Sec:Dome}.  	
	
	\begin{figure*}[t]
		\includegraphics[width=0.9\textwidth]{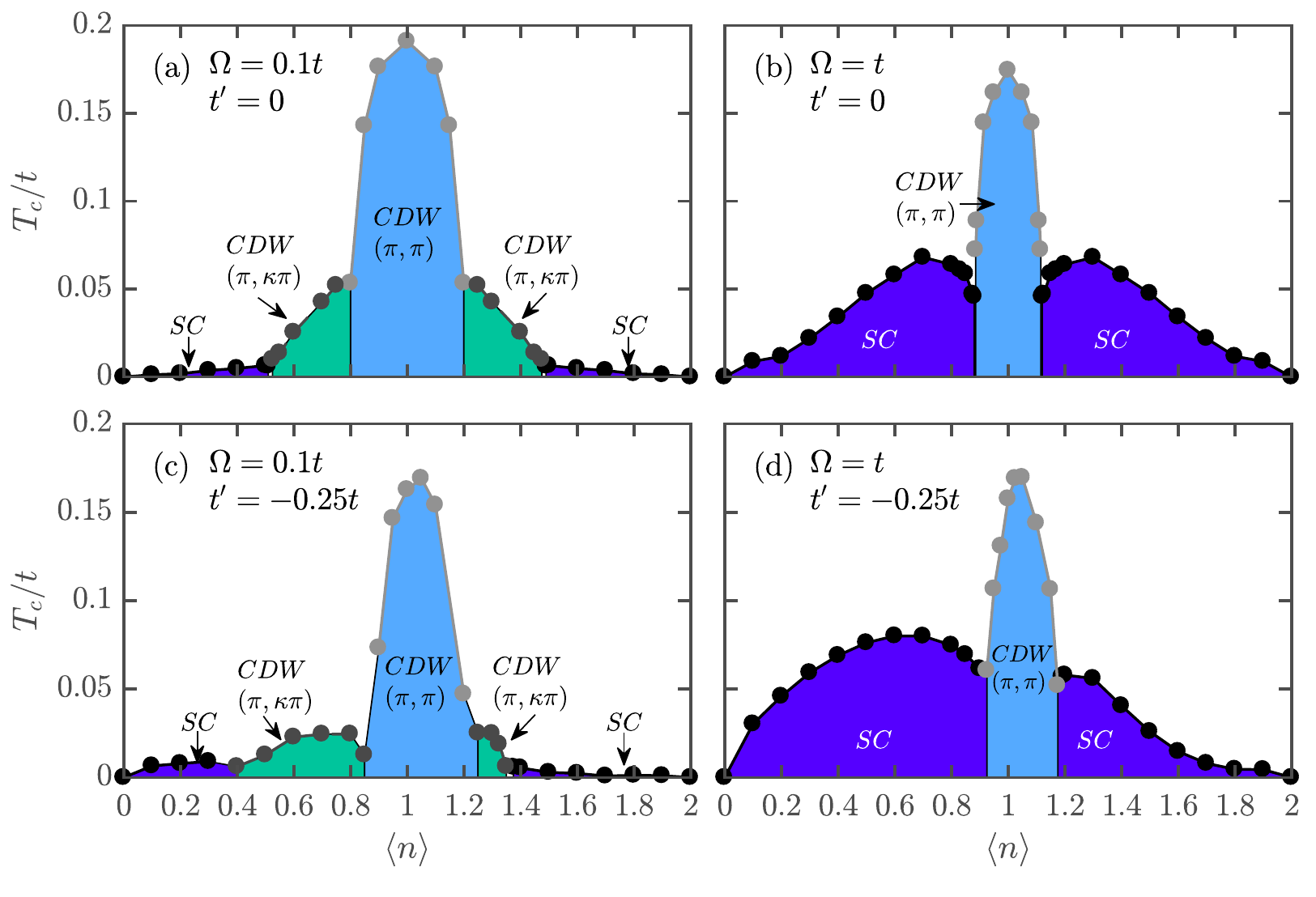}
		
		\caption{(Color online) Comparison of the temperature-filling phase diagrams for Holstein model with or
			without NNN hopping $t'$. (a) $\Omega=0.1t$, $t'=0$; (b) $\Omega=t$, $t'=0$; (c) $\Omega=0.1t$, $ t'=-0.25t $;
			and (d) $\Omega=0.1t$, $t'=-0.25t$. Results were obtained on a $N = 128^2$ site lattice and for a
			dimensionless coupling $\lambda = 0.3$.}
		\label{fig:PD_NNN_2x2}
	\end{figure*}
       
        \subsection{The effects of longer-range hopping on the phase diagram}\label{Sec:NNN_hopping}
        Previous studies of the Holstein model~\cite{Vekic1992} and the attractive Hubbard model~\cite{Hirsch1986a} found that enhanced pairing occurs when the Fermi level $E_\text{F}$ lies near the Van~Hove singularity once NNN hopping is included.  With this motivation, we now turn our attention to the effects of $t^\prime$ on the phase diagram.  In this case, the lack of particle-hole symmetry requires us to consider the temperature-filling phase diagram across the full range of electronic fillings, as shown in Fig. \ref{fig:PD_NNN_2x2}. We consider two representative values $\Omega = 0.1t$ [Figs. \ref{fig:PD_NNN_2x2}(a) and \ref{fig:PD_NNN_2x2}(c)] and $\Omega = t$ [Figs. \ref{fig:PD_NNN_2x2}(b) and \ref{fig:PD_NNN_2x2}(d)] and fix $t^\prime = -0.25t$, chosen to reflect a phase factor of opposite sign commonly encountered in diagonal hopping scenarios.  We note that a choice of opposite sign merely creates a mirror image (with respect to $n=1$) of depicted the phase diagrams, i.e., changing $n\to 2-n$ in the $x$-axis.

        Many of the effects of $t^\prime$ can be understood from its influence on the bare electronic structure.  The Van~Hove singularity in the bare 2D electronic density of states (DOS) shifts below (above) the middle of the band when $t^\prime<0$ ($t^\prime>0$) and, moreover, the Fermi surface inherits curvature that weakens nesting near half-filling. These changes, which are not mutually exclusive, are collectively associated with the suppression of charge-density correlations near half-filling.  Upon hole doping, the Fermi surface for $t^\prime < 0$ moves  towards the Van~Hove singularity, thus increasing the electron energy degeneracy where strong pairing correlations are already present and enhancing  the superconducting correlations on the hole-doped side of the phase diagram. Conversely, the superconducting  state on the electron-doped side of the phase diagram is suppressed.   These changes are most prominent for large $\Omega$ and clearly seen in the contrast between the top (for $ t^{\prime}=0 $) and the bottom row (for $ t^{\prime}\neq0 $) in Fig.~\ref{fig:PD_NNN_2x2}. 
        \begin{figure*}[t]
		\includegraphics[width=0.9\textwidth]{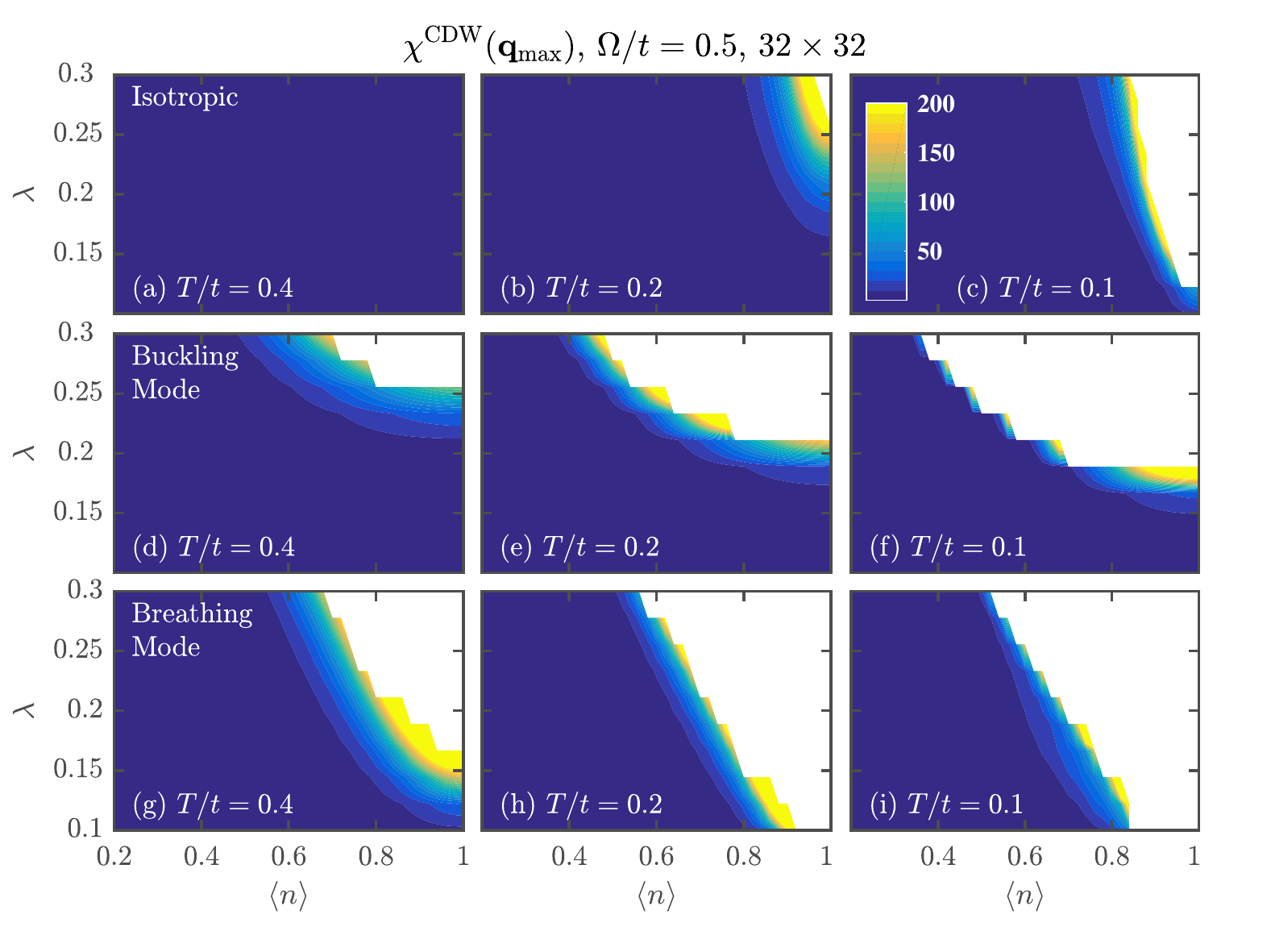}

		\caption{(Color online) The charge susceptibility for three decreasing temperatures plotted using contours
			over a window of $\lambda$ vs $n$.  Panels (a)--(c) show the Holstein (isotropic) coupling case, where
			a strong CDW susceptibility peaked at $\bq_\text{max} = (\pi,\pi)$ emerges near and at half-filling. Panels
			(d)--(f) show results for the buckling case, which reveal the onset of susceptibility peaked at $\bq_\text{max} = (0,0)$. This is not representative of
			a CDW with long range order and could be an indication of phase separation. The last row of panels (g)--(i) show
			results for the breathing mode coupling, which strongly favors a CDW phase peaked at
			$\bq_\text{max} = (\pi,\pi)$ for larger regions of the parameter space. Introducing the breathing and
			buckling mode $\bq$-dependence into the {\ep} coupling has significant influence over the competition between
			pairing correlations and charge-density correlations.}
		\label{fig:Xcdw_Iso_BK_BR_3x3}
	\end{figure*}
	
	\begin{figure*}
		\includegraphics[width=0.9\textwidth]{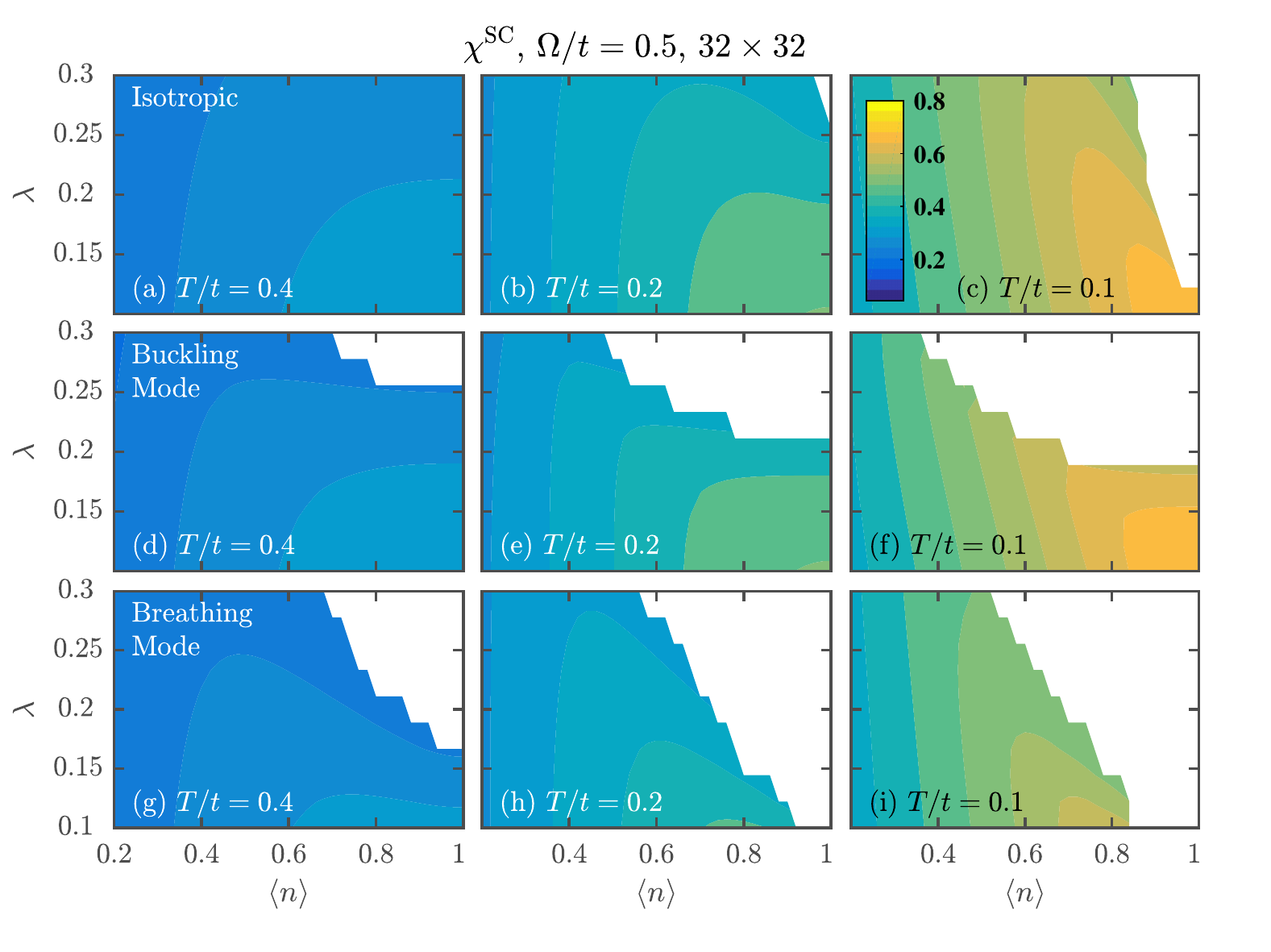}
		
		\caption{(Color online) The pairing susceptibility for three decreasing temperatures plotted using contours
			over a window of $\lambda$ vs $n$.  Similar to Fig.~\ref{fig:Xcdw_Iso_BK_BR_3x3}, panels (a)--(c) correspond
			to the Holstein (isotropic) coupling case, panels (d)--(f) for the buckling mode coupling, and panels
			(g)--(i) for the breathing mode coupling. Each temperature progression shares qualitatively similar features
			with pairing correlations largely suppressed at these temperatures. However, it should be noted that the
			Holstein case is unique. With the CDW region so closely confined to the half-filling region,
			pairing correlations can develop at higher temperatures.  A superconducting
			phase does emerge for each of the $g(\bq)$ cases, but the strong charge density correlations in the buckling
			and breathing mode cases push the SC boundary to lower values of $n$.}
		\label{fig:Xsc_Iso_BK_BR_3x3}
	\end{figure*}

        The effects of a longer-range hopping are more subtle for the smaller phonon frequency $\Omega = 0.1t$.  Here, $t^\prime$ lowers the transition temperatures of the commensurate and incommensurate CDW phases both for all dopings. At the same time, the boundary of the incommensurate CDW region is modified such that it extends towards smaller values of $n$ with hole doping. Likewise, the SC phase on the hole-doped side is confined to a smaller region of fillings but with an enhanced superconducting $T_c$. On the electron-doped side, the superconducting region also extends over a broader range of fillings but the overall $T_c$ is suppressed.  
        
	Similar asymmetric features in the phase diagram can result from other modifications/extensions of the Holstein model.  For instance, adding a small anharmonic term to the Holstein Hamiltonian also results in an asymmetric phase diagram with a larger SC phase and suppressed CDW phase on one side of half-filling~\cite{Freericks1996a}.  In that case, the changes in the phase diagram are the result of the modified phonon potential, which constrains the phonon displacements and enhances superconductivity. A more recent study focused on the electron effective mass $m^*$~\cite{Chandler2016}; however, in the 2D weak coupling case, the changes to $m^*$ resulting from $t^\prime \ne 0$ are quite small and effectively negligible. We also note that NNN  hopping can influence the competition between charge density and pairing correlations. In some real materials (e.g., transition metal dichalcogenides) it can be difficult to determine whether CDW and SC correlations are working together or competing for order~\cite{Heil2017}. In this simple model calculation, we find evidence for the latter through the absence of {\it both} large pairing and charge density correlations together near a critical temperature, as well as a decreasing $T_c^\text{SC}$ near the SC-CDW phase boundary.
  
	\subsection{Momentum Dependent Electron-Phonon Coupling}
	The traditional Holstein model can be easily extended by introducing fermionic ($\bk$) and bosonic ($\bq$) momentum  dependence to the {\ep} coupling constant $g\to g(\bk,\bq)$. As mentioned in section (\ref{subsec:qDependentInteractions}), our FFT-based algorithm can easily incorporate models where the vertex depends on the boson wavevector $g(\bq)$.         In general, we have found that convergence of the self-consistent equations is more challenging for such models near phase transitions, and it can become difficult to develop a full temperature dependence for the susceptibilities across all fillings relative to the isotropic coupling case. Nevertheless, we have obtained results for several popular models for the high-T$_c$ cuprates with a momentum dependent $g(\bq)$, which we present in this section.  
	
	The behavior of the susceptibilities $\chi^{\text{CDW}}$ and $\chi^{\text{SC}}$ for a given $g(\bq)$ is examined across a small range of $\lambda$ values in Fig.~\ref{fig:Xcdw_Iso_BK_BR_3x3} and Fig.~\ref{fig:Xsc_Iso_BK_BR_3x3}, respectively.  All of the results were obtained using a 32$\times$32 lattice with $\Omega/t=0.5$, and NN-hopping only ($t^\prime=0$). The first, second, and third rows of the figures correspond to isotropic, buckling, and breathing models, respectively (see Sec. \ref{subsec:qDependentInteractions}).  The individual columns show results for temperatures $T/t=0.4$, $0.2$, and $0.1$ from left to right.  The white regions indicate parameter ranges where no data is plotted.  The vertical axis in Fig.~\ref{fig:Xcdw_Iso_BK_BR_3x3} and Fig.~\ref{fig:Xsc_Iso_BK_BR_3x3} contain data points for ten values of $ \lambda $ separated by increments of $\Delta\lambda=0.02\bar{2}$.  Although the spacing in filling points along the horizontal axis is comparable ($ \Delta n = 0.025 $), the plot range is larger, making it appear as though $ \Delta\lambda $ was disproportionately coarse.  As a result, we see that the boundary separating the susceptibility contours from the white region is jagged in appearance.  Although the boundary is expected, it should be smooth in the limit where $ \Delta\lambda\to 0 $.  Therefore, no physical meaning should be attributed to the uneven nature of the boundary.  Despite this unwanted cosmetic detail, the white region beyond the (colored) contours is, to a good approximation, where the system would have settled into a charge ordered phase peaked at some vector $\bq_\text{max}$. 
	
    For the isotropic and breathing cases, we find that the CDW correlations are strongest at $\bq_\text{max}=(\pi,\pi)$. These results are expected since $\bq = (\pi,\pi)$ corresponds to a strong nesting condition near half-filling and $g(\bq)$ for the breathing mode is largest at this wavevector.  For the buckling case, we find the CDW correlations are strongest at  $\bq_\text{max}=(0,0)$. This is also not surprising since $g(\bq)$ for this model places most of the scattering weight on $\bq_\text{max}=(0,0)$ and none on the nesting vector (at half-filling) $(\pi,\pi)$.  We interpret large charge correlations at $\bq_\text{max}=(0,0)$ as a reflection of a tendency towards phase separation since we are considering a single-orbital model without the possibility for any intracell charge order. 

\begin{figure}[h]
		\includegraphics[width=\columnwidth]{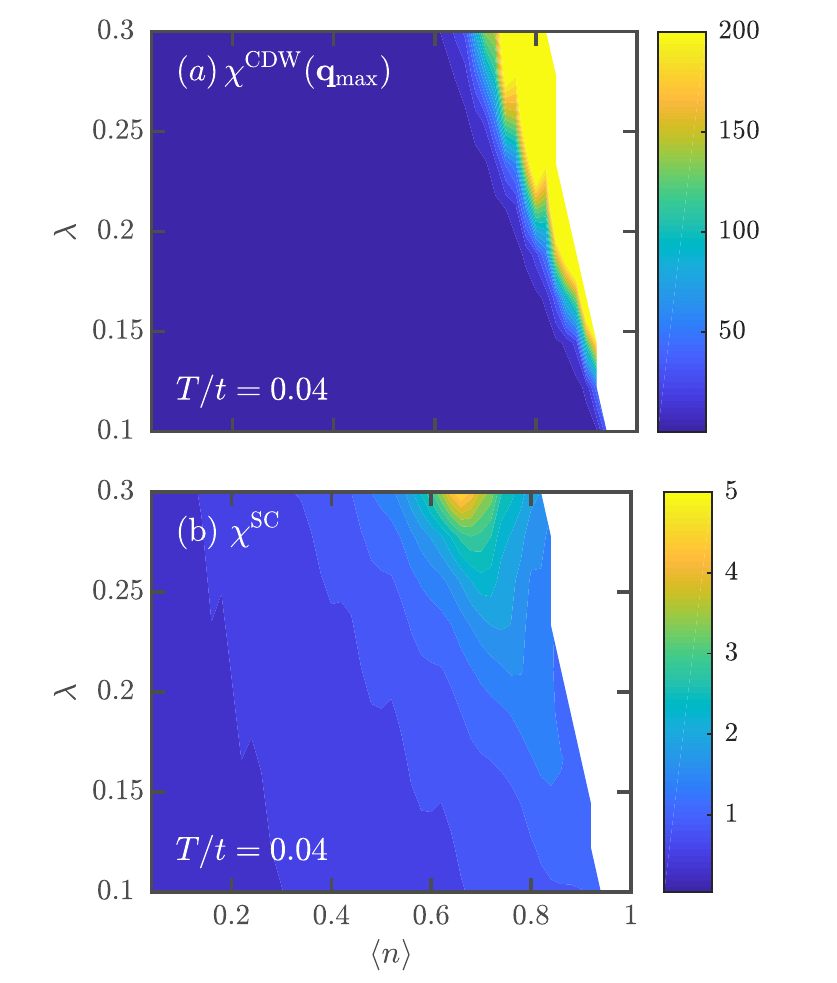}
		
		\caption{(Color online) For the isotropic coupling case with $\Omega=0.5t$, $T=0.04t$, and $t' = 0$, we
			plot the (a) charge-density wave susceptibility $\chi^\text{CDW}(\bq_\text{max})$ and (b) singlet-pairing
			susceptibility $\chi^\text{SC}$ using a $32\times32$ lattice.  At this lower temperature we see the pairing
			correlations becoming more significant around $\lambda=0.3$ and $n \approx 0.66$, which corresponds to a point near the top of the $T_{c}^{\text{SC}}$-dome in Fig.~\ref{fig:PD_Holstein}.}
		\label{fig:Xcdw_Xsc_Iso_2x1}
	\end{figure}

 	By comparing the results at different temperatures it becomes clear that the momentum dependent models yield higher $ T_{c}^\text{CDW} $'s than the isotropic case. For the Hosltein coupling [Figs~\ref{fig:Xcdw_Iso_BK_BR_3x3}(a)--(c)] the CDW correlations build somewhat tightly around half-filling whereas they extend much further as a function of both $n$ and $\lambda$ once a momentum dependent coupling is introduced.  A momentum dependent coupling also influences the $s$-wave pairing tendencies.  As shown in Fig.~\ref{fig:Xsc_Iso_BK_BR_3x3}, the contour lines for large values of $\chi_{c}^\text{SC}$ extend over larger $n$ and $\lambda$ in the Holstein case. We have performed other calculations confirming that the $s$-wave superconducting correlations generally shift to more dilute fillings and lower $T$ once the {\ep} vertex depends on $\bq$.  It should be emphasized that the contours in Fig.~\ref{fig:Xsc_Iso_BK_BR_3x3} can be somewhat deceptive since the temperatures at which $\chi^\text{SC}$ are plotted are well above $T_{c}^\text{SC}$.  For instance, $\chi^\text{SC}$ seems to be peaked at occupancies of $ n\sim 0.8 - 1$ and couplings $\lambda\sim0.1$ in Fig.~\ref{fig:Xsc_Iso_BK_BR_3x3}, but this peak actually shifts to lower fillings and larger $ \lambda $ values as the temperature is lowered further.  When we decrease the temperature so that $T\approx T_{c}^\text{SC}$, we see a rapid growth of $\chi^\text{SC}$ near the phase boundary as shown in Fig.~\ref{fig:Xcdw_Xsc_Iso_2x1}(b).  This figure reveals that the highest $T_{c}^\text{SC}$ occurs for larger $\lambda$ (for the range shown) and the features leading to the SC dome in Fig.~\ref{fig:PD_Holstein}(b) can be seen.  If we were to superimpose Fig.~\ref{fig:Xcdw_Xsc_Iso_2x1}(b) onto Fig.~\ref{fig:Xcdw_Xsc_Iso_2x1}(a), the correlations for pairing and a CDW would form a valley in-between the domes for each phase, reminiscent of the phase diagrams presented earlier. 
	
	\subsection{Renormalized Phonon Dispersions at Half-filling}\label{Sec:phonon_dispersion}
\begin{figure}[h]
		\includegraphics[width=\columnwidth]{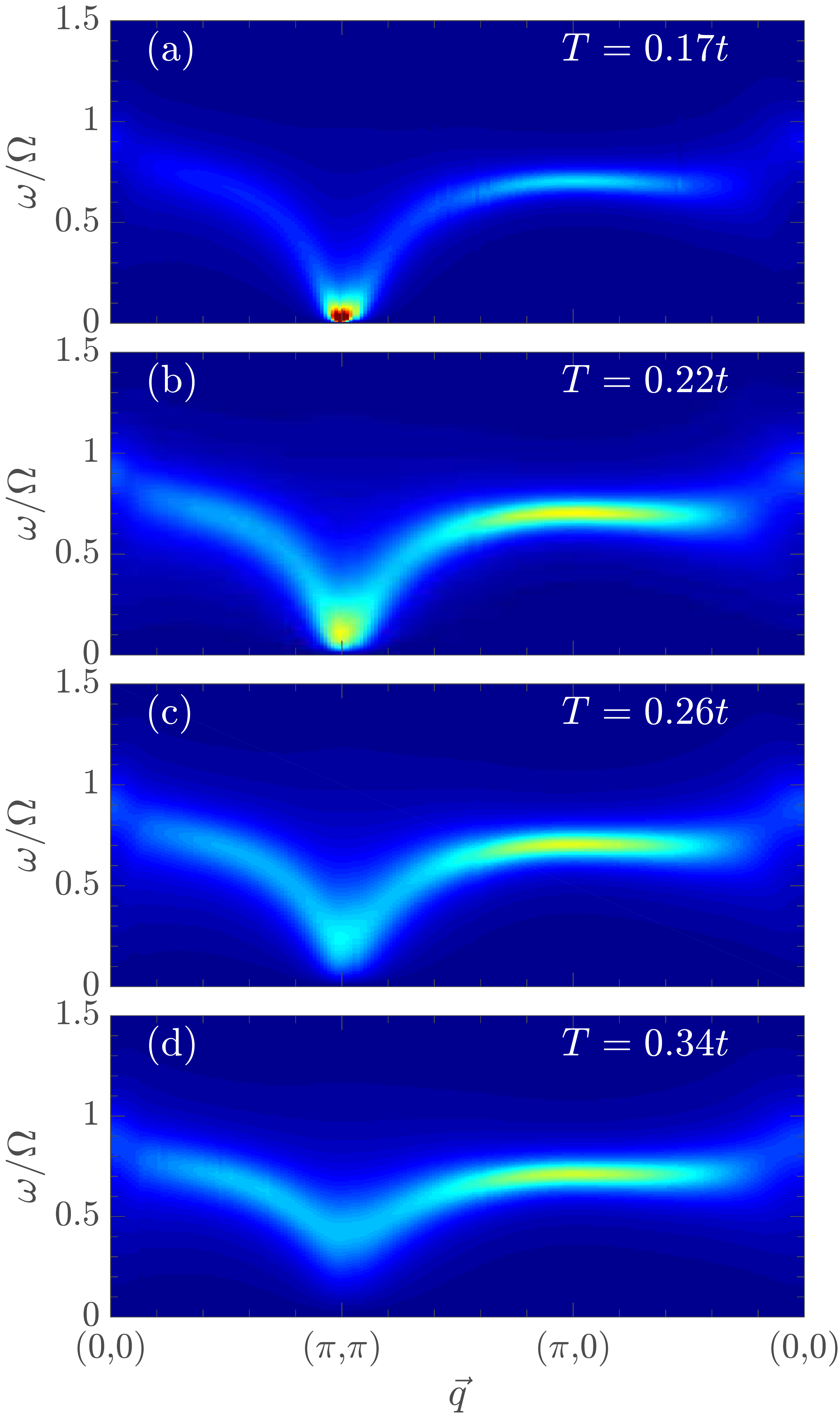}
		
		\caption{(Color Online) The temperature dependence of the phonon spectral function $B(\bq,\omega)=-\Im
			D(\bq,\ii\nu_m \to \omega+\ii 0^{+})/\pi$ for the half-filled Holstein model with $\lambda = 0.1875$. Results
			were obtained on an $N = 128^2$ cluster and the analytic continuation was performed using Pad{\'e} approximants.  
			The actual CDW transition temperature is $ T_{c}^{\text{CDW}}=0.144t $.   }
		\label{fig:Bqw}
	\end{figure}

        Until now we have largely focused on the electronic properties of the model. It is also instructive, however,  to consider the renormalization of the phononic properties in proximity to the CDW phase. In this section, we present the spectral properties of the phonons, which are obtained by analytically continuing the phonon Green's function to the real axis using Pad{\'e} approximants~\cite{Vidberg1977}. Fig.~\ref{fig:Bqw} shows the phonon spectral function $B(\bq,\omega)=-\Im D(\bq,\ii\nu_m \to \omega+\ii 0^{+})/\pi$ as a function of temperature along the high-symmetry path of the first Brillouin zone. Here, we are considering the case of a momentum-independent Holstein coupling, half-filling $n = 1.0 $, $\lambda = 0.19$, and $\Omega/t = 1.0$.  These spectra compare well with the results obtained from determinant QMC simulations~\cite{Li2015a} carried out for comparable values of $\lambda$ but on a smaller lattice ($\lambda = 0.25$, $8\times 8$).  
	
        Our first observation is that the overall energy of the phonon branch has softened significantly due to the electron-phonon coupling.  Due to the proximity of the CDW phase, the phonon spectral function also has a pronounced Kohn-like anomaly, where spectral weight becomes concentrated at $\bq = (\pi,\pi)$ and $ \omega\approx 0 $. As the temperature is lowered, and the charge correlations grow, spectral weight is  redistributed to lower energies and the Kohn anomaly becomes sharper.   Although it is not shown here, we have also found that in the case of the buckling mode coupling, the spectral weight indeed concentrates at $\bq = \mathbf{0}$. 

        We expect that the  renormalization of the phononic properties will influence the superconducting phase in nontrivial ways. For example, within weak coupling BCS theory, the superconducting transition temperature is given by $T_c \propto \Omega \exp(-1/\lambda)$. The phonon dispersion enters this expression twice, once in the prefactor and once in the dimensionless coupling constant $\lambda \propto 1/\Omega$. The softening of the phonon branch observed in Fig. \ref{fig:Bqw} will, therefore, simultaneously reduce the energy scale of the Cooper pairs and enhance the pairing strength. Which of these effects dominates is a nontrivial question, which is addressed in the following section.  
	
        \subsection{Origin of the Superconducting Dome}\label{Sec:Dome}
	As we noted in Sec. \ref{sec:PD}, we observe a non-monotonic behavior in the superconducting $T_c$, where the maximum value of $T_c$ occurs for fillings away from the boundary of the CDW phase.  This ``dome'' in the superconducting region of the phase diagram becomes more pronounced as the phonon frequency increases and arises from an interplay of the renormalized phononic and electronic  properties. 
  
\begin{figure}[h!]
	    	\includegraphics[width=0.9\columnwidth]{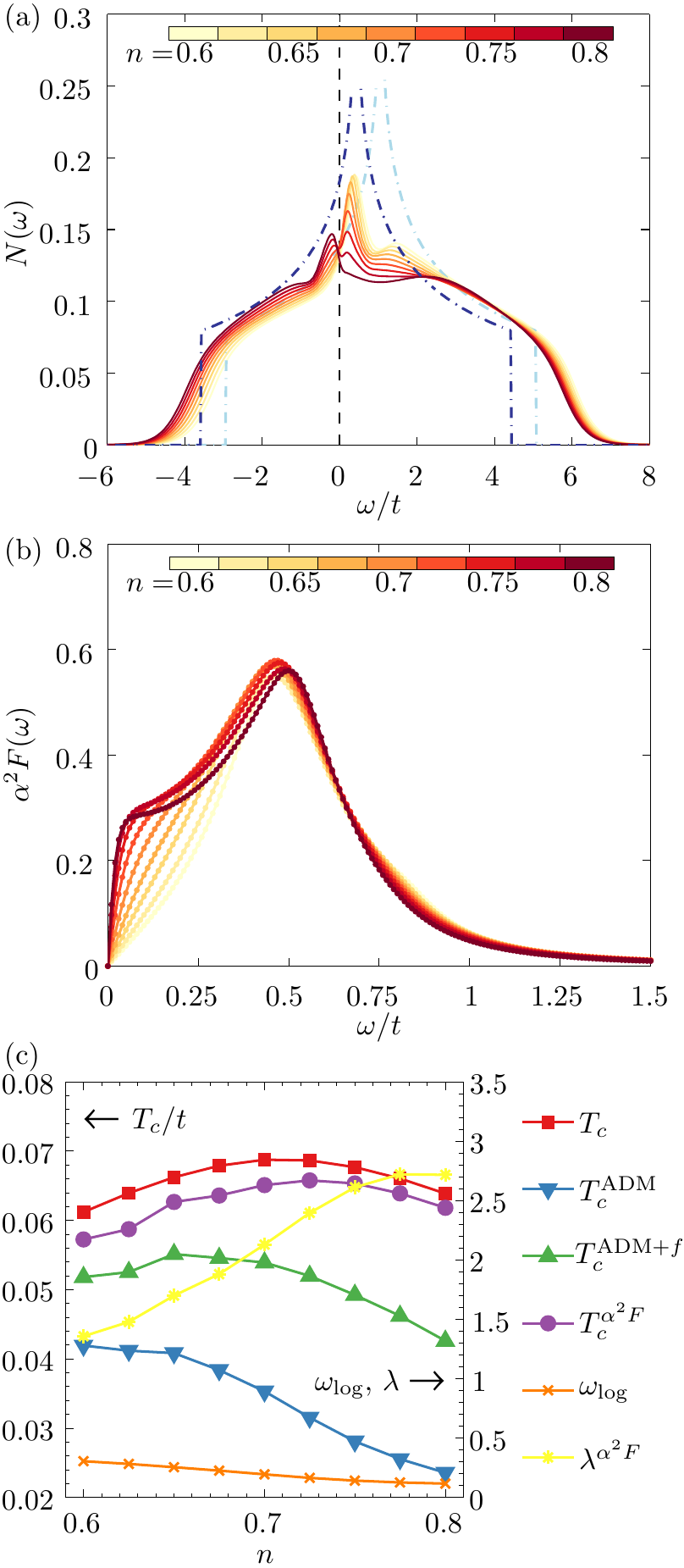}
			\caption{(Color online) (a) the electron density of states $ N(\omega) $, (b) the Eliashberg function $ \alpha^{2}F(\omega) $ multiplied by a factor of $ N_{\text{F}}/N_{\text{F}}^{\text{fd}} $, and (c) $ T_{c} $, $ \omega_{\text{log}} $, and $ \lambda $ as a function of $ n $.  A factor of $ N_{\text{F}}/N_{\text{F}}^{\text{fd}} $ is included in $ \alpha^{2}F(\omega) $ and $ \lambda^{\alpha^{2}F(\omega)} $.  The factor of $ N_{\text{F}}/N_{\text{F}}^{\text{fd}} $ is used to account for differences between the exact DOS and the DOS estimated by a sum of delta functions.}
			\label{fig:dos_a2F_lmdTc}
	\end{figure}

	To better understand how the dome appears, we examined several quantities commonly linked to pairing.  These include the electronic density of states (DOS) $ N(\omega) $ (per spin), the Eliashberg function $ \alpha^{2}F(\omega) $, the renormalized {\ep} coupling $ \lambda^{\alpha^{2}F} $, the logarithmic average frequency $ \omega_{\text{log}} $, and the superconducting critical temperature $ T_{c} $ (estimated by various approaches from the quantities examined here).  All of these quantities are calculated and shown in Fig.~\ref{fig:dos_a2F_lmdTc} over a range of electron occupancy $ n\in[0.6,0.8] $ using a fixed momentum $ \bk $-grid $ N=64\times 64 $,~\footnote{As indicated by the blue line in {\protect Fig.~\ref{fig:Tc_finite_size}}, this size is sufficiently large to discuss the superconductivity at the fillings considered.} a bare phonon frequency $ \Omega=t $, and a bare dimensionless {\ep} coupling $ \lambda=0.3 $.  Note that for panels \ref{fig:dos_a2F_lmdTc}(a) and \ref{fig:dos_a2F_lmdTc}(b), the calculations are performed on the imaginary frequency axis at a temperature $ T=0.07t $, ~\footnote{This temperature is slightly greater the {\protect$ T_{c} $} dome.  We have checked by solving the eigenvalue problem that the pairing vertex at this {\protect $ T $} is enhanced proportional to the {\protect $ T_{c} $} at each respective filling considered.} and then analytically continued onto the real frequency axis using Pad{\'e} approximants~\cite{Vidberg1977} with a small imaginary part $ \eta=0.005t $.  Moreover, the frequency $ \omega $ in these two plots is measured with respect to the chemical potential $ \mu $.  The calculation of $ \mu $ occurs on the imaginary axis, and any shifts in $ \mu $ stemming from the Pad\'{e} procedure, if they exist, are negligibly small since the filling from integrating the DOS over $\omega$ is essentially unchanged.
	
	The DOS is calculated by summing the electron spectral function $ A(\bk,\omega)=-\text{Im}[G(\bk,\omega)]/\pi $ 
	over momentum and is given by 
	\begin{equation}\label{Eq:N(w)_DOS}
	N(\omega) = \frac{1}{N}\sum_{\bk}A(\bk,\omega),
	\end{equation} 
	where $G(\bk,\omega) \equiv G(\bk,\ii\omega_{n}\rightarrow\omega+\ii 0^{+}) $.  In Fig.~\ref{fig:dos_a2F_lmdTc}(a), the solid curves represent $ N(\omega) $ at various fillings,   with the bright (dark) colors corresponding to smaller (larger) values of $ n $.  The two dash-dotted curves are the noninteracting DOS for the lowest filling $ n=0.6 $ (light blue) and the highest filling $ n=0.8 $ (dark blue) and are obtained from the exact result at $ T=0 $ (elliptic integral function for a cosine band structure).  Although not shown here, a plot of the non-interacting DOS for $ T\neq 0 $ using a finite $ \bk $-grid would also exhibit a broadened profile similar to $ N(\omega) $; however, unlike the interacting case, the peak position would remain at half-filling (i.e. the Van~Hove singularity).  The effect of filling in the	non-interacting case is just a rigid band shift of $ \mu $, while the shape of the interacting DOS, has a strong dependence on the filling.  In fact, the interacting DOS at the $ \omega $ corresponding to half-filling is strongly suppressed to a small hump for $ n=0.6 $, and even disappears for $ n=0.8 $.   Lastly, and perhaps most importantly, the DOS near and at the Fermi level $ N_{\text{F}}=N(0)$ has a non-monotonic dependence on the filling $ n $.  More specifically, the peak actually shifts from $ \omega>0 $ for $ n\sim0.6 $ to $ \omega<0 $ for $ n\sim0.8 $.  In other words, this shift of the peak in $ N(\omega) $	approximately follows the $ T_{c} $ dome over $ n $ as it moves from $ \omega>0 $ to $ \omega<0 $.  
	
	The isotropic Eliashberg function $ \alpha^{2}F(\omega) $, also known as the electron-phonon spectral function, can be obtained by taking a 
	double Fermi-surface average of the product between the squared {\ep} coupling $ |g(\bq)|^{2} $, the 
	phonon spectral function $ B(\bq,\omega) $, and the DOS at the Fermi-level $ N_{\text{F}}=N^{-1}\sum_{\bk}\delta(\xi_{\bk^{\prime}}) $, given by
	\begin{equation}
	\alpha^{2}F(\omega) = \langle\langle  N_{\text{F}}|g(\bk-\bk^{\prime})|^{2}B(\bk-\bk^{\prime},\omega)    \rangle\rangle_{\text{FS}},
	\end{equation}
	where the double Fermi surface average is defined by
	\begin{equation}\label{Eq:Dbl_FS_avg}
	\langle\langle f(\bk-\bk^{\prime}) \rangle\rangle_{\text{FS}}=\frac{\displaystyle\sum_{\bk}\sum_{\bk^{\prime}}f(\bk-\bk^{\prime})\delta(\xi_{\bk})\delta(\xi_{\bk^{\prime}})}{\displaystyle\sum_{\bk}\sum_{\bk^{\prime}}\delta(\xi_{\bk})\delta(\xi_{\bk^{\prime}})}.
	\end{equation}
	Here, the phonon spectral function $ B(\bq,\omega) $ is calculated from the renormalized phonon Green's function as defined in Section \ref{Sec:phonon_dispersion}.
	The delta functions appearing in Eq.~(\ref{Eq:Dbl_FS_avg}) and in other calculations are not true Dirac delta functions but are instead ``smeared'' delta functions, denoted as $ \tilde{\delta}(x) $.  These approximate delta functions are used when the system is constrained to a finite $ \bk $-grid.  We use a Fermi-Dirac smearing given by
	\begin{equation*}
	\tilde{\delta}_{\text{fd}}(x) = -\frac{\D}{\D x}\left(\frac{1}{\e^{x/\sigma}+1}\right)=\frac{1}{4\sigma\cosh^{2}\left(\frac{x}{2\sigma}\right)},
	\end{equation*}  
	which we use to obtain a DOS at the Fermi-level $ N_{\text{F}}^{\text{fd}}=N^{-1}\sum_{\bk}\tilde{\delta}_{\text{fd}}(\xi_{\bk}) $ with the broadening parameter $ \sigma=T $.
	The family of curves in Fig.~\ref{fig:dos_a2F_lmdTc}(b) shows the change in $ \alpha^{2}F(\omega) $ over a range in $ n $.  We have included factor of $ N_{\text{F}}/N_{\text{F}}^{\text{fd}} $ in $ \alpha^{2}F(\omega) $ to account for the differences between the exact DOS 
	
	Notice that $ \alpha^{2}F(\omega) $ in Fig.~\ref{fig:dos_a2F_lmdTc}(b) (for all fillings shown) is peaked at a frequency lower than the bare frequency $ \Omega=1.0t $, indicating that the phonon branch has renormalized. 
	Recall from Fig.~\ref{fig:Bqw} that phonon spectral weight shifts to lower frequency as the temperature is lowered, indicating corrections to the dispersion stemming from the formation of a CDW.  For fillings relevant to the $ T_c $-dome, the dispersion also shows signatures of competing CDW order developing, but now it is \emph{incommensurate}.  This implies that the system can approach a superconducting phase and simultaneously show signatures of CDW driven renormalization of the phonon dispersion.  Referring back to Fig.~\ref{fig:PD_comparison}, the $ T_{c}^{\text{CDW}} $ for the incommensurate CDW ($ \Omega=0.1t $) is smaller than the $ T_{c}^{\text{SC}} $-dome at $ \Omega=1.0t $.  Thus, at larger values of $ \Omega $, the incommensurate CDW correlations and their effects are present but never fully develop before the system becomes superconducting.
	    
	The dimensionless {\ep} coupling constant $ \lambda^{\alpha^{2}F} $ can be derived from the Eliashberg 
	function via
	\begin{equation}\label{Eq:lam^alpha2F}
	\lambda^{\alpha^{2}F} = 2\int_{0}^{\infty}\D \omega \frac{\alpha^{2}F(\omega)}{\omega}.  
	\end{equation} 
	This quantity measures the effective electron-phonon coupling after the phonon dispersion and the electron spectrum are renormalized by the interaction. It is plotted in Fig.~\ref{fig:dos_a2F_lmdTc}(c) and includes the same factor of $ N_{\text{F}}/N_{\text{F}}^{\text{fd}} $ introduced for $ \alpha^{2}F(\omega) $.  The approximately monotonic increase in this coupling across the range of filling $ n $ is generally favorable for pairing and thus also $ T_{c} $.  In the same figure, we also show the filling dependence on the logarithmic average frequency $ \omega_{\text{log}} $, given by
	\begin{equation}\label{Eq:wlog}
	\omega_{\text{log}}= \exp\left(\frac{2}{\lambda}\int_{0}^{\infty}\frac{\D\omega}{\omega}\alpha^{2}F(\omega)\ln(\omega)\right).
	\end{equation}
	The value of $ \omega_{\text{log}} $ monotonically decreases across the filling range, reflecting the softening of the phonon 
    branch as the CDW correlations develop.  Since $T_{c}\propto\omega_{\text{log}}$, we might expect a reduction in $ T_{c} $, however, there will be additional interplay with the change in $ \lambda^{\alpha^{2}F} $. 
	
	To supplement our results for the superconducting $ T_{c} $ obtained within the Migdal approximation we have also estimated the superconducting critical temperatures using three approaches commonly found in the literature.  First, we obtained $ T_{c}^{\alpha^{2}F} $ by solving the linearized gap equation [Eq.~(\ref{Eq:linGapEqnw})] using the computed $\alpha^2F(\omega)$ as input.  In Fig.~\ref{fig:dos_a2F_lmdTc}(c), the data for $ T_{c}^{\alpha^{2}F} $ most closely follows the $ T_{c} $ found within the Migdal approximation, exhibiting dome-like behavior.  
	If we exclude the corrective factor of $ N_{\text{F}}/N_{\text{F}}^{\text{fd}} $ in $ \alpha^{2}F(\omega), $ which accounts for the renormalization of the electron spectral function at the Fermi-level, the calculated $ T_{c}^{\alpha^{2}F} $ exhibits a monotonic increasing dependence on the filling $ n $ within the range considered here.   
	For the second and third estimates, we used the Allen-Dynes-modified McMillan (ADM) formula [Eq.~(\ref{Eq:ADM})]~\cite{Allen1975}, which we denote $ T_{c}^{\text{ADM}} $, and the ADM$ +f $ formula [Eq.~(\ref{Eq:ADMf})] to find $ T_{c}^{\text{ADM}+f} $.  Both of these formulas underestimate the critical temperature significantly, and only the $ T_{c}^{\text{ADM}+f} $ results exhibit non-monotonicity.  The wide discrepancy between the various methods for calculating $ T_{c} $ should be taken in to account when estimating the superconducting transition temperature from simplified, Fermi-surface averaged, isotropic Migdal-Eliashberg equations.  
	
	From this analysis, we conclude that the $ T_{c}^{\text{SC}} $ dome is tied to the competition of three renormalized quantities: the monotonic rise in $ \lambda^{\alpha^{2}F} $ which enhances pairing, the decrease in $ \omega_{\text{log}} $ which weakens the energy scale of pairing, and the non-monotonic filling dependence of the DOS around the Fermi-level.  In particular, it is interesting to note how the quasiparticle properties are renormalized at different fillings and how this affects the value of $ T_{c} $. While it is sometimes claimed in the literature that a superconducting dome is indicative of an unconventional pairing mechanism~\cite{Mazin2015}, our results indicate that this is not necessarily the case. A superconducting dome can be obtained in proximity to competing phases and this behavior will be more common in materials with narrow bandwidths (i.e., large values of $\Omega/t$). 
	
	\begin{figure}[h!]
	   	\includegraphics[width=0.9\columnwidth]{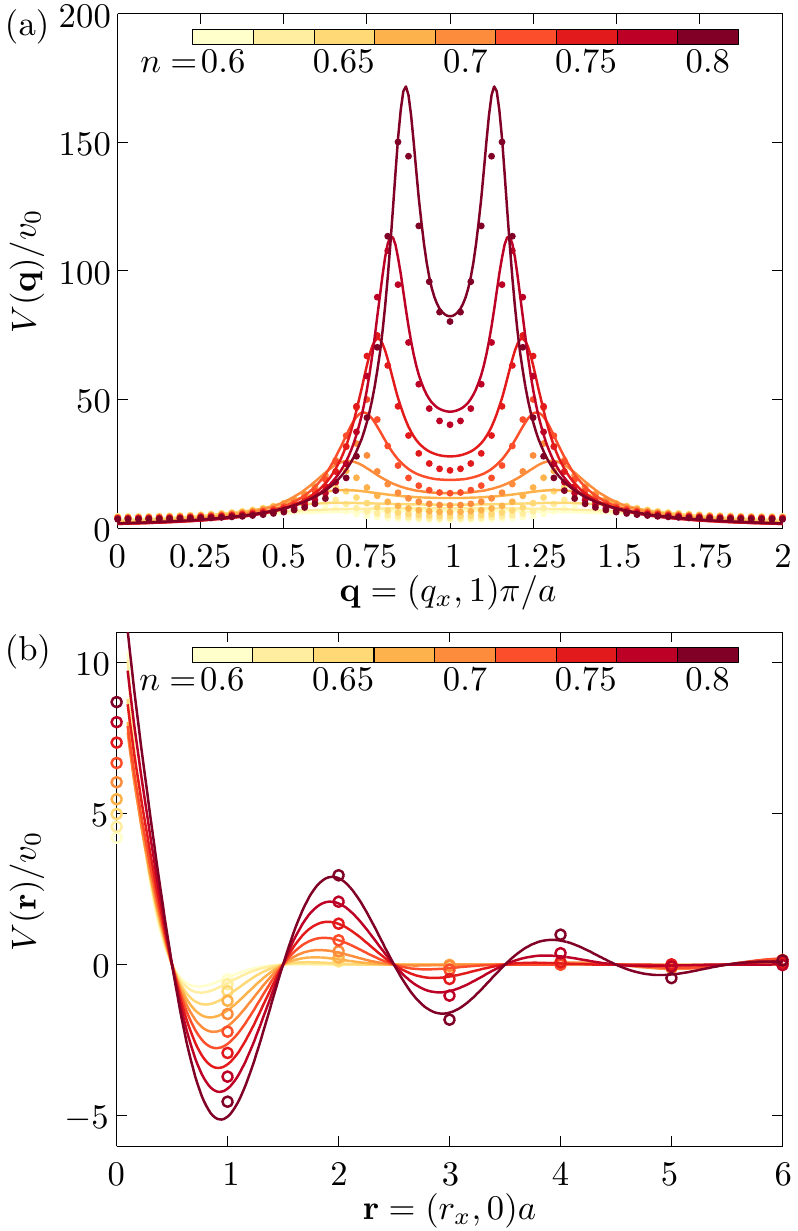}
		\caption{(Color online) The effective interaction $ V $ normalized by the bare interaction parameter $ v_{0}=2g^{2}/\Omega $ for various fillings $ n $ in (a) momentum space $ V(\bq,\ii\nu_{m}=0) $, and (b) along the $ r_{x} $-direction in real space $ V(r_{x},r_{y}=0,\ii\nu_{m}=0) $.  In (a), the data points are obtained from self-consistent calculations and the solid lines are from the corresponding fit to Lorentzian functions with the method of least squares.  In (b), the open circles are from the discrete Fourier transform of $ V(\bq) $ (dots) in panel (a); the lines are a cosine Fourier integral transform of the fitted Lorentzian function in panel (a).  The sign convention for the effective interaction in this figure is $ V>0 $ ($ V<0 $) for attractive (repulsive) interaction.  The colors of the lines or symbols indicate the value of the filling with bright (dark) color for small (large) filling $ n $.  The bare phonon frequency $ \Omega=t $ and the temperature is fixed at $ T=0.07t $ which is close to the superconducting critical temperatures for the chosen range of electron filling $ n $. }
		\label{fig:vq_vr}
	\end{figure} 
	
	\subsubsection*{Effective Interaction}

The bare effective electron-electron interaction for the Holstein model is purely local in real space and consequently uniform in momentum space. However, in the fully self-consistent approach, the effective interaction is related to the renormalized phonon propagator as shown by the double-wiggly line in the Migdal self-energy in Fig.~\ref{fig:Dressed_GF}(a).  Thus the formation of a momentum dependent renormalized phonon branch indicates that the effective electron-electron interaction must develop some real space structure and become nonlocal, and a deeper understanding of this interaction will shed light on the factors for enhancing or suppressing transition temperatures.  In this section we therefore examine the static effective interaction $ V(\bq)\equiv V(\bq,\ii\nu_{n}=0) $ in momentum space, and its (discrete) Fourier transform $ V(\br=\brm{R}_{i}) $ on the lattice.
	
	Defining the bare interaction parameter $ v(\bq)=2g^{2}/\Omega\equiv v_{0} $, we rewrite the bare {\ep} coupling
	constant $ \lambda=v_{0}/W $, where $ W $ is the band width.  
	The renormalized interaction can thus be related to the CDW susceptibility in Eq.~(\ref{Eq:Xcdw_q}) as  
	\begin{equation}\label{Eq:V(q)}
	V(\bq)=V(\bq,\ii\nu_{m}=0)=\frac{v_{0}}{1-v_{0}\chi_{0}(\bq)}
	\end{equation}  
	where $ \chi_{0}(\bq)\equiv\chi_{0}(\bq,\ii\nu_{m}=0) $.  In momentum space $ V(\bq) $ has either a single peak or quadruple peaks depending on whether the dominant CDW correlations  occur at the commensurate vector $ \bq_{\text{max}}=(\pi,\pi) $ or the incommensurate vectors given by $\bq_\text{max} = (\pi,\kappa\pi)$, and its symmetry-related points.  In this case, we are again interested in the range of fillings $ n\in[0.6,0.8] $ relevant to the SC-dome where we always observe  an incommensurate structure.  To better understand the analytical properties of the interaction, we have fit $ V(\bq) $ with a sum of four 2D Lorentzians
	\begin{equation}
	V(\bq) \approx \sum_{s=1}^{4}\frac{v_{\max}}{\xi^{2}|\bq-\bq_{\max,s}|^{2}+1},
	\end{equation}    
	where $ \bq_{\max,s} $ denotes all points related to $ \bq_{\max}=(\pi,\kappa\pi) $ by rotational symmetries, and $ \xi $, $ \kappa $, and $ v_{\max} $ are fitting parameters obtained by a least-squares fit to the data.  The symmetry of $ V(\bq) $ and its associated fit permit us to look along one cut of the 2D-momentum space $ \bq=(q_{x},1)\pi $ as shown in Fig.~\ref{fig:vq_vr}(a).  As the filling is increased for fixed $ \Omega=t $, and $ T=0.07t $, we see that the ordering vectors approach $ \bq_{\text{max}}=(\pi,\pi) $ and the effective interaction peak value $ v_{\max} $ becomes increasingly larger.
    
    The real-space structure of the effective interaction can be obtained form the Fourier transform 
    $ V(\br)=N^{-1}\sum_{\bq\in\text{BZ}}V(\bq)\e^{\ii\bq\cdot\br} $. Applying this procedure to the Lorentzian fits gives a  
    functional form 
    \begin{equation}\label{Eq:V(r)_fits}
    V(\br)\approx \frac{v_{\max}}{\pi\xi^{2}}R(r_{x},r_{y})K_{0}\left(\frac{|\br|}{\xi}\right)
    \end{equation}    
	where $ \br=(r_{x},r_{y}) $, $ R(r_{x},r_{y})=\cos(\pi r_{x} + \pi r_{y})[\cos(|\kappa-1|\pi r_{x}) + \cos(|\kappa-1|\pi r_{y})] $, and $ K_{\nu}(z) $ is the modified Bessel function of the second kind with $ \nu=0 $.  The effective interaction $ V(\br) $ obtained from our self-consistent Migdal calculations and its corresponding fits are plotted in Fig.~\ref{fig:vq_vr}(b) along the $ r_{x} $-axis, extending from the origin to a point where the effective interaction amplitudes taper off.  At low filling the interaction $ V(\br) $ is noticeably weaker and has a shorter estimated range $ \xi\approx0.5 $ (lattice spacing $ a=1 $) making it essentially local in extent.  As the filling increases, we see the emergence of oscillatory behavior and a more extended effective interaction.  Between $ 0.7\leq n \leq 0.8 $, on the decreasing side of the $ T_{c} $ dome, the correlation length approximately ranges from $ 2.3\leq\xi\leq5.7 $. 
	
	A recent paper~\cite{Langmann2018} by Langmann $ \textit{et al.} $ proposed that a superconducting dome is not necessarily associated with competing orders or exotic superconductivity, but instead can be the result of a finite-range potential.  Notably, they used a few phenomenological forms in real space for these finite-range potentials without self-consistently renormalizing them at different fillings across the dome.  Although we observe a finite-range potential and a $ T_{c} $-dome, we cannot deduce such a causation between them within our approach.  As was discussed in the previous section, the filling dependence of the interacting DOS $ N(\omega) $, the coupling $ \lambda^{\alpha^{2}F} $, and $ \omega_{\log} $ support the role of competing orders for our problem.  	   Moreover, the real-space structure of the effective interaction is directly linked to the formation of a $\bq$-dependent phonon dispersion due to the formation of long-range CDW correlations. 	
	
	\subsubsection*{Comparison to Previous Findings}
	Although we do not aim to present a comprehensive review of the literature surrounding the Holstein model and Holstein-like models, there are some other studies whose results can further contextualize the relevant physics in these models. For instance, some features of the temperature filling phase diagram have been explored by using different approaches such as Lang-Firsov canonical transformation~\cite{Grzybowski2007}, strong-coupling expansions~\cite{Freericks1996}, and QMC~\cite{Freericks1996a}. These studies are not explicitly focused on the weak-coupling regime where the Migdal approximation is most applicable, but there are some similarities worth noting.	

	In the Ref.~\onlinecite{Grzybowski2007}, a large CDW dome near half-filling and a small SC phase at lower fillings are found; however, in the phase diagram there are phase separated regions of non-ordered (NO) regions and a CDW, as well as CDW+SC regions. In the range of parameters used, we did not observe any indication of phase separation for the Holstein model.  A phase separation transition is an analog to the condensation transition between gas and liquid. In lattice models, it was first studied and observed in the extended Hubbard model~\cite{Gubernatis1985,Lin1986,Micnas1989} with a nearest-neighbor Hubbard interaction, later in the Hubbard model~\cite{Macridin2006,Aichhorn2007}, and in a {\ep} coupled case, the Hubbard-Holstein model~\cite{Castellani1995,Karakuzu2017}. Unlike CDW order, for a phase separation transition $\chi^\text{CDW}(\bq)$ diverges at $\bq = \mathbf{0}$. Therefore, for an anisotropic {\ep} coupling with strong forward scattering, it is possible to find a phase separation transition~\cite{Varelogiannis1998}. We note that we do see indications for this  physics for our buckling mode  case. 
	
	Two previously mentioned references~\cite{Freericks1996,Freericks1996a} focus on the addition of anharmonic phonon oscillations and find that anharmonicity enhances the overall size and spread of the SC region of the phase diagram.  However, the maximum $T_c$ is not enhanced significantly beyond that of the maximal values attained in the truly harmonic (Holstein) model. This result parallels the changes we discussed in section \ref{Sec:NNN_hopping} when diagonal hopping is permitted.  Although setting $ t^{\prime}\neq0 $ enhances the pairing (and thus $ T_{c}^{\text{SC}} $) relative to $ t^{\prime}=0 $, the improvement in the greatest value of $ T_{c}^{\text{SC}} $ is modestly small.
	
	\section{Summary and Conclusions} \label{sec:conclusions}
	We have presented a detailed analysis of the Holstein model within a fully self-consistent Migdal approximation, where both the renormalization of the electron and phonon properties are treated on an equal (approximate) footing. Using an efficient implementation based on fast Fourier transforms, we were able to simulate the model on lattice sizes much larger than those considered in the past. Our results revealed significant finite size effects when determining the CDW transition temperatures. This result should be kept in mind when simulating the Holstein model using numerically exact methods such as QMC that are limited to smaller lattice sizes.
	
	Comprehensive phase diagrams were mapped as a function of filling, revealing a mix of expected and unexpected features.  For large phonon frequencies, the dominant CDW ordering vector occurs at $\bq_\text{max} = (\pi, \pi)$; however, at smaller frequencies, superconducting $ T_{c} $'s are lowered sufficiently far to allow an incommensurate CDW phase to occupy a small range of fillings adjacent to the commensurate CDW phases near half-filling.  The $ s $-wave superconducting phase was present at lower values of the filling and was enhanced with increasing $ \Omega $.  Moreover, the filling dependence of $ T_{c}^{\text{SC}} $ was non-monotonic with a peak near $ n\approx0.7 $.  We addressed the possible factors responsible for this dome by studying the filling dependence of the interacting DOS, the renormalized {\ep} coupling, and $ \omega_{\log} $, and found that all three have competing effects on the pairing.  The DOS around the Fermi-surface exhibits non-monotonicity akin to $ T_{c}^{\text{SC}} $ across the same range of $ n $, while the renormalized {\ep} coupling and the $ \omega_{\log} $ increased and decreased over this range	respectively.  The latter two changes generally enhance and suppress pairing correlations, respectively, hence, they provide a measure of the competition.  Characterization of the effective electron-electron interaction $ V $ over both momentum and position space show that a very local (short range) interaction becomes further extended by several lattice constants as $ n $ approaches half-filling.  We cannot, however, attribute the origin of our superconducting $ T_{c} $ dome to this finite-range interaction in contrast to other approaches~\cite{Langmann2018}. 
	  
	The addition of a nonzero NNN hopping $t'<0\,(t'>0)$ to the electronic band promotes larger pairing correlations in the SC phase for $ n<1 $ ($ n>1 $) and suppresses the charge density correlations.  Asymmetry in the phase diagram across the full range of filling is consistent with the asymmetry in the DOS due to $ t'\neq 0 $ which makes more states available for pairing at fillings away from half-filling.  Moreover, the NNN hopping also weakens nesting at the Fermi surface, which in turn suppresses the CDW and thus reduces competition between the SC and CDW phases.  In addition, anisotropic {\ep} couplings and the corresponding correlations are compared.  Lastly, we showed that $g(\bq)$ modeling the breathing and buckling oxygen modes in the high-$ T_{c} $ cuprates induce a much larger range of charge density correlations and a suppression of superconductivity relative to the isotropic coupling.
	
	Implementation of the fully self-consistent Migdal-Eliashberg equations has been made freely available to the
	public~\cite{Note1}.
	
	\begin{acknowledgments}
		This work was supported by the Scientific Discovery through Advanced Computing (SciDAC) program funded by
		U.S. Department of Energy, Office of Science, Advanced Scientific Computing Research and Basic Energy
		Sciences, Division of Materials Sciences and Engineering.
		Y.~W. thanks A.-M.~S.~Tremblay for insightful discussions. Y.~W. acknowledges funding through the postdoctoral
		fellowship from Institut Quantique.		
	\end{acknowledgments}

	
	\appendix
	
	\section{Fast Fourier Transform of Physical Quantities} \label{sec:FFT}
	We consider the Fourier transform
	\begin{align}
	f(\br, \tau) = \frac{1}{N\beta} \sum_{\bk} \sum_{n=-\infty}^{\infty}
	\e^{\ii (\bk\cdot\br - \omega_n \tau)} f(\bk, \ii\omega_n),  \label{Eq:FFT}
	\end{align}
	where we choose the plane wave $\e^{\ii k\cdot x} = \e^{\ii (\bk\cdot\br-\omega t)} = \e^{\ii (\bk\cdot\br -
		\omega_n \tau)}$, and the inverse Fourier transform
	\begin{align}
	f(\bk, \ii\omega_n) = \sum_{\br} \int_{0}^{\beta} \D\tau\,
	\e^{-\ii (\bk\cdot\br - \omega_n \tau)} f(\br, \tau).  \label{Eq:iFFT}
	\end{align}
	Note that $\omega_n \tau = \omega t$ because the mappings between imaginary and real time-frequency
	variables are $\tau \to \ii t$ and $\ii\omega_n \to \omega$, respectively. Since the summations of $\bk$ and
	$\br$ range over the discretized first Brillouin zone and the lattice sites $\mathbf{R}_i$, respectively, the
	discrete Fourier transform using FFT is straightforward. Therefore, we need only discuss FFT between time and
	frequency domains.  As such, $\br$ and $\bk$ arguments will be suppressed in the function $f$ which stands for the
	Green's function $G$, self-energy $\Sigma$, the effective interaction $V$, or the irreducible susceptibility
	$\chi_0$.  As stated in the main text, $\omega_n = (2n+1)\pi/\beta$ is fermionic Matsubara frequency in $G(\ii\omega_n)$ and
	$\Sigma(\ii\omega_n)$, and $\omega_n = 2n\pi/\beta$ is bosonic Matsubara frequency in $V(\ii\omega_n)$ and
	$\chi_0(\ii\omega_n)$. In the self-consistent iterations, we perform a FFT on $G(\ii\omega_n)$ and $V(\ii\omega_n)$,
	and an iFFT on $\Sigma(\tau)$ and $\chi_0(\tau)$.
	
	For a practical calculation using the FFT we must use a finite number of Matsubara frequencies in the sum.  The
	uniform fermionic and bosonic Matsubara frequency grids $\omega_n = (2n+1)\pi/\beta$ and
	$\omega_n = 2n\pi/\beta$, where $-N_c \leq n \leq N_c -1$, are used in the sum for $G(\ii\omega_n)$ and
	$V(\ii\omega_n)$, respectively. Since $G(\ii\omega_n)\sim \mathcal{O}(\frac{1}{\omega_n})$ for large
	frequency, the Fourier transform featured in Eq.~(\ref{Eq:FFT}) has convergence issues for $G(\ii\omega_n)$ (see the
	discussion in the last part of Chapter~3 in Ref.~\onlinecite{Abrikosov1963}). Therefore, we subtract a
	function $G_0^\xi (\ii\omega_n) = 1/(\ii\omega_n - \xi)$ with the same large frequency dependence as
	$G(\ii\omega_n)$, use the fact that $\beta^{-1}\sum_{n=-\infty}^{\infty} \e^{-\ii \omega_n\tau} G_0^\xi
	(\ii\omega_n) = -\e^{(\beta-\tau)\xi}/(\e^{\beta\xi} + 1)$ for $\tau > 0$, and obtain the final formula for the
	Fourier transform
	\begin{align}
	G(\tau > 0) = \frac{1}{\beta} \sum_{n = -N_c}^{N_c - 1}
	\e^{-\ii \omega_n\tau} \tilde{G}(\ii\omega_n) - \frac{\e^{(\beta-\tau)\xi}}{\e^{\beta\xi} + 1},
	\end{align}
	where $\tilde{G}(\ii\omega_n) =  G(\ii\omega_n) - G_0^\xi (\ii\omega_n)$. Any constant value $\xi$ can be
	used, provided that $|\xi|\ll \omega_{n=N_c}$. In our calculations, we use $\xi = \xi_\bk$, i.e., the band
	dispersion. This choice is usually more accurate and requires smaller cut-off number $N_c$ than other
	choices. Setting $\xi = 0$, one recovers the familiar formula
	\begin{align}
	\lim_{\tau \to 0} G(\tau) = \frac{1}{\beta} \sum_{n=N_c}^{N_c - 1} G(\ii\omega_n)
	- \frac{1}{2} \sgn \tau.
	\end{align}
	which is Eq.~(17.36) on page 153 of Ref.~\onlinecite{Abrikosov1963}.
	
	For iFFT, we must reformulate the Fourier integral transform $f(\ii\omega_n) = \int_{0}^{\beta} \D\tau
	\,\e^{\ii\omega_n\tau} f(\tau)$ into a discrete Fourier transform. The pitfall and the method for doing this are
	discussed in Chapter~13.9 of Ref.~\onlinecite{Press1992} and in Chapter~2.10.2 of
	Ref.~\onlinecite{Davis1984}. As mentioned in the main text, the correct way to accomplish this task is to compute the
	Fourier integral exactly after interpolating $f(\tau)$ on the discrete $\tau$ grid using a continuous
	function such as spline or piecewise polynomial. Here, we choose the piecewise polynomial (Lagrange
	polynomial) and provide the explicit formula for the second (quadratic) order, which appears to be absent from
	literature.
	
	Denote $\delta = \beta/N_\tau$, the $\tau$ grid $\tau_l = (l-1)\delta$, where $1\leq l \leq N_\tau + 1$, $f_l
	= f(\tau_l)$, and $\tilde{f}_l = \e^{\ii\omega_n \tau_l} f(\tau_l)$. If the discontinuity exists at $\tau=0$
	and $\beta$, the end points should be understood as $0^+$ and $\beta^-$. The quadratic Lagrange polynomial
	used for interpolation is
	\begin{align}
	f(\tau)
	&\approx f_{l}  \frac{(\tau-\tau_{l+1})(\tau-\tau_{l+2})}{(\tau_{l}-\tau_{l+1})(\tau_{l}-\tau_{l+2})}   \notag \\
	&\quad + f_{l+1}\frac{(\tau-\tau_{l  })(\tau-\tau_{l+2})}{(\tau_{l+1}-\tau_{l})(\tau_{l+1}-\tau_{l+2})} \notag \\
	&\quad + f_{l+2}\frac{(\tau-\tau_{l  })(\tau-\tau_{l+1})}{(\tau_{l+2}-\tau_{l})(\tau_{l+2}-\tau_{l+1})},
	\end{align}
	where $\tau_l\leq \tau \leq \tau_{l+2}$. Using the above interpolating function, the final result for the
	Fourier integral transform is
	\begin{align}
	f(\ii\omega_n)
	&=       (c_1+c_2+c_3)\sum_{l=1}^{N_\tau} \tilde{f}_l -(p_3+c_3)f_\Delta          \notag\\
	&\quad + (s_1-c_2+p_3)\tilde{f}_1 + (s_2-c_3)\tilde{f}_2 + s_3\tilde{f}_3  \notag\\
	&\quad + p_1\tilde{f}_{N_\tau - 1} + (p_2-c_1)\tilde{f}_{N_\tau},
	\label{Eq:FFT2nd}
	\end{align}
	where
	
	\begin{align*}
	f_\Delta  &= f(\tau=0^{+}) - f(\tau=0^{-}) = \tilde{f}_1  - \tilde{f}_{N_\tau + 1},
	 \\
	c_1 &= \frac{\delta}{4}(I_{22} - 3I_{12} + 2I_{02}),
	\
	\,\,\,c_2  = -\frac{\delta \e^{-\ii\omega_n\delta}}{2}(I_{22} - 2I_{12}),
	\\
	c_3 &= \frac{\delta \e^{-2\ii\omega_n\delta}}{4}(I_{22} - I_{12}),
	\
	\,\,\,\,\,s_1  = \frac{\delta}{4}(I_{21} - 3I_{11} + 2I_{01}),
	\\
	s_2 &= -\frac{\delta \e^{-\ii\omega_n\delta}}{2}(I_{21} - 2I_{11}),
	\
	s_3  = \frac{\delta \e^{-2\ii\omega_n\delta}}{4}(I_{21} - I_{11}),
	\\
	p_1 &= \frac{\delta \e^{\ii\omega_n\delta}}{4}(I_{21} - I_{11}),
	\
	\quad\,\,\,\,\, p_2  = -\frac{\delta}{2}(I_{21} - I_{01}),
	\\
	p_3 &= \frac{\delta \e^{-\ii\omega_n\delta}}{4}(I_{21} + I_{11}),
	\end{align*}
	and $I_{vu} = \int_{0}^{u\delta}\D x\, \e^{\ii \omega_n x} x^{v}/\delta^{v+1}$, ($v\in \{0,1,2\}$), for which
	the closed analytical form can be found using integration by parts.  At this point, the sum in Eq.~(\ref{Eq:FFT2nd}) is
	suitable for evaluation by a FFT.
	
	Note that for $\omega_n=0$ (only for bosonic frequencies), we should use $I_{vu} = u^{v+1}/(v+1)$ and thus
	$(c_1,c_2,c_3)=(1,4,1)\delta/6$, $(s_1,s_2,s_3) = (p_3,p_2,p_1) =(5,8,-1)\delta/24$. Then
	Eq.~(\ref{Eq:FFT2nd}) becomes the usual composite Simpson's rule that applies to both even and odd number of
	intervals:
	\begin{align}
	f(\ii \omega_n=0)
	&=\delta \bigg[
	\frac{3}{8}f_{1} + \frac{7}{6}f_{2} + \frac{23}{24}f_{3} + \sum_{l=4}^{N_\tau - 2} f_l
	+ \frac{23}{24}f_{N_\tau - 1} \notag \\
	&\quad  + \frac{7}{6}f_{N_\tau} + \frac{3}{8}f_{N_\tau +1} \bigg],
	\end{align}
	which is a formula also found in Ref.~\onlinecite{Press1992}.\\
	
	\section{Linearized Isotropic Gap Equation and Modified McMillan $T_c$ Formulas}
	The linearized isotropic gap equation is given by~\cite{Carbotte1990,Allen1975}
	\begin{align}
	\Phi_n &=  \sum_{n'} K_{n,n'} \Phi_{n'}, \label{Eq:linGapEqnw}
	\end{align}
	where $K_{n,n'} = \pi T(\lambda_{n,n'} - \mu^{*})/(|\omega_{n'}|Z_{n'})$, $Z_n = 1 + (\pi T/\omega_n)
	\sum_{n'} \lambda_{n,n'} \omega_{n'} / |\omega_{n'}|$, and $\lambda_{n,n'}$ is given by the Eliashberg
	spectral function $\alpha^2 F(\omega)$
	\begin{align}
	\lambda_{n,n'} = \int_0^{\infty} \D\omega
	\frac{2\omega \alpha^2 F(\omega)}{(\omega_n - \omega_{n'})^2 + \omega^2}.
	\end{align}
	Here, $\omega_n$ and $\omega_{n'}$ are the fermionic Matsubara frequencies. The frequency dependent gap
	function is defined as $\Delta_n = \Phi_n/Z_n$. In this work, we set Coulomb pseudopotential $\mu^{*} = 0$
	when solving the linearized gap equation for $T_c$, which is defined as the temperature when the largest
	eigenvalue of the matrix $K_{n,n'}$ reaches unity from below.
	
	The Allen-Dynes-modified McMillan (ADM) $T_c$ formula~\cite{Allen1975} for weak coupling strength reads
	\begin{align}
	T_c &= \frac{\omega_\text{log}}{1.2} \exp
	\left( -\frac{1.04(1+\lambda)}{\lambda - \mu^*(1+0.62\lambda)}\right).
	\label{Eq:ADM}
	\end{align}
	
	For intermediate coupling strength, Allen and Dynes further modified the above formula with two additional
	coefficients $f_1$ and $f_2$. The corresponding $T_c$ formula~\cite{Allen1975} (ADM+$f$) is given by
	\begin{subequations}
		\label{Eq:ADMf}
		\begin{align}
		T_c &= \frac{f_1 f_2\omega_\text{log}}{1.2} \exp
		\left( -\frac{1.04(1+\lambda)}{\lambda - \mu^*(1+0.62\lambda)}\right),\\
		f_1 &= \left[ 1 + \left(\frac{\lambda}{2.46(1+3.8\mu^*)}\right)^{3/2} \right]^{1/3}, \\
		f_2 &= 1 + \frac{(\overline{\omega}_2/\omega_\text{log} - 1)\lambda^2}
		{\lambda^2 + [1.82(1+6.3\mu^*)(\overline{\omega}_2/\omega_\text{log})]^2}.
		\end{align}
	\end{subequations}
	Here, $\overline{\omega}_2 = \left[(2/\lambda) \int_{0}^{\infty} \omega \alpha^2 F(\omega) \D\omega
	\right]^{1/2}$ and $ \lambda=\lambda_{n,n}=\lambda^{\alpha^{2}F} $.  Again, it should be noted that we set $ \mu^{\ast}=0 $ when calculating $ T_{c} $.

	\bibliography{Migdal_v1,Migdal_v1Notes}

\begin{thebibliography}{112}%
\makeatletter
\providecommand \@ifxundefined [1]{%
 \@ifx{#1\undefined}
}%
\providecommand \@ifnum [1]{%
 \ifnum #1\expandafter \@firstoftwo
 \else \expandafter \@secondoftwo
 \fi
}%
\providecommand \@ifx [1]{%
 \ifx #1\expandafter \@firstoftwo
 \else \expandafter \@secondoftwo
 \fi
}%
\providecommand \natexlab [1]{#1}%
\providecommand \enquote  [1]{``#1''}%
\providecommand \bibnamefont  [1]{#1}%
\providecommand \bibfnamefont [1]{#1}%
\providecommand \citenamefont [1]{#1}%
\providecommand \href@noop [0]{\@secondoftwo}%
\providecommand \href [0]{\begingroup \@sanitize@url \@href}%
\providecommand \@href[1]{\@@startlink{#1}\@@href}%
\providecommand \@@href[1]{\endgroup#1\@@endlink}%
\providecommand \@sanitize@url [0]{\catcode `\\12\catcode `\$12\catcode
  `\&12\catcode `\#12\catcode `\^12\catcode `\_12\catcode `\%12\relax}%
\providecommand \@@startlink[1]{}%
\providecommand \@@endlink[0]{}%
\providecommand \url  [0]{\begingroup\@sanitize@url \@url }%
\providecommand \@url [1]{\endgroup\@href {#1}{\urlprefix }}%
\providecommand \urlprefix  [0]{URL }%
\providecommand \Eprint [0]{\href }%
\providecommand \doibase [0]{http://dx.doi.org/}%
\providecommand \selectlanguage [0]{\@gobble}%
\providecommand \bibinfo  [0]{\@secondoftwo}%
\providecommand \bibfield  [0]{\@secondoftwo}%
\providecommand \translation [1]{[#1]}%
\providecommand \BibitemOpen [0]{}%
\providecommand \bibitemStop [0]{}%
\providecommand \bibitemNoStop [0]{.\EOS\space}%
\providecommand \EOS [0]{\spacefactor3000\relax}%
\providecommand \BibitemShut  [1]{\csname bibitem#1\endcsname}%
\let\auto@bib@innerbib\@empty
\bibitem [{\citenamefont {Devreese}(1996)}]{Polaron}%
  \BibitemOpen
  \bibfield  {author} {\bibinfo {author} {\bibfnamefont {J.~T.}\ \bibnamefont
  {Devreese}},\ }\href@noop {} {\bibfield  {journal} {\bibinfo  {journal}
  {Encyclopedia of Applied Physics}\ }\textbf {\bibinfo {volume} {14}},\
  \bibinfo {pages} {383} (\bibinfo {year} {1996})}\BibitemShut {NoStop}%
\bibitem [{\citenamefont {Gr\"uner}(1988)}]{Gruner1988}%
  \BibitemOpen
  \bibfield  {author} {\bibinfo {author} {\bibfnamefont {G.}~\bibnamefont
  {Gr\"uner}},\ }\href {\doibase 10.1103/RevModPhys.60.1129} {\bibfield
  {journal} {\bibinfo  {journal} {Rev. Mod. Phys.}\ }\textbf {\bibinfo {volume}
  {60}},\ \bibinfo {pages} {1129} (\bibinfo {year} {1988})}\BibitemShut
  {NoStop}%
\bibitem [{\citenamefont {Bardeen}\ \emph {et~al.}(1957)\citenamefont
  {Bardeen}, \citenamefont {Cooper},\ and\ \citenamefont
  {Schrieffer}}]{Bardeen1957}%
  \BibitemOpen
  \bibfield  {author} {\bibinfo {author} {\bibfnamefont {J.}~\bibnamefont
  {Bardeen}}, \bibinfo {author} {\bibfnamefont {L.~N.}\ \bibnamefont {Cooper}},
  \ and\ \bibinfo {author} {\bibfnamefont {J.~R.}\ \bibnamefont {Schrieffer}},\
  }\href {\doibase 10.1103/PhysRev.108.1175} {\bibfield  {journal} {\bibinfo
  {journal} {Phys. Rev.}\ }\textbf {\bibinfo {volume} {108}},\ \bibinfo {pages}
  {1175} (\bibinfo {year} {1957})}\BibitemShut {NoStop}%
\bibitem [{\citenamefont {Scalapino}\ \emph {et~al.}(1966)\citenamefont
  {Scalapino}, \citenamefont {Schrieffer},\ and\ \citenamefont
  {Wilkins}}]{Scalapino1966}%
  \BibitemOpen
  \bibfield  {author} {\bibinfo {author} {\bibfnamefont {D.~J.}\ \bibnamefont
  {Scalapino}}, \bibinfo {author} {\bibfnamefont {J.~R.}\ \bibnamefont
  {Schrieffer}}, \ and\ \bibinfo {author} {\bibfnamefont {J.~W.}\ \bibnamefont
  {Wilkins}},\ }\href {\doibase 10.1103/PhysRev.148.263} {\bibfield  {journal}
  {\bibinfo  {journal} {Phys. Rev.}\ }\textbf {\bibinfo {volume} {148}},\
  \bibinfo {pages} {263} (\bibinfo {year} {1966})}\BibitemShut {NoStop}%
\bibitem [{\citenamefont {Bardeen}(1973)}]{Bardeen1973}%
  \BibitemOpen
  \bibfield  {author} {\bibinfo {author} {\bibfnamefont {J.}~\bibnamefont
  {Bardeen}},\ }\href {\doibase 10.1126/science.181.4106.1209} {\bibfield
  {journal} {\bibinfo  {journal} {Science}\ }\textbf {\bibinfo {volume}
  {181}},\ \bibinfo {pages} {1209} (\bibinfo {year} {1973})}\BibitemShut
  {NoStop}%
\bibitem [{\citenamefont {Marsiglio}\ and\ \citenamefont
  {Carbotte}(2008)}]{Marsiglio2008}%
  \BibitemOpen
  \bibfield  {author} {\bibinfo {author} {\bibfnamefont {F.}~\bibnamefont
  {Marsiglio}}\ and\ \bibinfo {author} {\bibfnamefont {J.~P.}\ \bibnamefont
  {Carbotte}},\ }\enquote {\bibinfo {title} {Electron-phonon
  superconductivity},}\ in\ \href {\doibase 10.1007/978-3-540-73253-2_3} {\emph
  {\bibinfo {booktitle} {Superconductivity: Conventional and Unconventional
  Superconductors}}},\ \bibinfo {editor} {edited by\ \bibinfo {editor}
  {\bibfnamefont {K.~H.}\ \bibnamefont {Bennemann}}\ and\ \bibinfo {editor}
  {\bibfnamefont {J.~B.}\ \bibnamefont {Ketterson}}}\ (\bibinfo  {publisher}
  {Springer Berlin Heidelberg},\ \bibinfo {address} {Berlin, Heidelberg},\
  \bibinfo {year} {2008})\ pp.\ \bibinfo {pages} {73--162}\BibitemShut
  {NoStop}%
\bibitem [{\citenamefont {Hamlin}(2015)}]{Hamlin2015}%
  \BibitemOpen
  \bibfield  {author} {\bibinfo {author} {\bibfnamefont {J.}~\bibnamefont
  {Hamlin}},\ }\href {\doibase 10.1016/j.physc.2015.02.032} {\bibfield
  {journal} {\bibinfo  {journal} {Physica C}\ }\textbf {\bibinfo {volume}
  {514}},\ \bibinfo {pages} {59} (\bibinfo {year} {2015})}\BibitemShut
  {NoStop}%
\bibitem [{\citenamefont {Buzea}\ and\ \citenamefont
  {Yamashita}(2001)}]{Buzea2001}%
  \BibitemOpen
  \bibfield  {author} {\bibinfo {author} {\bibfnamefont {C.}~\bibnamefont
  {Buzea}}\ and\ \bibinfo {author} {\bibfnamefont {T.}~\bibnamefont
  {Yamashita}},\ }\href {http://stacks.iop.org/0953-2048/14/i=11/a=201}
  {\bibfield  {journal} {\bibinfo  {journal} {Superconductor Science and
  Technology}\ }\textbf {\bibinfo {volume} {14}},\ \bibinfo {pages} {R115}
  (\bibinfo {year} {2001})}\BibitemShut {NoStop}%
\bibitem [{\citenamefont {Nagamatsu}\ \emph {et~al.}(2001)\citenamefont
  {Nagamatsu}, \citenamefont {Nakagawa}, \citenamefont {Muranaka},
  \citenamefont {Zenitani},\ and\ \citenamefont {Akimitsu}}]{Nagamatsu2001}%
  \BibitemOpen
  \bibfield  {author} {\bibinfo {author} {\bibfnamefont {J.}~\bibnamefont
  {Nagamatsu}}, \bibinfo {author} {\bibfnamefont {N.}~\bibnamefont {Nakagawa}},
  \bibinfo {author} {\bibfnamefont {T.}~\bibnamefont {Muranaka}}, \bibinfo
  {author} {\bibfnamefont {Y.}~\bibnamefont {Zenitani}}, \ and\ \bibinfo
  {author} {\bibfnamefont {J.}~\bibnamefont {Akimitsu}},\ }\href {\doibase
  10.1038/35065039} {\bibfield  {journal} {\bibinfo  {journal} {Nature}\
  }\textbf {\bibinfo {volume} {410}},\ \bibinfo {pages} {63} (\bibinfo {year}
  {2001})}\BibitemShut {NoStop}%
\bibitem [{\citenamefont {Drozdov}\ \emph {et~al.}(2015)\citenamefont
  {Drozdov}, \citenamefont {Eremets}, \citenamefont {Troyan}, \citenamefont
  {Ksenofontov},\ and\ \citenamefont {Shylin}}]{Drozdov2015}%
  \BibitemOpen
  \bibfield  {author} {\bibinfo {author} {\bibfnamefont {A.~P.}\ \bibnamefont
  {Drozdov}}, \bibinfo {author} {\bibfnamefont {M.~I.}\ \bibnamefont
  {Eremets}}, \bibinfo {author} {\bibfnamefont {I.~A.}\ \bibnamefont {Troyan}},
  \bibinfo {author} {\bibfnamefont {V.}~\bibnamefont {Ksenofontov}}, \ and\
  \bibinfo {author} {\bibfnamefont {S.~I.}\ \bibnamefont {Shylin}},\ }\href
  {\doibase 10.1038/nature14964} {\bibfield  {journal} {\bibinfo  {journal}
  {Nature}\ }\textbf {\bibinfo {volume} {525}},\ \bibinfo {pages} {73}
  (\bibinfo {year} {2015})}\BibitemShut {NoStop}%
\bibitem [{\citenamefont {Mazin}(2015{\natexlab{a}})}]{Mazin2015a}%
  \BibitemOpen
  \bibfield  {author} {\bibinfo {author} {\bibfnamefont {I.~I.}\ \bibnamefont
  {Mazin}},\ }\href {\doibase 10.1038/nature15203} {\bibfield  {journal}
  {\bibinfo  {journal} {Nature}\ }\textbf {\bibinfo {volume} {525}},\ \bibinfo
  {pages} {40} (\bibinfo {year} {2015}{\natexlab{a}})}\BibitemShut {NoStop}%
\bibitem [{\citenamefont {{Drozdov}}\ \emph {et~al.}(2018)\citenamefont
  {{Drozdov}}, \citenamefont {{Minkov}}, \citenamefont {{Besedin}},
  \citenamefont {{Kong}}, \citenamefont {{Kuzovnikov}}, \citenamefont
  {{Knyazev}},\ and\ \citenamefont {{Eremets}}}]{Drozdov2018}%
  \BibitemOpen
  \bibfield  {author} {\bibinfo {author} {\bibfnamefont {A.~P.}\ \bibnamefont
  {{Drozdov}}}, \bibinfo {author} {\bibfnamefont {V.~S.}\ \bibnamefont
  {{Minkov}}}, \bibinfo {author} {\bibfnamefont {S.~P.}\ \bibnamefont
  {{Besedin}}}, \bibinfo {author} {\bibfnamefont {P.~P.}\ \bibnamefont
  {{Kong}}}, \bibinfo {author} {\bibfnamefont {M.~A.}\ \bibnamefont
  {{Kuzovnikov}}}, \bibinfo {author} {\bibfnamefont {D.~A.}\ \bibnamefont
  {{Knyazev}}}, \ and\ \bibinfo {author} {\bibfnamefont {M.~I.}\ \bibnamefont
  {{Eremets}}},\ }\href@noop {} {\bibfield  {journal} {\bibinfo  {journal}
  {ArXiv e-prints}\ } (\bibinfo {year} {2018})},\ \Eprint
  {http://arxiv.org/abs/1808.07039} {arXiv:1808.07039 [cond-mat.supr-con]}
  \BibitemShut {NoStop}%
\bibitem [{\citenamefont {{Kruglov}}\ \emph {et~al.}(2018)\citenamefont
  {{Kruglov}}, \citenamefont {{Semenok}}, \citenamefont {{Szcz{\c
  e}{\'s}niak}}, \citenamefont {{Mahdi Davari Esfahani}}, \citenamefont
  {{Kvashnin}},\ and\ \citenamefont {{Oganov}}}]{Kruglov2018}%
  \BibitemOpen
  \bibfield  {author} {\bibinfo {author} {\bibfnamefont {I.~A.}\ \bibnamefont
  {{Kruglov}}}, \bibinfo {author} {\bibfnamefont {D.~V.}\ \bibnamefont
  {{Semenok}}}, \bibinfo {author} {\bibfnamefont {R.}~\bibnamefont {{Szcz{\c
  e}{\'s}niak}}}, \bibinfo {author} {\bibfnamefont {M.}~\bibnamefont {{Mahdi
  Davari Esfahani}}}, \bibinfo {author} {\bibfnamefont {A.~G.}\ \bibnamefont
  {{Kvashnin}}}, \ and\ \bibinfo {author} {\bibfnamefont {A.~R.}\ \bibnamefont
  {{Oganov}}},\ }\href@noop {} {\bibfield  {journal} {\bibinfo  {journal}
  {ArXiv e-prints}\ } (\bibinfo {year} {2018})},\ \Eprint
  {http://arxiv.org/abs/1810.01113} {arXiv:1810.01113 [cond-mat.supr-con]}
  \BibitemShut {NoStop}%
\bibitem [{\citenamefont {M\"{o}ller}\ \emph {et~al.}(2017)\citenamefont
  {M\"{o}ller}, \citenamefont {Sawatzky}, \citenamefont {Franz},\ and\
  \citenamefont {Berciu}}]{Moller2017}%
  \BibitemOpen
  \bibfield  {author} {\bibinfo {author} {\bibfnamefont {M.~M.}\ \bibnamefont
  {M\"{o}ller}}, \bibinfo {author} {\bibfnamefont {G.~A.}\ \bibnamefont
  {Sawatzky}}, \bibinfo {author} {\bibfnamefont {M.}~\bibnamefont {Franz}}, \
  and\ \bibinfo {author} {\bibfnamefont {M.}~\bibnamefont {Berciu}},\ }\href
  {\doibase 10.1038/s41467-017-02442-y} {\bibfield  {journal} {\bibinfo
  {journal} {Nat. Commun.}\ }\textbf {\bibinfo {volume} {8}},\ \bibinfo {pages}
  {2267} (\bibinfo {year} {2017})}\BibitemShut {NoStop}%
\bibitem [{\citenamefont {Baroni}\ \emph {et~al.}(2001)\citenamefont {Baroni},
  \citenamefont {de~Gironcoli}, \citenamefont {Dal~Corso},\ and\ \citenamefont
  {Giannozzi}}]{Baroni2001}%
  \BibitemOpen
  \bibfield  {author} {\bibinfo {author} {\bibfnamefont {S.}~\bibnamefont
  {Baroni}}, \bibinfo {author} {\bibfnamefont {S.}~\bibnamefont
  {de~Gironcoli}}, \bibinfo {author} {\bibfnamefont {A.}~\bibnamefont
  {Dal~Corso}}, \ and\ \bibinfo {author} {\bibfnamefont {P.}~\bibnamefont
  {Giannozzi}},\ }\href {\doibase 10.1103/RevModPhys.73.515} {\bibfield
  {journal} {\bibinfo  {journal} {Rev. Mod. Phys.}\ }\textbf {\bibinfo {volume}
  {73}},\ \bibinfo {pages} {515} (\bibinfo {year} {2001})}\BibitemShut
  {NoStop}%
\bibitem [{\citenamefont {Giustino}(2017)}]{Giustino2017}%
  \BibitemOpen
  \bibfield  {author} {\bibinfo {author} {\bibfnamefont {F.}~\bibnamefont
  {Giustino}},\ }\href {\doibase 10.1103/RevModPhys.89.015003} {\bibfield
  {journal} {\bibinfo  {journal} {Rev. Mod. Phys.}\ }\textbf {\bibinfo {volume}
  {89}},\ \bibinfo {pages} {015003} (\bibinfo {year} {2017})}\BibitemShut
  {NoStop}%
\bibitem [{\citenamefont {Choi}\ \emph {et~al.}(2002)\citenamefont {Choi},
  \citenamefont {Roundy}, \citenamefont {Sun}, \citenamefont {Cohen},\ and\
  \citenamefont {Louie}}]{Choi2002a}%
  \BibitemOpen
  \bibfield  {author} {\bibinfo {author} {\bibfnamefont {H.~J.}\ \bibnamefont
  {Choi}}, \bibinfo {author} {\bibfnamefont {D.}~\bibnamefont {Roundy}},
  \bibinfo {author} {\bibfnamefont {H.}~\bibnamefont {Sun}}, \bibinfo {author}
  {\bibfnamefont {M.~L.}\ \bibnamefont {Cohen}}, \ and\ \bibinfo {author}
  {\bibfnamefont {S.~G.}\ \bibnamefont {Louie}},\ }\href {\doibase
  10.1103/PhysRevB.66.020513} {\bibfield  {journal} {\bibinfo  {journal} {Phys.
  Rev. B}\ }\textbf {\bibinfo {volume} {66}},\ \bibinfo {pages} {020513}
  (\bibinfo {year} {2002})}\BibitemShut {NoStop}%
\bibitem [{\citenamefont {L\"uders}\ \emph {et~al.}(2005)\citenamefont
  {L\"uders}, \citenamefont {Marques}, \citenamefont {Lathiotakis},
  \citenamefont {Floris}, \citenamefont {Profeta}, \citenamefont {Fast},
  \citenamefont {Continenza}, \citenamefont {Massidda},\ and\ \citenamefont
  {Gross}}]{Luders2005}%
  \BibitemOpen
  \bibfield  {author} {\bibinfo {author} {\bibfnamefont {M.}~\bibnamefont
  {L\"uders}}, \bibinfo {author} {\bibfnamefont {M.~A.~L.}\ \bibnamefont
  {Marques}}, \bibinfo {author} {\bibfnamefont {N.~N.}\ \bibnamefont
  {Lathiotakis}}, \bibinfo {author} {\bibfnamefont {A.}~\bibnamefont {Floris}},
  \bibinfo {author} {\bibfnamefont {G.}~\bibnamefont {Profeta}}, \bibinfo
  {author} {\bibfnamefont {L.}~\bibnamefont {Fast}}, \bibinfo {author}
  {\bibfnamefont {A.}~\bibnamefont {Continenza}}, \bibinfo {author}
  {\bibfnamefont {S.}~\bibnamefont {Massidda}}, \ and\ \bibinfo {author}
  {\bibfnamefont {E.~K.~U.}\ \bibnamefont {Gross}},\ }\href {\doibase
  10.1103/PhysRevB.72.024545} {\bibfield  {journal} {\bibinfo  {journal} {Phys.
  Rev. B}\ }\textbf {\bibinfo {volume} {72}},\ \bibinfo {pages} {024545}
  (\bibinfo {year} {2005})}\BibitemShut {NoStop}%
\bibitem [{\citenamefont {Marques}\ \emph {et~al.}(2005)\citenamefont
  {Marques}, \citenamefont {L\"uders}, \citenamefont {Lathiotakis},
  \citenamefont {Profeta}, \citenamefont {Floris}, \citenamefont {Fast},
  \citenamefont {Continenza}, \citenamefont {Gross},\ and\ \citenamefont
  {Massidda}}]{Marques2005}%
  \BibitemOpen
  \bibfield  {author} {\bibinfo {author} {\bibfnamefont {M.~A.~L.}\
  \bibnamefont {Marques}}, \bibinfo {author} {\bibfnamefont {M.}~\bibnamefont
  {L\"uders}}, \bibinfo {author} {\bibfnamefont {N.~N.}\ \bibnamefont
  {Lathiotakis}}, \bibinfo {author} {\bibfnamefont {G.}~\bibnamefont
  {Profeta}}, \bibinfo {author} {\bibfnamefont {A.}~\bibnamefont {Floris}},
  \bibinfo {author} {\bibfnamefont {L.}~\bibnamefont {Fast}}, \bibinfo {author}
  {\bibfnamefont {A.}~\bibnamefont {Continenza}}, \bibinfo {author}
  {\bibfnamefont {E.~K.~U.}\ \bibnamefont {Gross}}, \ and\ \bibinfo {author}
  {\bibfnamefont {S.}~\bibnamefont {Massidda}},\ }\href {\doibase
  10.1103/PhysRevB.72.024546} {\bibfield  {journal} {\bibinfo  {journal} {Phys.
  Rev. B}\ }\textbf {\bibinfo {volume} {72}},\ \bibinfo {pages} {024546}
  (\bibinfo {year} {2005})}\BibitemShut {NoStop}%
\bibitem [{\citenamefont {Floris}\ \emph {et~al.}(2005)\citenamefont {Floris},
  \citenamefont {Profeta}, \citenamefont {Lathiotakis}, \citenamefont
  {L\"uders}, \citenamefont {Marques}, \citenamefont {Franchini}, \citenamefont
  {Gross}, \citenamefont {Continenza},\ and\ \citenamefont
  {Massidda}}]{Floris2005}%
  \BibitemOpen
  \bibfield  {author} {\bibinfo {author} {\bibfnamefont {A.}~\bibnamefont
  {Floris}}, \bibinfo {author} {\bibfnamefont {G.}~\bibnamefont {Profeta}},
  \bibinfo {author} {\bibfnamefont {N.~N.}\ \bibnamefont {Lathiotakis}},
  \bibinfo {author} {\bibfnamefont {M.}~\bibnamefont {L\"uders}}, \bibinfo
  {author} {\bibfnamefont {M.~A.~L.}\ \bibnamefont {Marques}}, \bibinfo
  {author} {\bibfnamefont {C.}~\bibnamefont {Franchini}}, \bibinfo {author}
  {\bibfnamefont {E.~K.~U.}\ \bibnamefont {Gross}}, \bibinfo {author}
  {\bibfnamefont {A.}~\bibnamefont {Continenza}}, \ and\ \bibinfo {author}
  {\bibfnamefont {S.}~\bibnamefont {Massidda}},\ }\href {\doibase
  10.1103/PhysRevLett.94.037004} {\bibfield  {journal} {\bibinfo  {journal}
  {Phys. Rev. Lett.}\ }\textbf {\bibinfo {volume} {94}},\ \bibinfo {pages}
  {037004} (\bibinfo {year} {2005})}\BibitemShut {NoStop}%
\bibitem [{\citenamefont {Margine}\ and\ \citenamefont
  {Giustino}(2013)}]{Margine2013}%
  \BibitemOpen
  \bibfield  {author} {\bibinfo {author} {\bibfnamefont {E.~R.}\ \bibnamefont
  {Margine}}\ and\ \bibinfo {author} {\bibfnamefont {F.}~\bibnamefont
  {Giustino}},\ }\href {\doibase 10.1103/PhysRevB.87.024505} {\bibfield
  {journal} {\bibinfo  {journal} {Phys. Rev. B}\ }\textbf {\bibinfo {volume}
  {87}},\ \bibinfo {pages} {024505} (\bibinfo {year} {2013})}\BibitemShut
  {NoStop}%
\bibitem [{\citenamefont {Holstein}(1959)}]{Holstein1959}%
  \BibitemOpen
  \bibfield  {author} {\bibinfo {author} {\bibfnamefont {T.}~\bibnamefont
  {Holstein}},\ }\href {\doibase 10.1016/0003-4916(59)90002-8} {\bibfield
  {journal} {\bibinfo  {journal} {Ann. Phys.}\ }\textbf {\bibinfo {volume}
  {8}},\ \bibinfo {pages} {325} (\bibinfo {year} {1959})}\BibitemShut {NoStop}%
\bibitem [{\citenamefont {Han}\ \emph {et~al.}(2002)\citenamefont {Han},
  \citenamefont {Lin},\ and\ \citenamefont {Wang}}]{Han2002}%
  \BibitemOpen
  \bibfield  {author} {\bibinfo {author} {\bibfnamefont {R.}~\bibnamefont
  {Han}}, \bibinfo {author} {\bibfnamefont {Z.}~\bibnamefont {Lin}}, \ and\
  \bibinfo {author} {\bibfnamefont {K.}~\bibnamefont {Wang}},\ }\href {\doibase
  10.1103/PhysRevB.65.174303} {\bibfield  {journal} {\bibinfo  {journal} {Phys.
  Rev. B}\ }\textbf {\bibinfo {volume} {65}},\ \bibinfo {pages} {174303}
  (\bibinfo {year} {2002})}\BibitemShut {NoStop}%
\bibitem [{\citenamefont {Berciu}(2007)}]{Berciu2007}%
  \BibitemOpen
  \bibfield  {author} {\bibinfo {author} {\bibfnamefont {M.}~\bibnamefont
  {Berciu}},\ }\href {\doibase 10.1103/PhysRevB.75.081101} {\bibfield
  {journal} {\bibinfo  {journal} {Phys. Rev. B}\ }\textbf {\bibinfo {volume}
  {75}},\ \bibinfo {pages} {081101} (\bibinfo {year} {2007})}\BibitemShut
  {NoStop}%
\bibitem [{\citenamefont {Zhang}\ \emph {et~al.}(2009)\citenamefont {Zhang},
  \citenamefont {Liu}, \citenamefont {Chen}, \citenamefont {Wang},\ and\
  \citenamefont {Wang}}]{Zhang2009b}%
  \BibitemOpen
  \bibfield  {author} {\bibinfo {author} {\bibfnamefont {Y.-Y.}\ \bibnamefont
  {Zhang}}, \bibinfo {author} {\bibfnamefont {T.}~\bibnamefont {Liu}}, \bibinfo
  {author} {\bibfnamefont {Q.-H.}\ \bibnamefont {Chen}}, \bibinfo {author}
  {\bibfnamefont {X.}~\bibnamefont {Wang}}, \ and\ \bibinfo {author}
  {\bibfnamefont {K.-L.}\ \bibnamefont {Wang}},\ }\href
  {http://stacks.iop.org/0953-8984/21/i=41/a=415601} {\bibfield  {journal}
  {\bibinfo  {journal} {J. Phys.: Condens. Matter}\ }\textbf {\bibinfo {volume}
  {21}},\ \bibinfo {pages} {415601} (\bibinfo {year} {2009})}\BibitemShut
  {NoStop}%
\bibitem [{\citenamefont {Mahan}(1990)}]{Mahan1990}%
  \BibitemOpen
  \bibfield  {author} {\bibinfo {author} {\bibfnamefont {G.~D.}\ \bibnamefont
  {Mahan}},\ }\href@noop {} {\emph {\bibinfo {title} {Many-Particle
  Physics}}},\ \bibinfo {edition} {2nd}\ ed.\ (\bibinfo  {publisher} {Plenum
  Press},\ \bibinfo {address} {New York, NY},\ \bibinfo {year}
  {1990})\BibitemShut {NoStop}%
\bibitem [{\citenamefont {Grzybowski}\ and\ \citenamefont
  {Micnas}(2007)}]{Grzybowski2007}%
  \BibitemOpen
  \bibfield  {author} {\bibinfo {author} {\bibfnamefont {P.}~\bibnamefont
  {Grzybowski}}\ and\ \bibinfo {author} {\bibfnamefont {P.}~\bibnamefont
  {Micnas}},\ }\href@noop {} {\bibfield  {journal} {\bibinfo  {journal} {Acta
  Phys. Pol. A}\ }\textbf {\bibinfo {volume} {111}},\ \bibinfo {pages} {455}
  (\bibinfo {year} {2007})}\BibitemShut {NoStop}%
\bibitem [{\citenamefont {Migdal}(1958)}]{Migdal1958}%
  \BibitemOpen
  \bibfield  {author} {\bibinfo {author} {\bibfnamefont {A.~B.}\ \bibnamefont
  {Migdal}},\ }\href@noop {} {\bibfield  {journal} {\bibinfo  {journal} {Sov.
  Phys. JETP}\ }\textbf {\bibinfo {volume} {7}},\ \bibinfo {pages} {996}
  (\bibinfo {year} {1958})},\ \bibinfo {note} {[Zh. Eksp. Teor. Fiz.
  \textbf{34}, 1438 (1958)]}\BibitemShut {NoStop}%
\bibitem [{\citenamefont {Eliashberg}(1960)}]{Eliashberg1960}%
  \BibitemOpen
  \bibfield  {author} {\bibinfo {author} {\bibfnamefont {G.~M.}\ \bibnamefont
  {Eliashberg}},\ }\href@noop {} {\bibfield  {journal} {\bibinfo  {journal}
  {Sov. Phys. JETP}\ }\textbf {\bibinfo {volume} {11}},\ \bibinfo {pages} {696}
  (\bibinfo {year} {1960})},\ \bibinfo {note} {[Zh. Eksp. Teor. Fiz.
  \textbf{38}, 966 (1960)]}\BibitemShut {NoStop}%
\bibitem [{\citenamefont {Engelsberg}\ and\ \citenamefont
  {Schrieffer}(1963)}]{Engelsberg1963}%
  \BibitemOpen
  \bibfield  {author} {\bibinfo {author} {\bibfnamefont {S.}~\bibnamefont
  {Engelsberg}}\ and\ \bibinfo {author} {\bibfnamefont {J.~R.}\ \bibnamefont
  {Schrieffer}},\ }\href {\doibase 10.1103/PhysRev.131.993} {\bibfield
  {journal} {\bibinfo  {journal} {Phys. Rev.}\ }\textbf {\bibinfo {volume}
  {131}},\ \bibinfo {pages} {993} (\bibinfo {year} {1963})}\BibitemShut
  {NoStop}%
\bibitem [{\citenamefont {Marsiglio}(1990)}]{Marsiglio1990}%
  \BibitemOpen
  \bibfield  {author} {\bibinfo {author} {\bibfnamefont {F.}~\bibnamefont
  {Marsiglio}},\ }\href {\doibase 10.1103/PhysRevB.42.2416} {\bibfield
  {journal} {\bibinfo  {journal} {Phys. Rev. B}\ }\textbf {\bibinfo {volume}
  {42}},\ \bibinfo {pages} {2416} (\bibinfo {year} {1990})}\BibitemShut
  {NoStop}%
\bibitem [{\citenamefont {Berger}\ \emph {et~al.}(1995)\citenamefont {Berger},
  \citenamefont {Val\'a\v{s}ek},\ and\ \citenamefont {von~der
  Linden}}]{Berger1995}%
  \BibitemOpen
  \bibfield  {author} {\bibinfo {author} {\bibfnamefont {E.}~\bibnamefont
  {Berger}}, \bibinfo {author} {\bibfnamefont {P.}~\bibnamefont
  {Val\'a\v{s}ek}}, \ and\ \bibinfo {author} {\bibfnamefont {W.}~\bibnamefont
  {von~der Linden}},\ }\href {\doibase 10.1103/PhysRevB.52.4806} {\bibfield
  {journal} {\bibinfo  {journal} {Phys. Rev. B}\ }\textbf {\bibinfo {volume}
  {52}},\ \bibinfo {pages} {4806} (\bibinfo {year} {1995})}\BibitemShut
  {NoStop}%
\bibitem [{\citenamefont {Abrikosov}\ \emph {et~al.}(1963)\citenamefont
  {Abrikosov}, \citenamefont {Gor'kov},\ and\ \citenamefont
  {Dzyaloshinski}}]{Abrikosov1963}%
  \BibitemOpen
  \bibfield  {author} {\bibinfo {author} {\bibfnamefont {A.}~\bibnamefont
  {Abrikosov}}, \bibinfo {author} {\bibfnamefont {L.}~\bibnamefont {Gor'kov}},
  \ and\ \bibinfo {author} {\bibfnamefont {I.}~\bibnamefont {Dzyaloshinski}},\
  }\href@noop {} {\emph {\bibinfo {title} {Methods of Quantum Field Theory in
  Statistical Physics}}}\ (\bibinfo  {publisher} {Prentice-Hall},\ \bibinfo
  {address} {Englewood Cliffs, NJ},\ \bibinfo {year} {1963})\BibitemShut
  {NoStop}%
\bibitem [{\citenamefont {Berciu}(2006)}]{Berciu2006}%
  \BibitemOpen
  \bibfield  {author} {\bibinfo {author} {\bibfnamefont {M.}~\bibnamefont
  {Berciu}},\ }\href {\doibase 10.1103/PhysRevLett.97.036402} {\bibfield
  {journal} {\bibinfo  {journal} {Phys. Rev. Lett.}\ }\textbf {\bibinfo
  {volume} {97}},\ \bibinfo {pages} {036402} (\bibinfo {year}
  {2006})}\BibitemShut {NoStop}%
\bibitem [{\citenamefont {Ebrahimnejad}\ and\ \citenamefont
  {Berciu}(2012)}]{Ebrahimnejad2012}%
  \BibitemOpen
  \bibfield  {author} {\bibinfo {author} {\bibfnamefont {H.}~\bibnamefont
  {Ebrahimnejad}}\ and\ \bibinfo {author} {\bibfnamefont {M.}~\bibnamefont
  {Berciu}},\ }\href {\doibase 10.1103/PhysRevB.86.205109} {\bibfield
  {journal} {\bibinfo  {journal} {Phys. Rev. B}\ }\textbf {\bibinfo {volume}
  {86}},\ \bibinfo {pages} {205109} (\bibinfo {year} {2012})}\BibitemShut
  {NoStop}%
\bibitem [{\citenamefont {Hirsch}\ and\ \citenamefont
  {Fradkin}(1982)}]{Hirsch1982}%
  \BibitemOpen
  \bibfield  {author} {\bibinfo {author} {\bibfnamefont {J.~E.}\ \bibnamefont
  {Hirsch}}\ and\ \bibinfo {author} {\bibfnamefont {E.}~\bibnamefont
  {Fradkin}},\ }\href {\doibase 10.1103/PhysRevLett.49.402} {\bibfield
  {journal} {\bibinfo  {journal} {Phys. Rev. Lett.}\ }\textbf {\bibinfo
  {volume} {49}},\ \bibinfo {pages} {402} (\bibinfo {year} {1982})}\BibitemShut
  {NoStop}%
\bibitem [{\citenamefont {Scalettar}\ \emph {et~al.}(1989)\citenamefont
  {Scalettar}, \citenamefont {Bickers},\ and\ \citenamefont
  {Scalapino}}]{Scalettar1989}%
  \BibitemOpen
  \bibfield  {author} {\bibinfo {author} {\bibfnamefont {R.~T.}\ \bibnamefont
  {Scalettar}}, \bibinfo {author} {\bibfnamefont {N.~E.}\ \bibnamefont
  {Bickers}}, \ and\ \bibinfo {author} {\bibfnamefont {D.~J.}\ \bibnamefont
  {Scalapino}},\ }\href {\doibase 10.1103/PhysRevB.40.197} {\bibfield
  {journal} {\bibinfo  {journal} {Phys. Rev. B}\ }\textbf {\bibinfo {volume}
  {40}},\ \bibinfo {pages} {197} (\bibinfo {year} {1989})}\BibitemShut
  {NoStop}%
\bibitem [{\citenamefont {Noack}\ \emph {et~al.}(1991)\citenamefont {Noack},
  \citenamefont {Scalapino},\ and\ \citenamefont {Scalettar}}]{Noack1991}%
  \BibitemOpen
  \bibfield  {author} {\bibinfo {author} {\bibfnamefont {R.~M.}\ \bibnamefont
  {Noack}}, \bibinfo {author} {\bibfnamefont {D.~J.}\ \bibnamefont
  {Scalapino}}, \ and\ \bibinfo {author} {\bibfnamefont {R.~T.}\ \bibnamefont
  {Scalettar}},\ }\href {\doibase 10.1103/PhysRevLett.66.778} {\bibfield
  {journal} {\bibinfo  {journal} {Phys. Rev. Lett.}\ }\textbf {\bibinfo
  {volume} {66}},\ \bibinfo {pages} {778} (\bibinfo {year} {1991})}\BibitemShut
  {NoStop}%
\bibitem [{\citenamefont {Veki\'{c}}\ \emph {et~al.}(1992)\citenamefont
  {Veki\'{c}}, \citenamefont {Noack},\ and\ \citenamefont {White}}]{Vekic1992}%
  \BibitemOpen
  \bibfield  {author} {\bibinfo {author} {\bibfnamefont {M.}~\bibnamefont
  {Veki\'{c}}}, \bibinfo {author} {\bibfnamefont {R.~M.}\ \bibnamefont
  {Noack}}, \ and\ \bibinfo {author} {\bibfnamefont {S.~R.}\ \bibnamefont
  {White}},\ }\href {\doibase 10.1103/PhysRevB.46.271} {\bibfield  {journal}
  {\bibinfo  {journal} {Phys. Rev. B}\ }\textbf {\bibinfo {volume} {46}},\
  \bibinfo {pages} {271} (\bibinfo {year} {1992})}\BibitemShut {NoStop}%
\bibitem [{\citenamefont {Niyaz}\ \emph {et~al.}(1993)\citenamefont {Niyaz},
  \citenamefont {Gubernatis}, \citenamefont {Scalettar},\ and\ \citenamefont
  {Fong}}]{Niyaz1993}%
  \BibitemOpen
  \bibfield  {author} {\bibinfo {author} {\bibfnamefont {P.}~\bibnamefont
  {Niyaz}}, \bibinfo {author} {\bibfnamefont {J.~E.}\ \bibnamefont
  {Gubernatis}}, \bibinfo {author} {\bibfnamefont {R.~T.}\ \bibnamefont
  {Scalettar}}, \ and\ \bibinfo {author} {\bibfnamefont {C.~Y.}\ \bibnamefont
  {Fong}},\ }\href {\doibase 10.1103/PhysRevB.48.16011} {\bibfield  {journal}
  {\bibinfo  {journal} {Phys. Rev. B}\ }\textbf {\bibinfo {volume} {48}},\
  \bibinfo {pages} {16011} (\bibinfo {year} {1993})}\BibitemShut {NoStop}%
\bibitem [{\citenamefont {Hohenadler}\ \emph {et~al.}(2004)\citenamefont
  {Hohenadler}, \citenamefont {Evertz},\ and\ \citenamefont {von~der
  Linden}}]{Hohenadler2004}%
  \BibitemOpen
  \bibfield  {author} {\bibinfo {author} {\bibfnamefont {M.}~\bibnamefont
  {Hohenadler}}, \bibinfo {author} {\bibfnamefont {H.~G.}\ \bibnamefont
  {Evertz}}, \ and\ \bibinfo {author} {\bibfnamefont {W.}~\bibnamefont {von~der
  Linden}},\ }\href {\doibase 10.1103/PhysRevB.69.024301} {\bibfield  {journal}
  {\bibinfo  {journal} {Phys. Rev. B}\ }\textbf {\bibinfo {volume} {69}},\
  \bibinfo {pages} {024301} (\bibinfo {year} {2004})}\BibitemShut {NoStop}%
\bibitem [{\citenamefont {Huang}\ \emph {et~al.}(2003)\citenamefont {Huang},
  \citenamefont {Hanke}, \citenamefont {Arrigoni},\ and\ \citenamefont
  {Scalapino}}]{Huang2003}%
  \BibitemOpen
  \bibfield  {author} {\bibinfo {author} {\bibfnamefont {Z.~B.}\ \bibnamefont
  {Huang}}, \bibinfo {author} {\bibfnamefont {W.}~\bibnamefont {Hanke}},
  \bibinfo {author} {\bibfnamefont {E.}~\bibnamefont {Arrigoni}}, \ and\
  \bibinfo {author} {\bibfnamefont {D.~J.}\ \bibnamefont {Scalapino}},\ }\href
  {\doibase 10.1103/PhysRevB.68.220507} {\bibfield  {journal} {\bibinfo
  {journal} {Phys. Rev. B}\ }\textbf {\bibinfo {volume} {68}},\ \bibinfo
  {pages} {220507} (\bibinfo {year} {2003})}\BibitemShut {NoStop}%
\bibitem [{\citenamefont {Johnston}\ \emph {et~al.}(2013)\citenamefont
  {Johnston}, \citenamefont {Nowadnick}, \citenamefont {Kung}, \citenamefont
  {Moritz}, \citenamefont {Scalettar},\ and\ \citenamefont
  {Devereaux}}]{Johnston2013}%
  \BibitemOpen
  \bibfield  {author} {\bibinfo {author} {\bibfnamefont {S.}~\bibnamefont
  {Johnston}}, \bibinfo {author} {\bibfnamefont {E.~A.}\ \bibnamefont
  {Nowadnick}}, \bibinfo {author} {\bibfnamefont {Y.~F.}\ \bibnamefont {Kung}},
  \bibinfo {author} {\bibfnamefont {B.}~\bibnamefont {Moritz}}, \bibinfo
  {author} {\bibfnamefont {R.~T.}\ \bibnamefont {Scalettar}}, \ and\ \bibinfo
  {author} {\bibfnamefont {T.~P.}\ \bibnamefont {Devereaux}},\ }\href {\doibase
  10.1103/PhysRevB.87.235133} {\bibfield  {journal} {\bibinfo  {journal} {Phys.
  Rev. B}\ }\textbf {\bibinfo {volume} {87}},\ \bibinfo {pages} {235133}
  (\bibinfo {year} {2013})}\BibitemShut {NoStop}%
\bibitem [{\citenamefont {Li}\ and\ \citenamefont {Johnston}(2015)}]{Li2015b}%
  \BibitemOpen
  \bibfield  {author} {\bibinfo {author} {\bibfnamefont {S.}~\bibnamefont
  {Li}}\ and\ \bibinfo {author} {\bibfnamefont {S.}~\bibnamefont {Johnston}},\
  }\href {http://stacks.iop.org/0295-5075/109/i=2/a=27007} {\bibfield
  {journal} {\bibinfo  {journal} {EPL (Europhysics Letters)}\ }\textbf
  {\bibinfo {volume} {109}},\ \bibinfo {pages} {27007} (\bibinfo {year}
  {2015})}\BibitemShut {NoStop}%
\bibitem [{\citenamefont {Esterlis}\ \emph {et~al.}(2018)\citenamefont
  {Esterlis}, \citenamefont {Nosarzewski}, \citenamefont {Huang}, \citenamefont
  {Moritz}, \citenamefont {Devereaux}, \citenamefont {Scalapino},\ and\
  \citenamefont {Kivelson}}]{Esterlis2018}%
  \BibitemOpen
  \bibfield  {author} {\bibinfo {author} {\bibfnamefont {I.}~\bibnamefont
  {Esterlis}}, \bibinfo {author} {\bibfnamefont {B.}~\bibnamefont
  {Nosarzewski}}, \bibinfo {author} {\bibfnamefont {E.~W.}\ \bibnamefont
  {Huang}}, \bibinfo {author} {\bibfnamefont {B.}~\bibnamefont {Moritz}},
  \bibinfo {author} {\bibfnamefont {T.~P.}\ \bibnamefont {Devereaux}}, \bibinfo
  {author} {\bibfnamefont {D.~J.}\ \bibnamefont {Scalapino}}, \ and\ \bibinfo
  {author} {\bibfnamefont {S.~A.}\ \bibnamefont {Kivelson}},\ }\href {\doibase
  10.1103/PhysRevB.97.140501} {\bibfield  {journal} {\bibinfo  {journal} {Phys.
  Rev. B}\ }\textbf {\bibinfo {volume} {97}},\ \bibinfo {pages} {140501}
  (\bibinfo {year} {2018})}\BibitemShut {NoStop}%
\bibitem [{\citenamefont {Costa}\ \emph {et~al.}(2018)\citenamefont {Costa},
  \citenamefont {Blommel}, \citenamefont {Chiu}, \citenamefont {Batrouni},\
  and\ \citenamefont {Scalettar}}]{Costa2018}%
  \BibitemOpen
  \bibfield  {author} {\bibinfo {author} {\bibfnamefont {N.~C.}\ \bibnamefont
  {Costa}}, \bibinfo {author} {\bibfnamefont {T.}~\bibnamefont {Blommel}},
  \bibinfo {author} {\bibfnamefont {W.-T.}\ \bibnamefont {Chiu}}, \bibinfo
  {author} {\bibfnamefont {G.}~\bibnamefont {Batrouni}}, \ and\ \bibinfo
  {author} {\bibfnamefont {R.~T.}\ \bibnamefont {Scalettar}},\ }\href {\doibase
  10.1103/PhysRevLett.120.187003} {\bibfield  {journal} {\bibinfo  {journal}
  {Phys. Rev. Lett.}\ }\textbf {\bibinfo {volume} {120}},\ \bibinfo {pages}
  {187003} (\bibinfo {year} {2018})}\BibitemShut {NoStop}%
\bibitem [{\citenamefont {Alder}\ \emph {et~al.}(1997)\citenamefont {Alder},
  \citenamefont {Runge},\ and\ \citenamefont {Scalettar}}]{Alder1997}%
  \BibitemOpen
  \bibfield  {author} {\bibinfo {author} {\bibfnamefont {B.~J.}\ \bibnamefont
  {Alder}}, \bibinfo {author} {\bibfnamefont {K.~J.}\ \bibnamefont {Runge}}, \
  and\ \bibinfo {author} {\bibfnamefont {R.~T.}\ \bibnamefont {Scalettar}},\
  }\href {\doibase 10.1103/PhysRevLett.79.3022} {\bibfield  {journal} {\bibinfo
   {journal} {Phys. Rev. Lett.}\ }\textbf {\bibinfo {volume} {79}},\ \bibinfo
  {pages} {3022} (\bibinfo {year} {1997})}\BibitemShut {NoStop}%
\bibitem [{\citenamefont {Ohgoe}\ and\ \citenamefont
  {Imada}(2014)}]{Ohgoe2014}%
  \BibitemOpen
  \bibfield  {author} {\bibinfo {author} {\bibfnamefont {T.}~\bibnamefont
  {Ohgoe}}\ and\ \bibinfo {author} {\bibfnamefont {M.}~\bibnamefont {Imada}},\
  }\href {\doibase 10.1103/PhysRevB.89.195139} {\bibfield  {journal} {\bibinfo
  {journal} {Phys. Rev. B}\ }\textbf {\bibinfo {volume} {89}},\ \bibinfo
  {pages} {195139} (\bibinfo {year} {2014})}\BibitemShut {NoStop}%
\bibitem [{\citenamefont {Ohgoe}\ and\ \citenamefont
  {Imada}(2017)}]{Ohgoe2017}%
  \BibitemOpen
  \bibfield  {author} {\bibinfo {author} {\bibfnamefont {T.}~\bibnamefont
  {Ohgoe}}\ and\ \bibinfo {author} {\bibfnamefont {M.}~\bibnamefont {Imada}},\
  }\href {\doibase 10.1103/PhysRevLett.119.197001} {\bibfield  {journal}
  {\bibinfo  {journal} {Phys. Rev. Lett.}\ }\textbf {\bibinfo {volume} {119}},\
  \bibinfo {pages} {197001} (\bibinfo {year} {2017})}\BibitemShut {NoStop}%
\bibitem [{\citenamefont {Freericks}\ \emph {et~al.}(1993)\citenamefont
  {Freericks}, \citenamefont {Jarrell},\ and\ \citenamefont
  {Scalapino}}]{Freericks1993}%
  \BibitemOpen
  \bibfield  {author} {\bibinfo {author} {\bibfnamefont {J.~K.}\ \bibnamefont
  {Freericks}}, \bibinfo {author} {\bibfnamefont {M.}~\bibnamefont {Jarrell}},
  \ and\ \bibinfo {author} {\bibfnamefont {D.~J.}\ \bibnamefont {Scalapino}},\
  }\href {\doibase 10.1103/PhysRevB.48.6302} {\bibfield  {journal} {\bibinfo
  {journal} {Phys. Rev. B}\ }\textbf {\bibinfo {volume} {48}},\ \bibinfo
  {pages} {6302} (\bibinfo {year} {1993})}\BibitemShut {NoStop}%
\bibitem [{\citenamefont {Ciuchi}\ \emph {et~al.}(1997)\citenamefont {Ciuchi},
  \citenamefont {de~Pasquale}, \citenamefont {Fratini},\ and\ \citenamefont
  {Feinberg}}]{Ciuchi1997}%
  \BibitemOpen
  \bibfield  {author} {\bibinfo {author} {\bibfnamefont {S.}~\bibnamefont
  {Ciuchi}}, \bibinfo {author} {\bibfnamefont {F.}~\bibnamefont {de~Pasquale}},
  \bibinfo {author} {\bibfnamefont {S.}~\bibnamefont {Fratini}}, \ and\
  \bibinfo {author} {\bibfnamefont {D.}~\bibnamefont {Feinberg}},\ }\href
  {\doibase 10.1103/PhysRevB.56.4494} {\bibfield  {journal} {\bibinfo
  {journal} {Phys. Rev. B}\ }\textbf {\bibinfo {volume} {56}},\ \bibinfo
  {pages} {4494} (\bibinfo {year} {1997})}\BibitemShut {NoStop}%
\bibitem [{\citenamefont {Freericks}\ \emph {et~al.}(1998)\citenamefont
  {Freericks}, \citenamefont {Zlati\'{c}}, \citenamefont {Chung},\ and\
  \citenamefont {Jarrell}}]{Freericks1998}%
  \BibitemOpen
  \bibfield  {author} {\bibinfo {author} {\bibfnamefont {J.~K.}\ \bibnamefont
  {Freericks}}, \bibinfo {author} {\bibfnamefont {V.}~\bibnamefont
  {Zlati\'{c}}}, \bibinfo {author} {\bibfnamefont {W.}~\bibnamefont {Chung}}, \
  and\ \bibinfo {author} {\bibfnamefont {M.}~\bibnamefont {Jarrell}},\ }\href
  {\doibase 10.1103/PhysRevB.58.11613} {\bibfield  {journal} {\bibinfo
  {journal} {Phys. Rev. B}\ }\textbf {\bibinfo {volume} {58}},\ \bibinfo
  {pages} {11613} (\bibinfo {year} {1998})}\BibitemShut {NoStop}%
\bibitem [{\citenamefont {Meyer}\ \emph {et~al.}(2002)\citenamefont {Meyer},
  \citenamefont {Hewson},\ and\ \citenamefont {Bulla}}]{Meyer2002}%
  \BibitemOpen
  \bibfield  {author} {\bibinfo {author} {\bibfnamefont {D.}~\bibnamefont
  {Meyer}}, \bibinfo {author} {\bibfnamefont {A.~C.}\ \bibnamefont {Hewson}}, \
  and\ \bibinfo {author} {\bibfnamefont {R.}~\bibnamefont {Bulla}},\ }\href
  {\doibase 10.1103/PhysRevLett.89.196401} {\bibfield  {journal} {\bibinfo
  {journal} {Phys. Rev. Lett.}\ }\textbf {\bibinfo {volume} {89}},\ \bibinfo
  {pages} {196401} (\bibinfo {year} {2002})}\BibitemShut {NoStop}%
\bibitem [{\citenamefont {Capone}\ and\ \citenamefont
  {Ciuchi}(2003)}]{Capone2003}%
  \BibitemOpen
  \bibfield  {author} {\bibinfo {author} {\bibfnamefont {M.}~\bibnamefont
  {Capone}}\ and\ \bibinfo {author} {\bibfnamefont {S.}~\bibnamefont
  {Ciuchi}},\ }\href {\doibase 10.1103/PhysRevLett.91.186405} {\bibfield
  {journal} {\bibinfo  {journal} {Phys. Rev. Lett.}\ }\textbf {\bibinfo
  {volume} {91}},\ \bibinfo {pages} {186405} (\bibinfo {year}
  {2003})}\BibitemShut {NoStop}%
\bibitem [{\citenamefont {Hague}\ and\ \citenamefont
  {d'Ambrumenil}(2008)}]{Hague2008}%
  \BibitemOpen
  \bibfield  {author} {\bibinfo {author} {\bibfnamefont {J.~P.}\ \bibnamefont
  {Hague}}\ and\ \bibinfo {author} {\bibfnamefont {N.}~\bibnamefont
  {d'Ambrumenil}},\ }\href {\doibase 10.1007/s10909-008-9800-z} {\bibfield
  {journal} {\bibinfo  {journal} {J. Low Temp. Phys.}\ }\textbf {\bibinfo
  {volume} {151}},\ \bibinfo {pages} {1149} (\bibinfo {year}
  {2008})}\BibitemShut {NoStop}%
\bibitem [{\citenamefont {Bauer}\ \emph {et~al.}(2011)\citenamefont {Bauer},
  \citenamefont {Han},\ and\ \citenamefont {Gunnarsson}}]{Bauer2011}%
  \BibitemOpen
  \bibfield  {author} {\bibinfo {author} {\bibfnamefont {J.}~\bibnamefont
  {Bauer}}, \bibinfo {author} {\bibfnamefont {J.~E.}\ \bibnamefont {Han}}, \
  and\ \bibinfo {author} {\bibfnamefont {O.}~\bibnamefont {Gunnarsson}},\
  }\href {\doibase 10.1103/PhysRevB.84.184531} {\bibfield  {journal} {\bibinfo
  {journal} {Phys. Rev. B}\ }\textbf {\bibinfo {volume} {84}},\ \bibinfo
  {pages} {184531} (\bibinfo {year} {2011})}\BibitemShut {NoStop}%
\bibitem [{\citenamefont {Murakami}\ \emph {et~al.}(2014)\citenamefont
  {Murakami}, \citenamefont {Werner}, \citenamefont {Tsuji},\ and\
  \citenamefont {Aoki}}]{Murakami2014}%
  \BibitemOpen
  \bibfield  {author} {\bibinfo {author} {\bibfnamefont {Y.}~\bibnamefont
  {Murakami}}, \bibinfo {author} {\bibfnamefont {P.}~\bibnamefont {Werner}},
  \bibinfo {author} {\bibfnamefont {N.}~\bibnamefont {Tsuji}}, \ and\ \bibinfo
  {author} {\bibfnamefont {H.}~\bibnamefont {Aoki}},\ }\href {\doibase
  10.1103/PhysRevLett.113.266404} {\bibfield  {journal} {\bibinfo  {journal}
  {Phys. Rev. Lett.}\ }\textbf {\bibinfo {volume} {113}},\ \bibinfo {pages}
  {266404} (\bibinfo {year} {2014})}\BibitemShut {NoStop}%
\bibitem [{\citenamefont {Tezuka}\ \emph {et~al.}(2007)\citenamefont {Tezuka},
  \citenamefont {Arita},\ and\ \citenamefont {Aoki}}]{Tezuka2007}%
  \BibitemOpen
  \bibfield  {author} {\bibinfo {author} {\bibfnamefont {M.}~\bibnamefont
  {Tezuka}}, \bibinfo {author} {\bibfnamefont {R.}~\bibnamefont {Arita}}, \
  and\ \bibinfo {author} {\bibfnamefont {H.}~\bibnamefont {Aoki}},\ }\href
  {\doibase 10.1103/PhysRevB.76.155114} {\bibfield  {journal} {\bibinfo
  {journal} {Phys. Rev. B}\ }\textbf {\bibinfo {volume} {76}},\ \bibinfo
  {pages} {155114} (\bibinfo {year} {2007})}\BibitemShut {NoStop}%
\bibitem [{\citenamefont {Eliashberg}(1961)}]{Eliashberg1961}%
  \BibitemOpen
  \bibfield  {author} {\bibinfo {author} {\bibfnamefont {G.~M.}\ \bibnamefont
  {Eliashberg}},\ }\href@noop {} {\bibfield  {journal} {\bibinfo  {journal}
  {Sov. Phys. JETP}\ }\textbf {\bibinfo {volume} {12}},\ \bibinfo {pages}
  {1000} (\bibinfo {year} {1961})}\BibitemShut {NoStop}%
\bibitem [{\citenamefont {Aperis}\ \emph {et~al.}(2015)\citenamefont {Aperis},
  \citenamefont {Maldonado},\ and\ \citenamefont {Oppeneer}}]{Aperis2015}%
  \BibitemOpen
  \bibfield  {author} {\bibinfo {author} {\bibfnamefont {A.}~\bibnamefont
  {Aperis}}, \bibinfo {author} {\bibfnamefont {P.}~\bibnamefont {Maldonado}}, \
  and\ \bibinfo {author} {\bibfnamefont {P.~M.}\ \bibnamefont {Oppeneer}},\
  }\href {\doibase 10.1103/PhysRevB.92.054516} {\bibfield  {journal} {\bibinfo
  {journal} {Phys. Rev. B}\ }\textbf {\bibinfo {volume} {92}},\ \bibinfo
  {pages} {054516} (\bibinfo {year} {2015})}\BibitemShut {NoStop}%
\bibitem [{\citenamefont {Ponc\'{e}}\ \emph {et~al.}(2016)\citenamefont
  {Ponc\'{e}}, \citenamefont {Margine}, \citenamefont {Verdi},\ and\
  \citenamefont {Giustino}}]{Ponce2016}%
  \BibitemOpen
  \bibfield  {author} {\bibinfo {author} {\bibfnamefont {S.}~\bibnamefont
  {Ponc\'{e}}}, \bibinfo {author} {\bibfnamefont {E.}~\bibnamefont {Margine}},
  \bibinfo {author} {\bibfnamefont {C.}~\bibnamefont {Verdi}}, \ and\ \bibinfo
  {author} {\bibfnamefont {F.}~\bibnamefont {Giustino}},\ }\href {\doibase
  10.1016/j.cpc.2016.07.028} {\bibfield  {journal} {\bibinfo  {journal}
  {Comput. Phys. Commun.}\ }\textbf {\bibinfo {volume} {209}},\ \bibinfo
  {pages} {116} (\bibinfo {year} {2016})}\BibitemShut {NoStop}%
\bibitem [{\citenamefont {Cuk}\ \emph {et~al.}(2004)\citenamefont {Cuk},
  \citenamefont {Baumberger}, \citenamefont {Lu}, \citenamefont {Ingle},
  \citenamefont {Zhou}, \citenamefont {Eisaki}, \citenamefont {Kaneko},
  \citenamefont {Hussain}, \citenamefont {Devereaux}, \citenamefont {Nagaosa},\
  and\ \citenamefont {Shen}}]{Cuk2004}%
  \BibitemOpen
  \bibfield  {author} {\bibinfo {author} {\bibfnamefont {T.}~\bibnamefont
  {Cuk}}, \bibinfo {author} {\bibfnamefont {F.}~\bibnamefont {Baumberger}},
  \bibinfo {author} {\bibfnamefont {D.~H.}\ \bibnamefont {Lu}}, \bibinfo
  {author} {\bibfnamefont {N.}~\bibnamefont {Ingle}}, \bibinfo {author}
  {\bibfnamefont {X.~J.}\ \bibnamefont {Zhou}}, \bibinfo {author}
  {\bibfnamefont {H.}~\bibnamefont {Eisaki}}, \bibinfo {author} {\bibfnamefont
  {N.}~\bibnamefont {Kaneko}}, \bibinfo {author} {\bibfnamefont
  {Z.}~\bibnamefont {Hussain}}, \bibinfo {author} {\bibfnamefont {T.~P.}\
  \bibnamefont {Devereaux}}, \bibinfo {author} {\bibfnamefont {N.}~\bibnamefont
  {Nagaosa}}, \ and\ \bibinfo {author} {\bibfnamefont {Z.-X.}\ \bibnamefont
  {Shen}},\ }\href {\doibase 10.1103/PhysRevLett.93.117003} {\bibfield
  {journal} {\bibinfo  {journal} {Phys. Rev. Lett.}\ }\textbf {\bibinfo
  {volume} {93}},\ \bibinfo {pages} {117003} (\bibinfo {year}
  {2004})}\BibitemShut {NoStop}%
\bibitem [{\citenamefont {Shai}\ \emph {et~al.}(2013)\citenamefont {Shai},
  \citenamefont {Adamo}, \citenamefont {Shen}, \citenamefont {Brooks},
  \citenamefont {Harter}, \citenamefont {Monkman}, \citenamefont {Burganov},
  \citenamefont {Schlom},\ and\ \citenamefont {Shen}}]{Shai2013}%
  \BibitemOpen
  \bibfield  {author} {\bibinfo {author} {\bibfnamefont {D.~E.}\ \bibnamefont
  {Shai}}, \bibinfo {author} {\bibfnamefont {C.}~\bibnamefont {Adamo}},
  \bibinfo {author} {\bibfnamefont {D.~W.}\ \bibnamefont {Shen}}, \bibinfo
  {author} {\bibfnamefont {C.~M.}\ \bibnamefont {Brooks}}, \bibinfo {author}
  {\bibfnamefont {J.~W.}\ \bibnamefont {Harter}}, \bibinfo {author}
  {\bibfnamefont {E.~J.}\ \bibnamefont {Monkman}}, \bibinfo {author}
  {\bibfnamefont {B.}~\bibnamefont {Burganov}}, \bibinfo {author}
  {\bibfnamefont {D.~G.}\ \bibnamefont {Schlom}}, \ and\ \bibinfo {author}
  {\bibfnamefont {K.~M.}\ \bibnamefont {Shen}},\ }\href {\doibase
  10.1103/PhysRevLett.110.087004} {\bibfield  {journal} {\bibinfo  {journal}
  {Phys. Rev. Lett.}\ }\textbf {\bibinfo {volume} {110}},\ \bibinfo {pages}
  {087004} (\bibinfo {year} {2013})}\BibitemShut {NoStop}%
\bibitem [{\citenamefont {Yang}\ \emph {et~al.}(2016)\citenamefont {Yang},
  \citenamefont {Liu}, \citenamefont {Fan}, \citenamefont {Yao}, \citenamefont
  {Xiang}, \citenamefont {Zhang}, \citenamefont {Li}, \citenamefont {Li},
  \citenamefont {Liu}, \citenamefont {Shen},\ and\ \citenamefont
  {Jiang}}]{Yang2016}%
  \BibitemOpen
  \bibfield  {author} {\bibinfo {author} {\bibfnamefont {H.~F.}\ \bibnamefont
  {Yang}}, \bibinfo {author} {\bibfnamefont {Z.~T.}\ \bibnamefont {Liu}},
  \bibinfo {author} {\bibfnamefont {C.~C.}\ \bibnamefont {Fan}}, \bibinfo
  {author} {\bibfnamefont {Q.}~\bibnamefont {Yao}}, \bibinfo {author}
  {\bibfnamefont {P.}~\bibnamefont {Xiang}}, \bibinfo {author} {\bibfnamefont
  {K.~L.}\ \bibnamefont {Zhang}}, \bibinfo {author} {\bibfnamefont {M.~Y.}\
  \bibnamefont {Li}}, \bibinfo {author} {\bibfnamefont {H.}~\bibnamefont {Li}},
  \bibinfo {author} {\bibfnamefont {J.~S.}\ \bibnamefont {Liu}}, \bibinfo
  {author} {\bibfnamefont {D.~W.}\ \bibnamefont {Shen}}, \ and\ \bibinfo
  {author} {\bibfnamefont {M.~H.}\ \bibnamefont {Jiang}},\ }\href {\doibase
  10.1103/PhysRevB.93.121102} {\bibfield  {journal} {\bibinfo  {journal} {Phys.
  Rev. B}\ }\textbf {\bibinfo {volume} {93}},\ \bibinfo {pages} {121102}
  (\bibinfo {year} {2016})}\BibitemShut {NoStop}%
\bibitem [{\citenamefont {Margine}\ \emph {et~al.}(2016)\citenamefont
  {Margine}, \citenamefont {Lambert},\ and\ \citenamefont
  {Giustino}}]{Margine2016}%
  \BibitemOpen
  \bibfield  {author} {\bibinfo {author} {\bibfnamefont {E.~R.}\ \bibnamefont
  {Margine}}, \bibinfo {author} {\bibfnamefont {H.}~\bibnamefont {Lambert}}, \
  and\ \bibinfo {author} {\bibfnamefont {F.}~\bibnamefont {Giustino}},\ }\href
  {http://dx.doi.org/10.1038/srep21414} {\bibfield  {journal} {\bibinfo
  {journal} {Scientific Reports}\ }\textbf {\bibinfo {volume} {6}},\ \bibinfo
  {pages} {21414} (\bibinfo {year} {2016})}\BibitemShut {NoStop}%
\bibitem [{\citenamefont {Verdi}\ \emph {et~al.}(2017)\citenamefont {Verdi},
  \citenamefont {Caruso},\ and\ \citenamefont {Giustino}}]{Verdi2017}%
  \BibitemOpen
  \bibfield  {author} {\bibinfo {author} {\bibfnamefont {C.}~\bibnamefont
  {Verdi}}, \bibinfo {author} {\bibfnamefont {F.}~\bibnamefont {Caruso}}, \
  and\ \bibinfo {author} {\bibfnamefont {F.}~\bibnamefont {Giustino}},\ }\href
  {http://dx.doi.org/10.1038/ncomms15769} {\bibfield  {journal} {\bibinfo
  {journal} {Nature Communications}\ }\textbf {\bibinfo {volume} {8}},\
  \bibinfo {pages} {15769} (\bibinfo {year} {2017})}\BibitemShut {NoStop}%
\bibitem [{\citenamefont {Swartz}\ \emph {et~al.}(2018)\citenamefont {Swartz},
  \citenamefont {Inoue}, \citenamefont {Merz}, \citenamefont {Hikita},
  \citenamefont {Raghu}, \citenamefont {Devereaux}, \citenamefont {Johnston},\
  and\ \citenamefont {Hwang}}]{Swartz2018}%
  \BibitemOpen
  \bibfield  {author} {\bibinfo {author} {\bibfnamefont {A.~G.}\ \bibnamefont
  {Swartz}}, \bibinfo {author} {\bibfnamefont {H.}~\bibnamefont {Inoue}},
  \bibinfo {author} {\bibfnamefont {T.~A.}\ \bibnamefont {Merz}}, \bibinfo
  {author} {\bibfnamefont {Y.}~\bibnamefont {Hikita}}, \bibinfo {author}
  {\bibfnamefont {S.}~\bibnamefont {Raghu}}, \bibinfo {author} {\bibfnamefont
  {T.~P.}\ \bibnamefont {Devereaux}}, \bibinfo {author} {\bibfnamefont
  {S.}~\bibnamefont {Johnston}}, \ and\ \bibinfo {author} {\bibfnamefont
  {H.~Y.}\ \bibnamefont {Hwang}},\ }\href {\doibase 10.1073/pnas.1713916115}
  {\bibfield  {journal} {\bibinfo  {journal} {Proceedings of the National
  Academy of Sciences}\ }\textbf {\bibinfo {volume} {82}},\ \bibinfo {pages}
  {224304} (\bibinfo {year} {2018})}\BibitemShut {NoStop}%
\bibitem [{\citenamefont {Grimaldi}\ \emph {et~al.}(1995)\citenamefont
  {Grimaldi}, \citenamefont {Pietronero},\ and\ \citenamefont
  {Str\"assler}}]{Grimaldi1995}%
  \BibitemOpen
  \bibfield  {author} {\bibinfo {author} {\bibfnamefont {C.}~\bibnamefont
  {Grimaldi}}, \bibinfo {author} {\bibfnamefont {L.}~\bibnamefont
  {Pietronero}}, \ and\ \bibinfo {author} {\bibfnamefont {S.}~\bibnamefont
  {Str\"assler}},\ }\href {\doibase 10.1103/PhysRevLett.75.1158} {\bibfield
  {journal} {\bibinfo  {journal} {Phys. Rev. Lett.}\ }\textbf {\bibinfo
  {volume} {75}},\ \bibinfo {pages} {1158} (\bibinfo {year}
  {1995})}\BibitemShut {NoStop}%
\bibitem [{\citenamefont {Gunnarsson}(1997)}]{Gunnarsson1997}%
  \BibitemOpen
  \bibfield  {author} {\bibinfo {author} {\bibfnamefont {O.}~\bibnamefont
  {Gunnarsson}},\ }\href {\doibase 10.1103/RevModPhys.69.575} {\bibfield
  {journal} {\bibinfo  {journal} {Rev. Mod. Phys.}\ }\textbf {\bibinfo {volume}
  {69}},\ \bibinfo {pages} {575} (\bibinfo {year} {1997})}\BibitemShut
  {NoStop}%
\bibitem [{\citenamefont {Subedi}\ and\ \citenamefont
  {Boeri}(2011)}]{Subedi2011}%
  \BibitemOpen
  \bibfield  {author} {\bibinfo {author} {\bibfnamefont {A.}~\bibnamefont
  {Subedi}}\ and\ \bibinfo {author} {\bibfnamefont {L.}~\bibnamefont {Boeri}},\
  }\href {\doibase 10.1103/PhysRevB.84.020508} {\bibfield  {journal} {\bibinfo
  {journal} {Phys. Rev. B}\ }\textbf {\bibinfo {volume} {84}},\ \bibinfo
  {pages} {020508} (\bibinfo {year} {2011})}\BibitemShut {NoStop}%
\bibitem [{\citenamefont {Casula}\ \emph {et~al.}(2011)\citenamefont {Casula},
  \citenamefont {Calandra}, \citenamefont {Profeta},\ and\ \citenamefont
  {Mauri}}]{Casula2011}%
  \BibitemOpen
  \bibfield  {author} {\bibinfo {author} {\bibfnamefont {M.}~\bibnamefont
  {Casula}}, \bibinfo {author} {\bibfnamefont {M.}~\bibnamefont {Calandra}},
  \bibinfo {author} {\bibfnamefont {G.}~\bibnamefont {Profeta}}, \ and\
  \bibinfo {author} {\bibfnamefont {F.}~\bibnamefont {Mauri}},\ }\href
  {\doibase 10.1103/PhysRevLett.107.137006} {\bibfield  {journal} {\bibinfo
  {journal} {Phys. Rev. Lett.}\ }\textbf {\bibinfo {volume} {107}},\ \bibinfo
  {pages} {137006} (\bibinfo {year} {2011})}\BibitemShut {NoStop}%
\bibitem [{\citenamefont {Gor{\textquoteright}kov}(2016)}]{Gorkov2016PNAS}%
  \BibitemOpen
  \bibfield  {author} {\bibinfo {author} {\bibfnamefont {L.~P.}\ \bibnamefont
  {Gor{\textquoteright}kov}},\ }\href {\doibase 10.1073/pnas.1604145113}
  {\bibfield  {journal} {\bibinfo  {journal} {Proceedings of the National
  Academy of Sciences}\ }\textbf {\bibinfo {volume} {113}},\ \bibinfo {pages}
  {4646} (\bibinfo {year} {2016})}\BibitemShut {NoStop}%
\bibitem [{\citenamefont {{W{\"o}lfle}}\ and\ \citenamefont
  {{Balatsky}}(2018)}]{Wolfle2018}%
  \BibitemOpen
  \bibfield  {author} {\bibinfo {author} {\bibfnamefont {P.}~\bibnamefont
  {{W{\"o}lfle}}}\ and\ \bibinfo {author} {\bibfnamefont {A.~V.}\ \bibnamefont
  {{Balatsky}}},\ }\href@noop {} {\bibfield  {journal} {\bibinfo  {journal}
  {ArXiv e-prints}\ } (\bibinfo {year} {2018})},\ \Eprint
  {http://arxiv.org/abs/1803.06993} {arXiv:1803.06993 [cond-mat.supr-con]}
  \BibitemShut {NoStop}%
\bibitem [{\citenamefont {Gor'kov}(2016)}]{Gorkov2016PRB}%
  \BibitemOpen
  \bibfield  {author} {\bibinfo {author} {\bibfnamefont {L.~P.}\ \bibnamefont
  {Gor'kov}},\ }\href {\doibase 10.1103/PhysRevB.93.054517} {\bibfield
  {journal} {\bibinfo  {journal} {Phys. Rev. B}\ }\textbf {\bibinfo {volume}
  {93}},\ \bibinfo {pages} {054517} (\bibinfo {year} {2016})}\BibitemShut
  {NoStop}%
\bibitem [{\citenamefont {Lee}\ \emph {et~al.}(2014)\citenamefont {Lee},
  \citenamefont {Schmitt}, \citenamefont {Moore}, \citenamefont {Johnston},
  \citenamefont {Cui}, \citenamefont {Li}, \citenamefont {Yi}, \citenamefont
  {Liu}, \citenamefont {Hashimoto}, \citenamefont {Zhang}, \citenamefont {Lu},
  \citenamefont {Devereaux}, \citenamefont {Lee},\ and\ \citenamefont
  {Shen}}]{Lee2014}%
  \BibitemOpen
  \bibfield  {author} {\bibinfo {author} {\bibfnamefont {J.~J.}\ \bibnamefont
  {Lee}}, \bibinfo {author} {\bibfnamefont {F.~T.}\ \bibnamefont {Schmitt}},
  \bibinfo {author} {\bibfnamefont {R.~G.}\ \bibnamefont {Moore}}, \bibinfo
  {author} {\bibfnamefont {S.}~\bibnamefont {Johnston}}, \bibinfo {author}
  {\bibfnamefont {Y.-T.}\ \bibnamefont {Cui}}, \bibinfo {author} {\bibfnamefont
  {W.}~\bibnamefont {Li}}, \bibinfo {author} {\bibfnamefont {M.}~\bibnamefont
  {Yi}}, \bibinfo {author} {\bibfnamefont {Z.~K.}\ \bibnamefont {Liu}},
  \bibinfo {author} {\bibfnamefont {M.}~\bibnamefont {Hashimoto}}, \bibinfo
  {author} {\bibfnamefont {Y.}~\bibnamefont {Zhang}}, \bibinfo {author}
  {\bibfnamefont {D.~H.}\ \bibnamefont {Lu}}, \bibinfo {author} {\bibfnamefont
  {T.~P.}\ \bibnamefont {Devereaux}}, \bibinfo {author} {\bibfnamefont {D.-H.}\
  \bibnamefont {Lee}}, \ and\ \bibinfo {author} {\bibfnamefont {Z.-X.}\
  \bibnamefont {Shen}},\ }\href {https://doi.org/10.1038/nature13894}
  {\bibfield  {journal} {\bibinfo  {journal} {Nature}\ }\textbf {\bibinfo
  {volume} {515}},\ \bibinfo {pages} {245} (\bibinfo {year}
  {2014})}\BibitemShut {NoStop}%
\bibitem [{\citenamefont {Benedetti}\ and\ \citenamefont
  {Zeyher}(1998)}]{Benedetti1998}%
  \BibitemOpen
  \bibfield  {author} {\bibinfo {author} {\bibfnamefont {P.}~\bibnamefont
  {Benedetti}}\ and\ \bibinfo {author} {\bibfnamefont {R.}~\bibnamefont
  {Zeyher}},\ }\href {\doibase 10.1103/PhysRevB.58.14320} {\bibfield  {journal}
  {\bibinfo  {journal} {Phys. Rev. B}\ }\textbf {\bibinfo {volume} {58}},\
  \bibinfo {pages} {14320} (\bibinfo {year} {1998})}\BibitemShut {NoStop}%
\bibitem [{\citenamefont {Alexandrov}(2001)}]{Alexandrov2001}%
  \BibitemOpen
  \bibfield  {author} {\bibinfo {author} {\bibfnamefont {A.~S.}\ \bibnamefont
  {Alexandrov}},\ }\href {http://stacks.iop.org/0295-5075/56/i=1/a=092}
  {\bibfield  {journal} {\bibinfo  {journal} {EPL (Europhysics Letters)}\
  }\textbf {\bibinfo {volume} {56}},\ \bibinfo {pages} {92} (\bibinfo {year}
  {2001})}\BibitemShut {NoStop}%
\bibitem [{\citenamefont {Serene}\ and\ \citenamefont
  {Hess}(1991)}]{Serene1991}%
  \BibitemOpen
  \bibfield  {author} {\bibinfo {author} {\bibfnamefont {J.~W.}\ \bibnamefont
  {Serene}}\ and\ \bibinfo {author} {\bibfnamefont {D.~W.}\ \bibnamefont
  {Hess}},\ }\href {\doibase 10.1103/PhysRevB.44.3391} {\bibfield  {journal}
  {\bibinfo  {journal} {Phys. Rev. B}\ }\textbf {\bibinfo {volume} {44}},\
  \bibinfo {pages} {3391} (\bibinfo {year} {1991})}\BibitemShut {NoStop}%
\bibitem [{\citenamefont {Deisz}\ \emph {et~al.}(2002)\citenamefont {Deisz},
  \citenamefont {Hess},\ and\ \citenamefont {Serene}}]{Deisz2002}%
  \BibitemOpen
  \bibfield  {author} {\bibinfo {author} {\bibfnamefont {J.~J.}\ \bibnamefont
  {Deisz}}, \bibinfo {author} {\bibfnamefont {D.~W.}\ \bibnamefont {Hess}}, \
  and\ \bibinfo {author} {\bibfnamefont {J.~W.}\ \bibnamefont {Serene}},\
  }\href {\doibase 10.1103/PhysRevB.66.014539} {\bibfield  {journal} {\bibinfo
  {journal} {Phys. Rev. B}\ }\textbf {\bibinfo {volume} {66}},\ \bibinfo
  {pages} {014539} (\bibinfo {year} {2002})}\BibitemShut {NoStop}%
\bibitem [{\citenamefont {Georges}\ \emph {et~al.}(1996)\citenamefont
  {Georges}, \citenamefont {Kotliar}, \citenamefont {Krauth},\ and\
  \citenamefont {Rozenberg}}]{Georges1996}%
  \BibitemOpen
  \bibfield  {author} {\bibinfo {author} {\bibfnamefont {A.}~\bibnamefont
  {Georges}}, \bibinfo {author} {\bibfnamefont {G.}~\bibnamefont {Kotliar}},
  \bibinfo {author} {\bibfnamefont {W.}~\bibnamefont {Krauth}}, \ and\ \bibinfo
  {author} {\bibfnamefont {M.~J.}\ \bibnamefont {Rozenberg}},\ }\href {\doibase
  10.1103/RevModPhys.68.13} {\bibfield  {journal} {\bibinfo  {journal} {Rev.
  Mod. Phys.}\ }\textbf {\bibinfo {volume} {68}},\ \bibinfo {pages} {13}
  (\bibinfo {year} {1996})}\BibitemShut {NoStop}%
\bibitem [{Note1()}]{Note1}%
  \BibitemOpen
  \bibinfo {note} {The {\unhbox \voidb@x \hbox {\protect \textsc {Matlab}}}
  code is released at \protect \url
  {https://github.com/johnstonResearchGroup/Migdal}}\BibitemShut {NoStop}%
\bibitem [{\citenamefont {Bulut}\ and\ \citenamefont
  {Scalapino}(1996)}]{Bulut1996}%
  \BibitemOpen
  \bibfield  {author} {\bibinfo {author} {\bibfnamefont {N.}~\bibnamefont
  {Bulut}}\ and\ \bibinfo {author} {\bibfnamefont {D.~J.}\ \bibnamefont
  {Scalapino}},\ }\href {\doibase 10.1103/PhysRevB.54.14971} {\bibfield
  {journal} {\bibinfo  {journal} {Phys. Rev. B}\ }\textbf {\bibinfo {volume}
  {54}},\ \bibinfo {pages} {14971} (\bibinfo {year} {1996})}\BibitemShut
  {NoStop}%
\bibitem [{\citenamefont {Sandvik}\ \emph {et~al.}(2004)\citenamefont
  {Sandvik}, \citenamefont {Scalapino},\ and\ \citenamefont
  {Bickers}}]{Sandvik2004}%
  \BibitemOpen
  \bibfield  {author} {\bibinfo {author} {\bibfnamefont {A.~W.}\ \bibnamefont
  {Sandvik}}, \bibinfo {author} {\bibfnamefont {D.~J.}\ \bibnamefont
  {Scalapino}}, \ and\ \bibinfo {author} {\bibfnamefont {N.~E.}\ \bibnamefont
  {Bickers}},\ }\href {\doibase 10.1103/PhysRevB.69.094523} {\bibfield
  {journal} {\bibinfo  {journal} {Phys. Rev. B}\ }\textbf {\bibinfo {volume}
  {69}},\ \bibinfo {pages} {094523} (\bibinfo {year} {2004})}\BibitemShut
  {NoStop}%
\bibitem [{\citenamefont {Johnston}\ \emph {et~al.}(2010)\citenamefont
  {Johnston}, \citenamefont {Vernay}, \citenamefont {Moritz}, \citenamefont
  {Shen}, \citenamefont {Nagaosa}, \citenamefont {Zaanen},\ and\ \citenamefont
  {Devereaux}}]{Johnston2010b}%
  \BibitemOpen
  \bibfield  {author} {\bibinfo {author} {\bibfnamefont {S.}~\bibnamefont
  {Johnston}}, \bibinfo {author} {\bibfnamefont {F.}~\bibnamefont {Vernay}},
  \bibinfo {author} {\bibfnamefont {B.}~\bibnamefont {Moritz}}, \bibinfo
  {author} {\bibfnamefont {Z.-X.}\ \bibnamefont {Shen}}, \bibinfo {author}
  {\bibfnamefont {N.}~\bibnamefont {Nagaosa}}, \bibinfo {author} {\bibfnamefont
  {J.}~\bibnamefont {Zaanen}}, \ and\ \bibinfo {author} {\bibfnamefont {T.~P.}\
  \bibnamefont {Devereaux}},\ }\href {\doibase 10.1103/PhysRevB.82.064513}
  {\bibfield  {journal} {\bibinfo  {journal} {Phys. Rev. B}\ }\textbf {\bibinfo
  {volume} {82}},\ \bibinfo {pages} {064513} (\bibinfo {year}
  {2010})}\BibitemShut {NoStop}%
\bibitem [{\citenamefont {{Schrodi}}\ \emph {et~al.}(2018)\citenamefont
  {{Schrodi}}, \citenamefont {{Aperis}},\ and\ \citenamefont
  {{Oppeneer}}}]{Schrodi2018}%
  \BibitemOpen
  \bibfield  {author} {\bibinfo {author} {\bibfnamefont {F.}~\bibnamefont
  {{Schrodi}}}, \bibinfo {author} {\bibfnamefont {A.}~\bibnamefont {{Aperis}}},
  \ and\ \bibinfo {author} {\bibfnamefont {P.~M.}\ \bibnamefont {{Oppeneer}}},\
  }\href@noop {} {\bibfield  {journal} {\bibinfo  {journal} {ArXiv e-prints}\ }
  (\bibinfo {year} {2018})},\ \Eprint {http://arxiv.org/abs/1810.01601}
  {arXiv:1810.01601 [cond-mat.supr-con]} \BibitemShut {NoStop}%
\bibitem [{\citenamefont {Anderson}(1965)}]{Anderson1965}%
  \BibitemOpen
  \bibfield  {author} {\bibinfo {author} {\bibfnamefont {D.~G.}\ \bibnamefont
  {Anderson}},\ }\href {\doibase 10.1145/321296.321305} {\bibfield  {journal}
  {\bibinfo  {journal} {J. ACM}\ }\textbf {\bibinfo {volume} {12}},\ \bibinfo
  {pages} {547} (\bibinfo {year} {1965})}\BibitemShut {NoStop}%
\bibitem [{\citenamefont {Eyert}(1996)}]{Eyert1996}%
  \BibitemOpen
  \bibfield  {author} {\bibinfo {author} {\bibfnamefont {V.}~\bibnamefont
  {Eyert}},\ }\href {\doibase 10.1006/jcph.1996.0059} {\bibfield  {journal}
  {\bibinfo  {journal} {J. Comput. Phys.}\ }\textbf {\bibinfo {volume} {124}},\
  \bibinfo {pages} {271} (\bibinfo {year} {1996})}\BibitemShut {NoStop}%
\bibitem [{\citenamefont {Walker}\ and\ \citenamefont {Ni}(2011)}]{Walker2011}%
  \BibitemOpen
  \bibfield  {author} {\bibinfo {author} {\bibfnamefont {H.~F.}\ \bibnamefont
  {Walker}}\ and\ \bibinfo {author} {\bibfnamefont {P.}~\bibnamefont {Ni}},\
  }\href {\doibase 10.1137/10078356X} {\bibfield  {journal} {\bibinfo
  {journal} {SIAM J. Numer. Anal.}\ }\textbf {\bibinfo {volume} {49}},\
  \bibinfo {pages} {1715} (\bibinfo {year} {2011})}\BibitemShut {NoStop}%
\bibitem [{\citenamefont {Kosterlitz}\ and\ \citenamefont
  {Thouless}(1973)}]{Kosterlitz1973}%
  \BibitemOpen
  \bibfield  {author} {\bibinfo {author} {\bibfnamefont {J.~M.}\ \bibnamefont
  {Kosterlitz}}\ and\ \bibinfo {author} {\bibfnamefont {D.~J.}\ \bibnamefont
  {Thouless}},\ }\href {http://stacks.iop.org/0022-3719/6/i=7/a=010} {\bibfield
   {journal} {\bibinfo  {journal} {J. Phys. C: Solid State Phys.}\ }\textbf
  {\bibinfo {volume} {6}},\ \bibinfo {pages} {1181} (\bibinfo {year}
  {1973})}\BibitemShut {NoStop}%
\bibitem [{\citenamefont {Hirsch}\ and\ \citenamefont
  {Scalapino}(1986)}]{Hirsch1986a}%
  \BibitemOpen
  \bibfield  {author} {\bibinfo {author} {\bibfnamefont {J.~E.}\ \bibnamefont
  {Hirsch}}\ and\ \bibinfo {author} {\bibfnamefont {D.~J.}\ \bibnamefont
  {Scalapino}},\ }\href {\doibase 10.1103/PhysRevLett.56.2732} {\bibfield
  {journal} {\bibinfo  {journal} {Phys. Rev. Lett.}\ }\textbf {\bibinfo
  {volume} {56}},\ \bibinfo {pages} {2732} (\bibinfo {year}
  {1986})}\BibitemShut {NoStop}%
\bibitem [{\citenamefont {Freericks}\ \emph {et~al.}(1996)\citenamefont
  {Freericks}, \citenamefont {Jarrell},\ and\ \citenamefont
  {Mahan}}]{Freericks1996a}%
  \BibitemOpen
  \bibfield  {author} {\bibinfo {author} {\bibfnamefont {J.~K.}\ \bibnamefont
  {Freericks}}, \bibinfo {author} {\bibfnamefont {M.}~\bibnamefont {Jarrell}},
  \ and\ \bibinfo {author} {\bibfnamefont {G.~D.}\ \bibnamefont {Mahan}},\
  }\href {\doibase 10.1103/PhysRevLett.77.4588} {\bibfield  {journal} {\bibinfo
   {journal} {Phys. Rev. Lett.}\ }\textbf {\bibinfo {volume} {77}},\ \bibinfo
  {pages} {4588} (\bibinfo {year} {1996})}\BibitemShut {NoStop}%
\bibitem [{\citenamefont {Chandler}\ \emph {et~al.}(2016)\citenamefont
  {Chandler}, \citenamefont {Prosko},\ and\ \citenamefont
  {Marsiglio}}]{Chandler2016}%
  \BibitemOpen
  \bibfield  {author} {\bibinfo {author} {\bibfnamefont {C.~J.}\ \bibnamefont
  {Chandler}}, \bibinfo {author} {\bibfnamefont {C.}~\bibnamefont {Prosko}}, \
  and\ \bibinfo {author} {\bibfnamefont {F.}~\bibnamefont {Marsiglio}},\ }\href
  {\doibase 10.1038/srep32591} {\bibfield  {journal} {\bibinfo  {journal} {Sci.
  Rep.}\ }\textbf {\bibinfo {volume} {6}},\ \bibinfo {pages} {32591} (\bibinfo
  {year} {2016})}\BibitemShut {NoStop}%
\bibitem [{\citenamefont {Heil}\ \emph {et~al.}(2017)\citenamefont {Heil},
  \citenamefont {Ponc\'e}, \citenamefont {Lambert}, \citenamefont {Schlipf},
  \citenamefont {Margine},\ and\ \citenamefont {Giustino}}]{Heil2017}%
  \BibitemOpen
  \bibfield  {author} {\bibinfo {author} {\bibfnamefont {C.}~\bibnamefont
  {Heil}}, \bibinfo {author} {\bibfnamefont {S.}~\bibnamefont {Ponc\'e}},
  \bibinfo {author} {\bibfnamefont {H.}~\bibnamefont {Lambert}}, \bibinfo
  {author} {\bibfnamefont {M.}~\bibnamefont {Schlipf}}, \bibinfo {author}
  {\bibfnamefont {E.~R.}\ \bibnamefont {Margine}}, \ and\ \bibinfo {author}
  {\bibfnamefont {F.}~\bibnamefont {Giustino}},\ }\href {\doibase
  10.1103/PhysRevLett.119.087003} {\bibfield  {journal} {\bibinfo  {journal}
  {Phys. Rev. Lett.}\ }\textbf {\bibinfo {volume} {119}},\ \bibinfo {pages}
  {087003} (\bibinfo {year} {2017})}\BibitemShut {NoStop}%
\bibitem [{\citenamefont {Vidberg}\ and\ \citenamefont
  {Serene}(1977)}]{Vidberg1977}%
  \BibitemOpen
  \bibfield  {author} {\bibinfo {author} {\bibfnamefont {H.~J.}\ \bibnamefont
  {Vidberg}}\ and\ \bibinfo {author} {\bibfnamefont {J.~W.}\ \bibnamefont
  {Serene}},\ }\href {\doibase 10.1007/BF00655090} {\bibfield  {journal}
  {\bibinfo  {journal} {J. Low Temp. Phys.}\ }\textbf {\bibinfo {volume}
  {29}},\ \bibinfo {pages} {179} (\bibinfo {year} {1977})}\BibitemShut
  {NoStop}%
\bibitem [{\citenamefont {Li}\ \emph {et~al.}(2015)\citenamefont {Li},
  \citenamefont {Nowadnick},\ and\ \citenamefont {Johnston}}]{Li2015a}%
  \BibitemOpen
  \bibfield  {author} {\bibinfo {author} {\bibfnamefont {S.}~\bibnamefont
  {Li}}, \bibinfo {author} {\bibfnamefont {E.~A.}\ \bibnamefont {Nowadnick}}, \
  and\ \bibinfo {author} {\bibfnamefont {S.}~\bibnamefont {Johnston}},\ }\href
  {\doibase 10.1103/PhysRevB.92.064301} {\bibfield  {journal} {\bibinfo
  {journal} {Phys. Rev. B}\ }\textbf {\bibinfo {volume} {92}},\ \bibinfo
  {pages} {064301} (\bibinfo {year} {2015})}\BibitemShut {NoStop}%
\bibitem [{Note2()}]{Note2}%
  \BibitemOpen
  \bibinfo {note} {As indicated by the blue line in {\protect Fig.~\ref
  {fig:Tc_finite_size}}, this size is sufficiently large to discuss the
  superconductivity at the fillings considered.}\BibitemShut {Stop}%
\bibitem [{Note3()}]{Note3}%
  \BibitemOpen
  \bibinfo {note} {This temperature is slightly greater the {\protect $ T_{c}
  $} dome. We have checked by solving the eigenvalue problem that the pairing
  vertex at this {\protect $ T $} is enhanced proportional to the {\protect $
  T_{c} $} at each respective filling considered.}\BibitemShut {Stop}%
\bibitem [{\citenamefont {Allen}\ and\ \citenamefont
  {Dynes}(1975)}]{Allen1975}%
  \BibitemOpen
  \bibfield  {author} {\bibinfo {author} {\bibfnamefont {P.~B.}\ \bibnamefont
  {Allen}}\ and\ \bibinfo {author} {\bibfnamefont {R.~C.}\ \bibnamefont
  {Dynes}},\ }\href {\doibase 10.1103/PhysRevB.12.905} {\bibfield  {journal}
  {\bibinfo  {journal} {Phys. Rev. B}\ }\textbf {\bibinfo {volume} {12}},\
  \bibinfo {pages} {905} (\bibinfo {year} {1975})}\BibitemShut {NoStop}%
\bibitem [{\citenamefont {Mazin}(2015{\natexlab{b}})}]{Mazin2015}%
  \BibitemOpen
  \bibfield  {author} {\bibinfo {author} {\bibfnamefont {I.~I.}\ \bibnamefont
  {Mazin}},\ }\href {http://dx.doi.org/10.1038/nmat4371} {\bibfield  {journal}
  {\bibinfo  {journal} {Nature Materials}\ }\textbf {\bibinfo {volume} {14}},\
  \bibinfo {pages} {755} (\bibinfo {year} {2015}{\natexlab{b}})}\BibitemShut
  {NoStop}%
\bibitem [{\citenamefont {{Langmann}}\ \emph {et~al.}(2018)\citenamefont
  {{Langmann}}, \citenamefont {{Triola}},\ and\ \citenamefont
  {{Balatsky}}}]{Langmann2018}%
  \BibitemOpen
  \bibfield  {author} {\bibinfo {author} {\bibfnamefont {E.}~\bibnamefont
  {{Langmann}}}, \bibinfo {author} {\bibfnamefont {C.}~\bibnamefont
  {{Triola}}}, \ and\ \bibinfo {author} {\bibfnamefont {A.~V.}\ \bibnamefont
  {{Balatsky}}},\ }\href@noop {} {\bibfield  {journal} {\bibinfo  {journal}
  {ArXiv e-prints}\ } (\bibinfo {year} {2018})},\ \Eprint
  {http://arxiv.org/abs/1810.03349} {arXiv:1810.03349 [cond-mat.supr-con]}
  \BibitemShut {NoStop}%
\bibitem [{\citenamefont {Freericks}\ and\ \citenamefont
  {Mahan}(1996)}]{Freericks1996}%
  \BibitemOpen
  \bibfield  {author} {\bibinfo {author} {\bibfnamefont {J.~K.}\ \bibnamefont
  {Freericks}}\ and\ \bibinfo {author} {\bibfnamefont {G.~D.}\ \bibnamefont
  {Mahan}},\ }\href {\doibase 10.1103/PhysRevB.54.9372} {\bibfield  {journal}
  {\bibinfo  {journal} {Phys. Rev. B}\ }\textbf {\bibinfo {volume} {54}},\
  \bibinfo {pages} {9372} (\bibinfo {year} {1996})}\BibitemShut {NoStop}%
\bibitem [{\citenamefont {Gubernatis}\ \emph {et~al.}(1985)\citenamefont
  {Gubernatis}, \citenamefont {Scalapino}, \citenamefont {Sugar},\ and\
  \citenamefont {Toussaint}}]{Gubernatis1985}%
  \BibitemOpen
  \bibfield  {author} {\bibinfo {author} {\bibfnamefont {J.~E.}\ \bibnamefont
  {Gubernatis}}, \bibinfo {author} {\bibfnamefont {D.~J.}\ \bibnamefont
  {Scalapino}}, \bibinfo {author} {\bibfnamefont {R.~L.}\ \bibnamefont
  {Sugar}}, \ and\ \bibinfo {author} {\bibfnamefont {W.~D.}\ \bibnamefont
  {Toussaint}},\ }\href {\doibase 10.1103/PhysRevB.32.103} {\bibfield
  {journal} {\bibinfo  {journal} {Phys. Rev. B}\ }\textbf {\bibinfo {volume}
  {32}},\ \bibinfo {pages} {103} (\bibinfo {year} {1985})}\BibitemShut
  {NoStop}%
\bibitem [{\citenamefont {Lin}\ and\ \citenamefont {Hirsch}(1986)}]{Lin1986}%
  \BibitemOpen
  \bibfield  {author} {\bibinfo {author} {\bibfnamefont {H.~Q.}\ \bibnamefont
  {Lin}}\ and\ \bibinfo {author} {\bibfnamefont {J.~E.}\ \bibnamefont
  {Hirsch}},\ }\href {\doibase 10.1103/PhysRevB.33.8155} {\bibfield  {journal}
  {\bibinfo  {journal} {Phys. Rev. B}\ }\textbf {\bibinfo {volume} {33}},\
  \bibinfo {pages} {8155} (\bibinfo {year} {1986})}\BibitemShut {NoStop}%
\bibitem [{\citenamefont {Micnas}\ \emph {et~al.}(1989)\citenamefont {Micnas},
  \citenamefont {Ranninger},\ and\ \citenamefont {Robaszkiewicz}}]{Micnas1989}%
  \BibitemOpen
  \bibfield  {author} {\bibinfo {author} {\bibfnamefont {R.}~\bibnamefont
  {Micnas}}, \bibinfo {author} {\bibfnamefont {J.}~\bibnamefont {Ranninger}}, \
  and\ \bibinfo {author} {\bibfnamefont {S.}~\bibnamefont {Robaszkiewicz}},\
  }\href {\doibase 10.1103/PhysRevB.39.11653} {\bibfield  {journal} {\bibinfo
  {journal} {Phys. Rev. B}\ }\textbf {\bibinfo {volume} {39}},\ \bibinfo
  {pages} {11653} (\bibinfo {year} {1989})}\BibitemShut {NoStop}%
\bibitem [{\citenamefont {Macridin}\ \emph {et~al.}(2006)\citenamefont
  {Macridin}, \citenamefont {Jarrell},\ and\ \citenamefont
  {Maier}}]{Macridin2006}%
  \BibitemOpen
  \bibfield  {author} {\bibinfo {author} {\bibfnamefont {A.}~\bibnamefont
  {Macridin}}, \bibinfo {author} {\bibfnamefont {M.}~\bibnamefont {Jarrell}}, \
  and\ \bibinfo {author} {\bibfnamefont {T.}~\bibnamefont {Maier}},\ }\href
  {\doibase 10.1103/PhysRevB.74.085104} {\bibfield  {journal} {\bibinfo
  {journal} {Phys. Rev. B}\ }\textbf {\bibinfo {volume} {74}},\ \bibinfo
  {pages} {085104} (\bibinfo {year} {2006})}\BibitemShut {NoStop}%
\bibitem [{\citenamefont {Aichhorn}\ \emph {et~al.}(2007)\citenamefont
  {Aichhorn}, \citenamefont {Arrigoni}, \citenamefont {Potthoff},\ and\
  \citenamefont {Hanke}}]{Aichhorn2007}%
  \BibitemOpen
  \bibfield  {author} {\bibinfo {author} {\bibfnamefont {M.}~\bibnamefont
  {Aichhorn}}, \bibinfo {author} {\bibfnamefont {E.}~\bibnamefont {Arrigoni}},
  \bibinfo {author} {\bibfnamefont {M.}~\bibnamefont {Potthoff}}, \ and\
  \bibinfo {author} {\bibfnamefont {W.}~\bibnamefont {Hanke}},\ }\href
  {\doibase 10.1103/PhysRevB.76.224509} {\bibfield  {journal} {\bibinfo
  {journal} {Phys. Rev. B}\ }\textbf {\bibinfo {volume} {76}},\ \bibinfo
  {pages} {224509} (\bibinfo {year} {2007})}\BibitemShut {NoStop}%
\bibitem [{\citenamefont {Castellani}\ \emph {et~al.}(1995)\citenamefont
  {Castellani}, \citenamefont {Di~Castro},\ and\ \citenamefont
  {Grilli}}]{Castellani1995}%
  \BibitemOpen
  \bibfield  {author} {\bibinfo {author} {\bibfnamefont {C.}~\bibnamefont
  {Castellani}}, \bibinfo {author} {\bibfnamefont {C.}~\bibnamefont
  {Di~Castro}}, \ and\ \bibinfo {author} {\bibfnamefont {M.}~\bibnamefont
  {Grilli}},\ }\href {\doibase 10.1103/PhysRevLett.75.4650} {\bibfield
  {journal} {\bibinfo  {journal} {Phys. Rev. Lett.}\ }\textbf {\bibinfo
  {volume} {75}},\ \bibinfo {pages} {4650} (\bibinfo {year}
  {1995})}\BibitemShut {NoStop}%
\bibitem [{\citenamefont {Karakuzu}\ \emph {et~al.}(2017)\citenamefont
  {Karakuzu}, \citenamefont {Tocchio}, \citenamefont {Sorella},\ and\
  \citenamefont {Becca}}]{Karakuzu2017}%
  \BibitemOpen
  \bibfield  {author} {\bibinfo {author} {\bibfnamefont {S.}~\bibnamefont
  {Karakuzu}}, \bibinfo {author} {\bibfnamefont {L.~F.}\ \bibnamefont
  {Tocchio}}, \bibinfo {author} {\bibfnamefont {S.}~\bibnamefont {Sorella}}, \
  and\ \bibinfo {author} {\bibfnamefont {F.}~\bibnamefont {Becca}},\ }\href
  {\doibase 10.1103/PhysRevB.96.205145} {\bibfield  {journal} {\bibinfo
  {journal} {Phys. Rev. B}\ }\textbf {\bibinfo {volume} {96}},\ \bibinfo
  {pages} {205145} (\bibinfo {year} {2017})}\BibitemShut {NoStop}%
\bibitem [{\citenamefont {Varelogiannis}(1998)}]{Varelogiannis1998}%
  \BibitemOpen
  \bibfield  {author} {\bibinfo {author} {\bibfnamefont {G.}~\bibnamefont
  {Varelogiannis}},\ }\href {\doibase 10.1103/PhysRevB.57.13743} {\bibfield
  {journal} {\bibinfo  {journal} {Phys. Rev. B}\ }\textbf {\bibinfo {volume}
  {57}},\ \bibinfo {pages} {13743} (\bibinfo {year} {1998})}\BibitemShut
  {NoStop}%
\bibitem [{\citenamefont {Press}\ \emph {et~al.}(1992)\citenamefont {Press},
  \citenamefont {Teukolsky}, \citenamefont {Vetterling},\ and\ \citenamefont
  {Flannery}}]{Press1992}%
  \BibitemOpen
  \bibfield  {author} {\bibinfo {author} {\bibfnamefont {W.}~\bibnamefont
  {Press}}, \bibinfo {author} {\bibfnamefont {S.}~\bibnamefont {Teukolsky}},
  \bibinfo {author} {\bibfnamefont {W.}~\bibnamefont {Vetterling}}, \ and\
  \bibinfo {author} {\bibfnamefont {B.}~\bibnamefont {Flannery}},\ }\href@noop
  {} {\emph {\bibinfo {title} {Numerical Recipes in FORTRAN: The Art of
  Scientific Computing}}},\ \bibinfo {edition} {2nd}\ ed.\ (\bibinfo
  {publisher} {Cambridge University Press},\ \bibinfo {address} {Cambridge},\
  \bibinfo {year} {1992})\BibitemShut {NoStop}%
\bibitem [{\citenamefont {Davis}\ and\ \citenamefont
  {Rabinowitz}(1984)}]{Davis1984}%
  \BibitemOpen
  \bibfield  {author} {\bibinfo {author} {\bibfnamefont {P.~J.}\ \bibnamefont
  {Davis}}\ and\ \bibinfo {author} {\bibfnamefont {P.}~\bibnamefont
  {Rabinowitz}},\ }\href@noop {} {\emph {\bibinfo {title} {Methods of Numerical
  Integration}}},\ \bibinfo {edition} {2nd}\ ed.\ (\bibinfo  {publisher}
  {Academic Press},\ \bibinfo {address} {San Diego, CA},\ \bibinfo {year}
  {1984})\BibitemShut {NoStop}%
\bibitem [{\citenamefont {Carbotte}(1990)}]{Carbotte1990}%
  \BibitemOpen
  \bibfield  {author} {\bibinfo {author} {\bibfnamefont {J.~P.}\ \bibnamefont
  {Carbotte}},\ }\href {\doibase 10.1103/RevModPhys.62.1027} {\bibfield
  {journal} {\bibinfo  {journal} {Rev. Mod. Phys.}\ }\textbf {\bibinfo {volume}
  {62}},\ \bibinfo {pages} {1027} (\bibinfo {year} {1990})}\BibitemShut
  {NoStop}%
\end{thebibliography}%
	
\end{document}